\documentclass[apj,twocolumn,twocolappendix,floatfix]{openjournal}

\DeclareRobustCommand{\VAN}[3]{#2}
\let\VANthebibliography\thebibliography
\def\thebibliography{\DeclareRobustCommand{\VAN}[3]{##3}\VANthebibliography}

\usepackage{amsmath}	
\usepackage{amssymb}	
\usepackage{booktabs}
\usepackage{lipsum}
\usepackage[T1]{fontenc}
\usepackage{graphicx}	
\usepackage[breaklinks,colorlinks,citecolor=blue,urlcolor=blue,linkcolor=blue]{hyperref}
\usepackage{cleveref}
\usepackage{multirow}
\usepackage{newtxtext,newtxmath}
\usepackage{orcidlink}
\usepackage{pifont}
\usepackage[nobottomtitles]{titlesec}
\usepackage{natbib}
\usepackage{xparse}
\usepackage{ifthen}
\titleformat{\section}{\filcenter\MakeUppercase}{\thesection.}{0.5em}{}

\usepackage{savesym}
\savesymbol{tablenum}
\usepackage{siunitx}
\restoresymbol{SIX}{tablenum}
\usepackage{needspace}
\usepackage{overpic}
\usepackage{xcolor}

\newcommand{\PHOEBE}{\texttt{PHOEBE}}
\newcommand{\iSpec}{\texttt{iSpec}}
\newcommand{\Gaia}{\textit{Gaia}}
\newcommand{\TESS}{\textit{TESS}}

\newcommand{\cmark}{\ding{51}}%
\newcommand{\xmark}{\ding{55}}%
\newcommand{\teffratio}{$T_{\rm{eff,2}}/T_{\rm{eff,1}}$}

\NewDocumentCommand{\primarymass}{m}{
\ifthenelse{\equal{#1}{3328584192518301184}}{$M_1=2.142^{+0.009}_{-0.009}\ M_\odot$}
{\ifthenelse{\equal{#1}{3157581134781556480}}{$M_1=2.68^{+0.02}_{-0.03}\ M_\odot$}
{\ifthenelse{\equal{#1}{5388654952421552768}}{$M_1=1.056^{+0.006}_{-0.007}\ M_\odot$}
{\ifthenelse{\equal{#1}{5347923063144824448}}{$M_1=1.41^{+0.02}_{-0.02}\ M_\odot$}
{\ifthenelse{\equal{#1}{6188279177469245952}}{$M_1=1.12^{+0.02}_{-0.02}\ M_\odot$}
{\ifthenelse{\equal{#1}{5966976692576953216}}{$M_1=2.20^{+0.01}_{-0.01}\ M_\odot$}
{\ifthenelse{\equal{#1}{1969468871480562560}}{$M_1=3.49^{+0.02}_{-0.02}\ M_\odot$}
{\ifthenelse{\equal{#1}{2002164086682203904}}{$M_1=2.321^{+0.006}_{-0.006}\ M_\odot$}
{Invalid option}}}}}}}}
}

\NewDocumentCommand{\secondarymass}{m}{
\ifthenelse{\equal{#1}{3328584192518301184}}{$M_2=2.082^{+0.008}_{-0.009}\ M_\odot$}
{\ifthenelse{\equal{#1}{3157581134781556480}}{$M_2=2.67^{+0.03}_{-0.03}\ M_\odot$}
{\ifthenelse{\equal{#1}{5388654952421552768}}{$M_2=1.050^{+0.007}_{-0.008}\ M_\odot$}
{\ifthenelse{\equal{#1}{5347923063144824448}}{$M_2=1.40^{+0.02}_{-0.02}\ M_\odot$}
{\ifthenelse{\equal{#1}{6188279177469245952}}{$M_2=1.01^{+0.01}_{-0.01}\ M_\odot$}
{\ifthenelse{\equal{#1}{5966976692576953216}}{$M_2=2.26^{+0.02}_{-0.02}\ M_\odot$}
{\ifthenelse{\equal{#1}{1969468871480562560}}{$M_2=3.33^{+0.01}_{-0.01}\ M_\odot$}
{\ifthenelse{\equal{#1}{2002164086682203904}}{$M_2=2.318^{+0.009}_{-0.01}\ M_\odot$}
{Invalid option}}}}}}}}
}

\NewDocumentCommand{\primaryradius}{m}{
\ifthenelse{\equal{#1}{3328584192518301184}}{$R_1=9.5^{+0.8}_{-0.9}\ R_\odot$}
{\ifthenelse{\equal{#1}{3157581134781556480}}{$R_1=12.3^{+0.2}_{-0.2}\ R_\odot$}
{\ifthenelse{\equal{#1}{5388654952421552768}}{$R_1=4.52^{+0.09}_{-0.08}\ R_\odot$}
{\ifthenelse{\equal{#1}{5347923063144824448}}{$R_1=11.7^{+0.1}_{-0.1}\ R_\odot$}
{\ifthenelse{\equal{#1}{6188279177469245952}}{$R_1=3.1^{+0.2}_{-0.1}\ R_\odot$}
{\ifthenelse{\equal{#1}{5966976692576953216}}{$R_1=8.5^{+0.2}_{-0.2}\ R_\odot$}
{\ifthenelse{\equal{#1}{1969468871480562560}}{$R_1=21.2^{+0.5}_{-0.5}\ R_\odot$}
{\ifthenelse{\equal{#1}{2002164086682203904}}{$R_1=10.9^{+0.1}_{-0.1}\ R_\odot$}
{Invalid option}}}}}}}}
}

\NewDocumentCommand{\secondaryradius}{m}{
\ifthenelse{\equal{#1}{3328584192518301184}}{$R_2=3.1^{+0.4}_{-0.4}\ R_\odot$}
{\ifthenelse{\equal{#1}{3157581134781556480}}{$R_2=12.1^{+0.2}_{-0.2}\ R_\odot$}
{\ifthenelse{\equal{#1}{5388654952421552768}}{$R_2=2.12^{+0.06}_{-0.06}\ R_\odot$}
{\ifthenelse{\equal{#1}{5347923063144824448}}{$R_2=9.7^{+0.2}_{-0.2}\ R_\odot$}
{\ifthenelse{\equal{#1}{6188279177469245952}}{$R_2=1.05^{+0.08}_{-0.07}\ R_\odot$}
{\ifthenelse{\equal{#1}{5966976692576953216}}{$R_2=9.7^{+0.1}_{-0.2}\ R_\odot$}
{\ifthenelse{\equal{#1}{1969468871480562560}}{$R_2=21.4^{+0.4}_{-0.5}\ R_\odot$}
{\ifthenelse{\equal{#1}{2002164086682203904}}{$R_2=10.5^{+0.1}_{-0.1}\ R_\odot$}
{Invalid option}}}}}}}}
}

\defcitealias{Rowan22}{R22}

\begin{document}
\title[Eclipsing Red Giants]{Precise and Accurate Mass and Radius Measurements of Fifteen Galactic Red Giants in Detached Eclipsing Binaries}


\author{\vspace{-1.3cm}D. M. Rowan\,\orcidlink{0000-0003-2431-981X}$^{1,2}$,
        K. Z. Stanek\,\orcidlink{0009-0001-1470-8400}$^{1,2}$,
        C. S. Kochanek\ $^{1,2}$,
        Todd A. Thompson\,\orcidlink{0000-0003-2377-9574}$^{1,2,3}$,
        T. Jayasinghe\,\orcidlink{0000-0002-6244-477X}$^{4}$\\
        J. Blaum\,\orcidlink{0000-0003-1142-3095}$^{5}$,
        B. J. Fulton\,\orcidlink{0000-0003-3504-5316}$^{6}$,
        I. Ilyin\,\orcidlink{0000-0002-0551-046X}$^{7}$,
        H. Isaacson\,\orcidlink{0000-0002-0531-1073}$^{5}$,
        N. LeBaron\,\orcidlink{0000-0002-2249-0595}$^{5}$,
        Jessica R. Lu\,\orcidlink{0000-0001-9611-0009}$^{5}$,
        David V. Martin\,\orcidlink{0000-0002-7595-6360}$^{8}$
}

\affiliation{$^{1}$Department of Astronomy, The Ohio State University, 140 West 18th Avenue, Columbus, OH, 43210, USA \\
             $^{2}$Center for Cosmology and Astroparticle Physics, The Ohio State University, 191 W. Woodruff Avenue, Columbus, OH, 43210, USA \\
             $^{3}$Department of Physics, The Ohio State University, Columbus, Ohio, 43210, USA \\
             $^{4}$Independent Researcher, San Jose, California, USA \\
             $^{5}$Department of Astronomy, University of California Berkeley, Berkeley CA 94720, USA \\
             $^{6}$NASA Exoplanet Science Institute/Caltech-IPAC, Pasadena, CA 91125, USA \\
             $^{7}$Leibniz Institute for Astrophysics Potsdam (AIP), An der Sternwarte 16, D-14482 Potsdam, Germany \\
             $^{8}$Department of Physics and Astronomy, Tufts University, Medford, MA 02155, USA
}

\begin{abstract}
Precise and accurate mass and radius measurements of evolved stars are crucial to calibrating stellar models. Stars in detached eclipsing binaries (EBs) are excellent potential calibrators because their stellar parameters can be measured with fractional uncertainties of a few percent, independent of stellar models. The All-Sky Automated Survey for Supernovae (ASAS-SN) has identified tens of thousands of EBs, $>$35,000 of which were included in the ASAS-SN eclipsing binaries catalog. Here, we select eight EBs from this sample that contain giants based on their \Gaia{} colors and absolute magnitudes. We use LBT/PEPSI, APF, and CHIRON to obtain multi-epoch spectra of these binaries and measure their radial velocities using two-dimensional cross-correlation methods. We simultaneously fit the ASAS-SN light curves and the radial velocities with \PHOEBE{} to derive accurate and precise masses and radii with fractional uncertainties of $\lesssim 3\%$. For four systems, we also include Transiting Exoplanet Survey Satellite (\TESS{}) light curves in our \PHOEBE{} models, which significantly improves the radius determinations. In seven of our systems, both components have evolved off of the main sequence, and one system has a giant star component with a main sequence, Sun-like companion. Finally, we compare our mass and radius measurements to single-star evolutionary tracks and distinguish between systems that are first ascent red giant branch stars and those that are likely core helium-burning stars. 
\end{abstract}

\maketitle

\section{Introduction}


Detached eclipsing binaries (EBs) can be used to obtain the most accurate and precise measurements of stellar masses and radii without the need for stellar models. The EB light curve can be used to determine the orbital period, inclination, and the radii of the two stars relative to the orbital semimajor axis. With the addition of radial velocities for both components, the physical radii and masses can be determined with fractional uncertainties of a few percent \citep{Andersen91}.

Detached EBs have long served as powerful observational constraints to develop stellar evolution models, characterize exoplanets, and calibrate other methods of mass estimation. \citet{Torres10} compiled a sample of 95 detached eclipsing binaries with mass and radius uncertainties $\lesssim 3\%$ and used these measurements to derive empirical relations between spectroscopic parameters and masses and radii. Their results have been widely used to make comparisons with stellar models \citep[e.g.,][]{Paxton13} and to characterize exoplanets \citep[e.g.,][]{Siverd12, Rodriguez21, Duck22}. 

Despite the substantial expansion of photometric and spectroscopic surveys in recent years, only a few hundred detached eclipsing binaries have been fully characterized to precisions of a few percent \citep{Southworth15, Maxted23}. Furthermore, the distribution of stars with dynamical mass and radius measurements is not uniform across the Hertzsprung-Russell diagram. In particular, only $\sim$16\% of the eclipsing binaries included in the \citet{Southworth15} catalog and only three stars in the \citet{Torres10} catalog are significantly evolved off the main sequence. The lack of eclipsing red giants can be understood as an observational bias, since longer period orbits are required to host detached red giant binaries and such systems will have lower eclipse probabilities \citep{Beck24}. The vast majority of the evolved binaries with precise mass and radius measurements are in the Magellanic Clouds \citep[e.g.,][]{Graczyk12, Pietrzynski11, Pietrzynski13, Graczyk14, Graczyk18, Graczyk20}, and these systems serve as distance indicators in addition to being benchmarks for comparing with stellar models at low metallicity. There have been some eclipsing red giants identified in \textit{Kepler} and the All-Sky Automated Survey \citep[ASAS,][]{Pojmanski97} in the Milky Way field \citep[e.g.,][]{Brogaard18, Brogaard22, Ratajczak21}, but more systems are needed to make direct comparisons to stellar models at a range of masses and metallicities. 

There are a number of physical processes in stars where mass and radius measurements can be used to constrain theoretical models. For example, convective overshoot in stars is expected to bring extra hydrogen from convective envelopes into the core, increasing the core mass and extending the main sequence lifetime. Dynamical mass and radius measurements for evolved stars can be used to determine the extent of this extra-mixing \citep[e.g.,][]{Garcia14, Claret16}. More generally, \citet{DelBurgo18} compared dynamical mass measurements to predicted masses from stellar models and found larger differences for subgiant and red giant stars than for main sequence stars. Once masses and radii are measured, stellar models can be used to infer stellar ages. Different methods of age determination can predict significantly different ages \citep[e.g.,][]{Gaulme22}, and dynamical mass and radius measurements from EBs are valuable for testing these models. Expanding the sample of evolved stars with dynamical mass measurements at a range of masses and metallicities will allow for more comprehensive comparisons to stellar models. 

\begin{table*}
    \centering
    \caption{Properties of the eight EBs on the giant branch. The orbital period is from \citetalias{Rowan22} and the distance is the \Gaia{} EDR3 \citet{BailerJones21} photogeometric distance. The absolute magnitude is based on this distance and {\tt mwdust} three-dimensional dust maps \citep{Bovy16}. The $N_{\rm{RV}}$ column is the number of RV measurements. Finally, the \TESS{} column reports which targets have eclipses observed in \TESS{} that are not cut off by an orbit/sector gap.}
    \sisetup{table-auto-round,
     group-digits=false}
    \setlength{\tabcolsep}{10pt}
    \begin{center}
        \begin{tabular}{l l S[table-format=3.4] S[table-format=2.4] S[table-format=2.1] S[table-format=2.5] c S[table-format=4.1] S[table-format=1.1] c}
\toprule
           {Source} & {Short Name} &       {RA} &      {DEC} &     {$G$} &  {Period} &  {$N_{\rm{RV}}$} &  {Distance} &   {$M_G$} & {\textit{TESS}} \\
{} & {} & {(deg)} & {(deg)} & {(mag)} & {(d)} & {} & {(pc)} & {(mag)} & {}\\ 
\midrule
3328584192518301184 &        J0611 &  92.830279 &   8.499277 & 12.308417 & 69.005794 &               16 & 2196.610110 &  0.246896 &          \cmark \\
3157581134781556480 &        J0656 & 104.077177 &   9.440764 & 12.152586 & 41.447208 &               13 & 3626.034180 & -0.792700 &          \xmark \\
5388654952421552768 &        J1108 & 167.003719 & -44.116354 & 11.693777 & 32.393783 &                7 &  685.921326 &  2.258671 &          \cmark \\
5347923063144824448 &        J1109 & 167.455428 & -52.172025 & 10.584274 & 31.751774 &                8 & 1129.595210 & -0.106012 &          \cmark \\
6188279177469245952 &        J1329 & 202.302775 & -28.556858 & 11.688045 & 37.335766 &                9 &  611.739136 &  2.564766 &          \xmark \\
5966976692576953216 &        J1705 & 256.394900 & -39.792751 & 11.632104 & 52.610909 &               10 & 1356.204830 &  0.194431 &          \xmark \\
1969468871480562560 &        J2107 & 316.861073 &  42.233722 & 11.877251 & 68.124909 &               12 & 2475.812010 & -1.517036 &          \xmark \\
2002164086682203904 &        J2236 & 339.119881 &  52.641188 & 11.628980 & 36.837454 &               13 & 2191.447750 & -0.519056 &          \cmark \\
\bottomrule
\end{tabular}

    \end{center}
    \label{tab:summary_table}
\end{table*}

Masses derived from eclipsing binaries can also be used to calibrate other mass estimation methods such as abundances \citep[e.g.,][]{Roberts24} or asteroseismology \citep[e.g.,][]{Hekker11}. The masses of evolved stars determined from eclipsing binaries provide benchmarks for asteroseismology where scaling relations are used to convert the oscillation frequencies into measure masses and radii \citep{Kjeldsen95}. \citet{Hekker10} used \textit{Kepler} photometry to identify oscillations in an eclipsing red giant and spectroscopic follow-up found that scaling relations were in agreement with the dynamical masses and radii \citep{Frandsen13, Themebl18}. \citet{Gaulme16} used a sample of 10 oscillating giants in \textit{Kepler} eclipsing binaries and found that radii and masses were typically overestimated by $\sim5\%$ and $\sim15\%$, respectively, when using asteroseismic scaling relations. Larger samples of oscillating red giant binaries have found similar mass and radius discrepancies \citep{Benbakoura21}, but there are some systems where the predictions from scaling relations have been found to match dynamical measurements if the reference frequency spacing $\Delta \nu$ is modified from the typical Solar value. Oscillating giants in EBs have been used to make modifications to standard scaling relations \citep{Kallinger18}, and larger samples of oscillating giants in eclipsing binaries are needed to better calibrate these scaling relations.

Large, all-sky surveys can be used as a starting point to considerably expand the sample of detached EBs with precise mass and radius measurements \citep[e.g.,][]{Prsa11,Prsa22}. More than 200,000 eclipsing binaries have been identified in light curves from the All-Sky Automated Survey for Supernovae \citep[ASAS-SN,][]{Shappee14, Kochanek17} using machine learning methods \citep{Jayasinghe19, Christy23}. \citet[][hereafter R22]{Rowan22} focused on the sample of detached eclipsing binaries from \citet{Jayasinghe19} and used PHysics Of Eclipsing BinariEs \citep[\PHOEBE{}][]{Prsa05, Conroy20} to model $>35,000$ of their light curves. The ASAS-SN EB catalog includes parameter measurements such as the orbital eccentricity and ratio of effective temperatures, as well as estimates of the evolutionary state of the photometric primary based on \Gaia{} photometry \citep{GaiaCollaboration21}, distances from \citet{BailerJones21}, and three-dimensional dust extinction maps from \citet{Bovy16}. More than 600 systems in the \citetalias{Rowan22} catalog were classified as red giants. 

Here, we have obtained spectroscopic follow-up of eight of these bright, double-lined spectroscopic (SB2) EBs. Section \S\ref{sec:target_selection} describes how the targets were selected and the ASAS-SN and \TESS{} light curves. Section \S\ref{sec:observations} describes the spectroscopic observations and the radial velocity measurements. We simultaneously fit the ASAS-SN light curves and radial velocity measurements in Section \S\ref{sec:phoebe}. In Section \S\ref{sec:tracks} we compare our derived masses and radii to evolutionary tracks, and we discuss the results for each target in \S\ref{sec:results}. 

\section{Target Selection and Photometric Observations}\label{sec:target_selection}

\begin{figure}
    \centering
    \includegraphics[width=\linewidth]{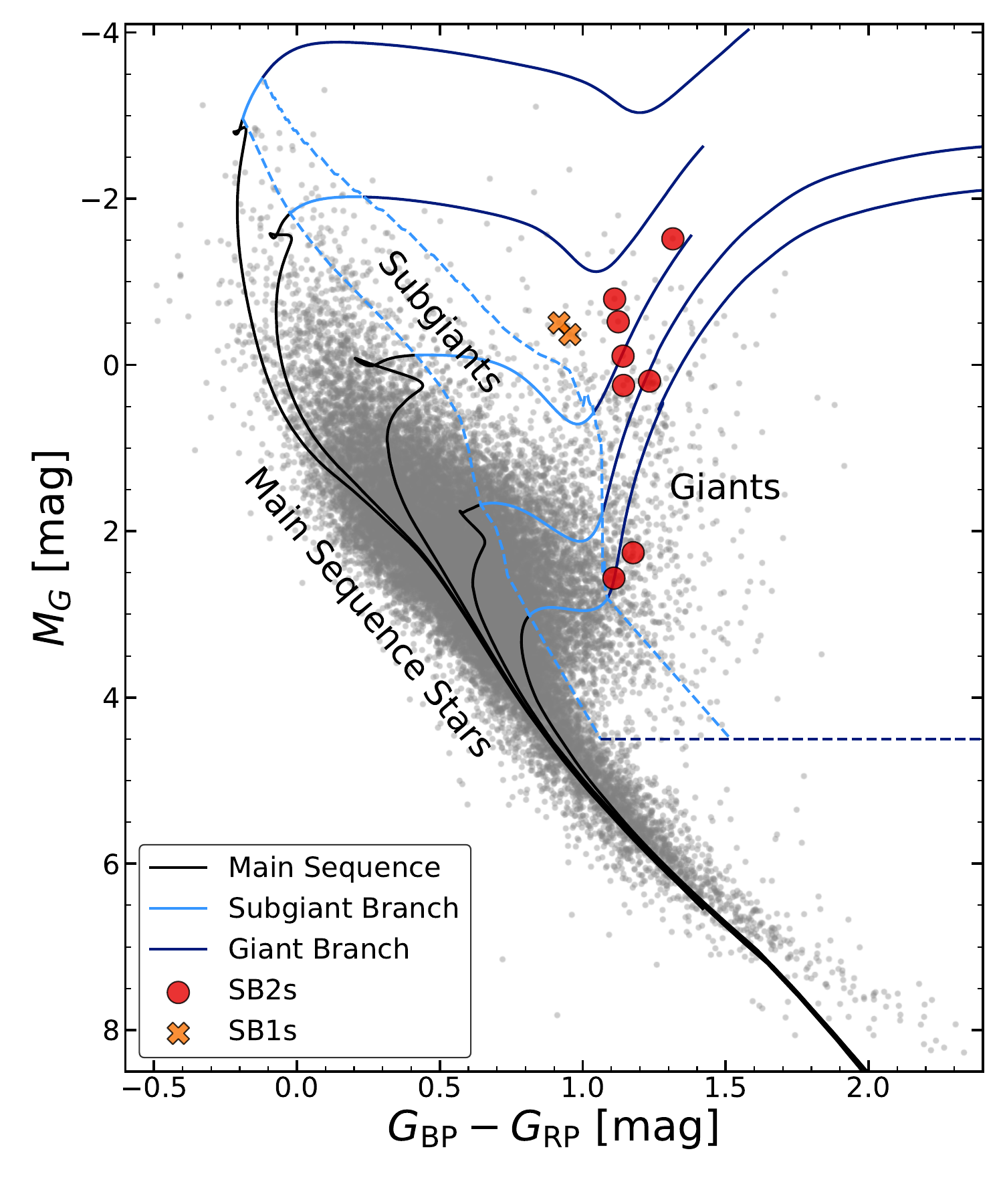}
    \caption{Extinction-corrected \Gaia{} color-magnitude diagram (CMD). The eight SB2s characterized here are marked in red. We also observed two systems that are SB1s (orange crosses). These are discussed in Appendix \ref{sec:sb1s}. These binaries are not resolved in \Gaia{}, so the CMD position represents the combined flux from both binary components. The gray background shows the ASAS-SN Eclipsing Binaries catalog from \citetalias{Rowan22}. The lines show how the binaries in the ASAS-SN catalog are classified as main sequence, subgiant, and giant binaries.}
    \label{fig:cmd}
\end{figure}

The ASAS-SN EB catalog\footnote{\url{https://asas-sn.osu.edu/binaries}}\citepalias{Rowan22} includes parameter estimates for more than 35,000 detached eclipsing binaries. The catalog also reports the evolutionary state of the photometric primary based on the extinction-corrected \Gaia{} color-magnitude diagram (CMD). We selected ten detached systems on the giant branch that are bright enough for easy spectroscopic follow-up ($V \lesssim 13$~mag). Figure \ref{fig:cmd} shows these targets on the \Gaia{} CMD and Table \ref{tab:summary_table} reports their parameters. Two systems were found to be single-lined spectroscopic binaries (SB1s). Since dynamical masses can only be determined for SB2s, we discuss these two systems in Appendix \ref{sec:sb1s} and focus the rest of our analysis and discussion on the eight SB2s.

\begin{figure*}
    \centering
    \includegraphics[width=\linewidth]{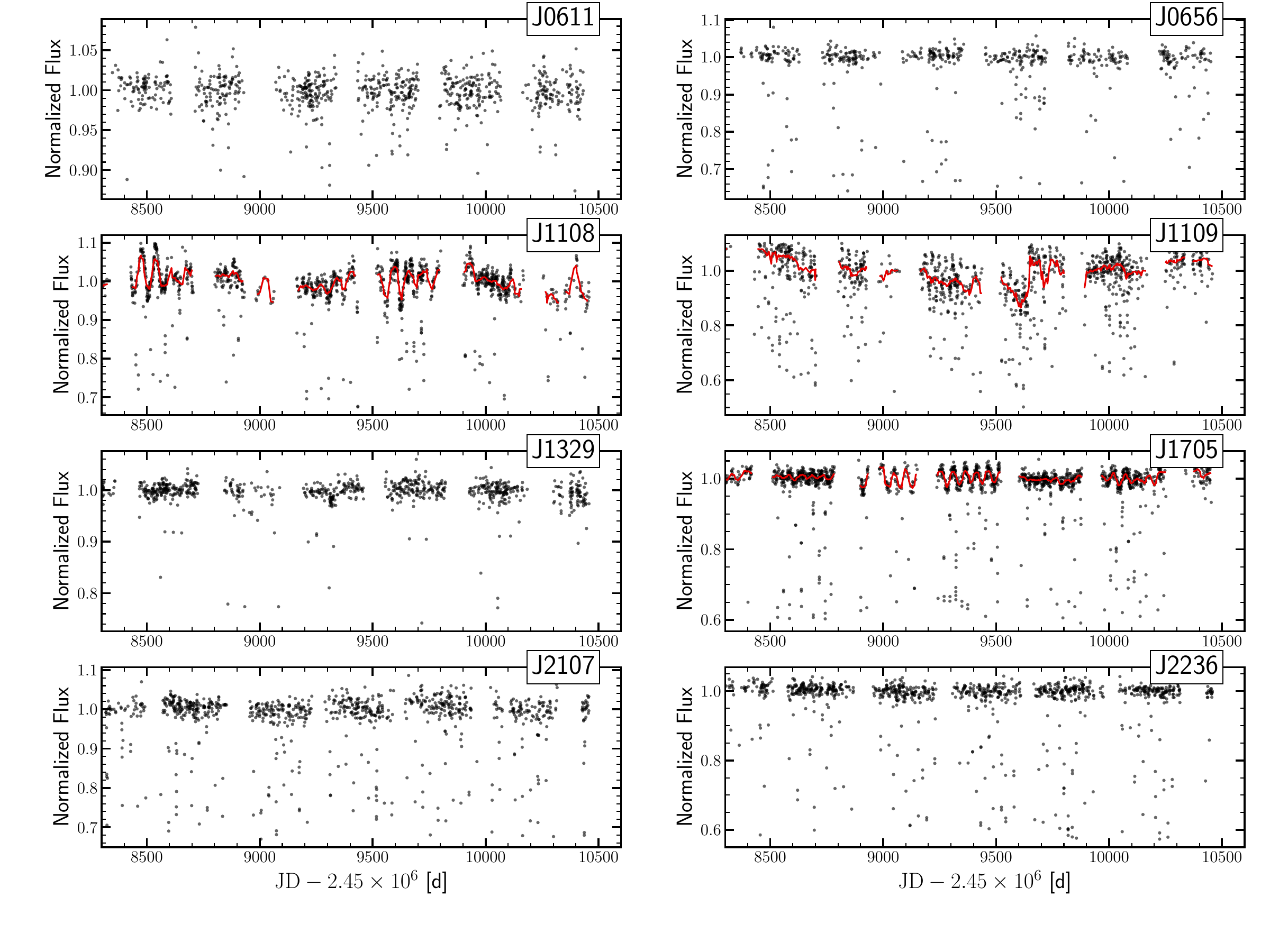}
    \caption{ASAS-SN $g$-band light curves of the eight eclipsing red giants. Three systems (J1108, J1109, J1705) show additional variations due to spots on the surface of one or both of the components. For these three systems we use {\tt wotan} to detrend the light curve using the biweight method. The detrending model is shown by the red curve.}
    \label{fig:unfolded_lightcurves}
\end{figure*}

We use light curves from the All-Sky Automated Survey for Supernovae \citep[ASAS-SN][]{Shappee14, Kochanek17}. These eclipsing binaries were identified and classified in the ASAS-SN Variable Stars Catalog \citep{Jayasinghe19} and further characterized in \citetalias{Rowan22}. ASAS-SN observed primarily in the $V$-band from 2012 to mid-2018. At the end of 2017, ASAS-SN switched to the $g$-band and added three additional telescope units. Here, we use only the ASAS-SN $g$-band data. The light curves are obtained from SkyPatrol V2 \citep{Hart23}. We do an additional inter-camera calibration using a damped random walk Gaussian process for interpolation to optimize the camera offsets \citep[e.g.,][]{Kozlowski10}.

Figure \ref{fig:unfolded_lightcurves} shows ASAS-SN light curves of the eight eclipsing binaries. It is immediately apparent that some of the targets show variability due to star spots in addition to the eclipses. This produces additional scatter in the out-of-eclipse variability in the phase-folded light curve. We detrend the light curves using a biweight filter as implemented in {\tt wotan} \citep{Hippke19} with a window matching the orbital period of the binary. The model trends are shown in red in Figure \ref{fig:unfolded_lightcurves} where they are used.

\begin{figure}
    \centering
    \includegraphics[width=\linewidth]{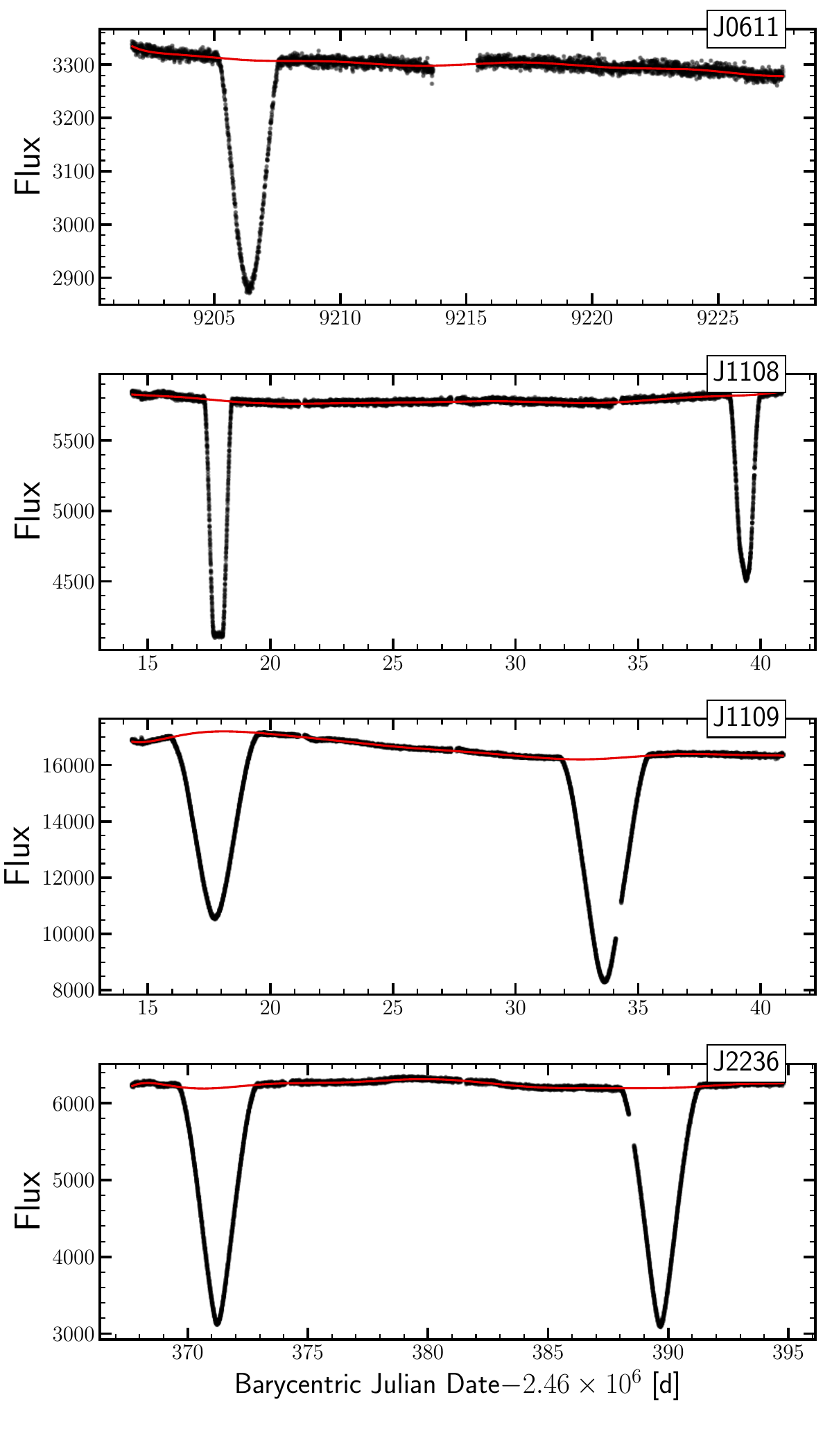}
    \caption{\TESS{} aperture flux light curves generated from full-frame images with the {\tt TGLC} pipeline \citep{Han23} for four of the red giant EB systems. We use {\tt wotan} to detrend the light curve, masking out times during the eclipses. The trends, shown in red, are likely a combination of \TESS{} systematic effects and stellar variability.}
    \label{fig:tglc_tess}
\end{figure}

\begin{figure}
    \centering
    \includegraphics[width=0.9\linewidth]{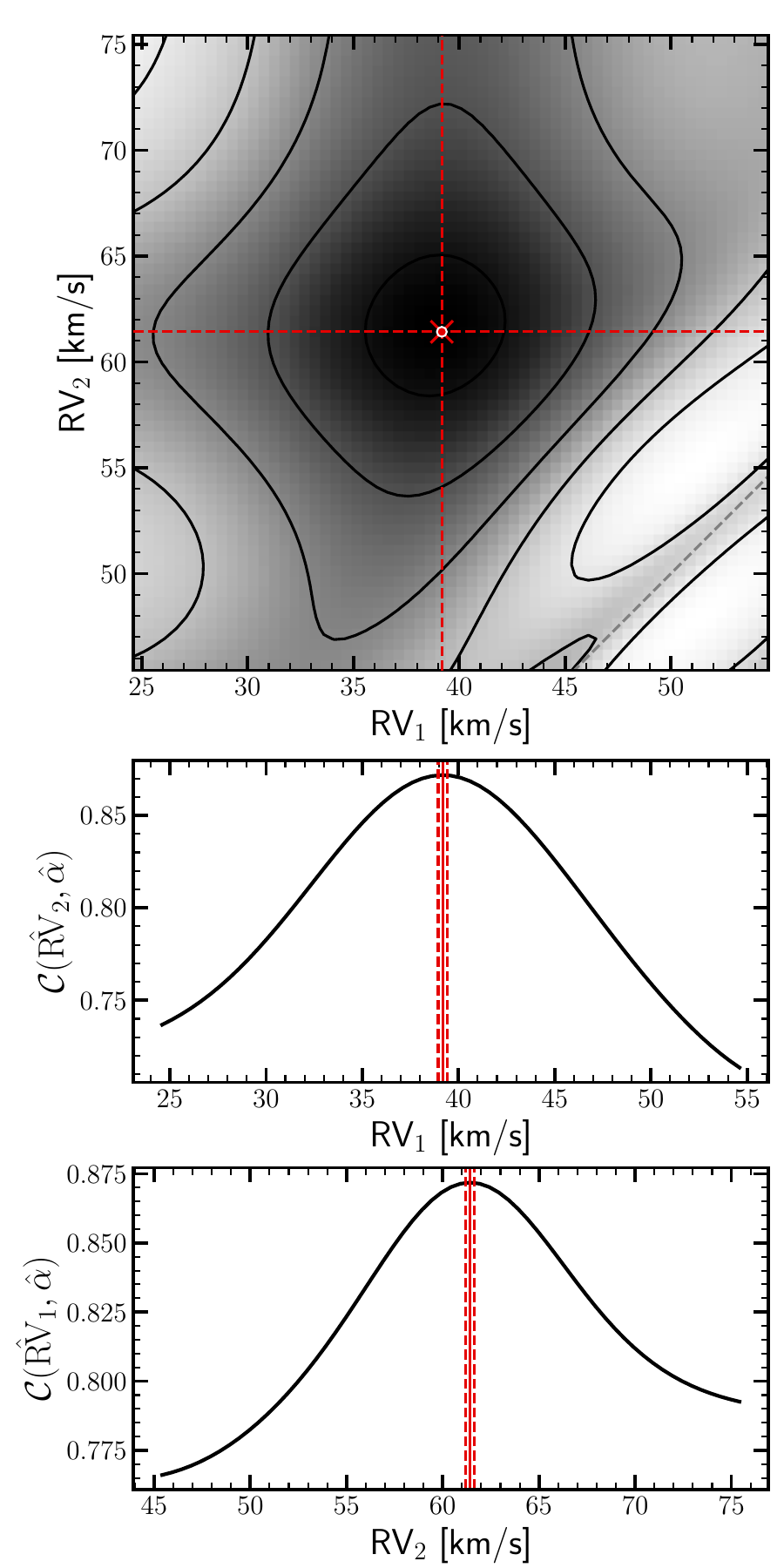}
    \caption{Example of RV determination for a spectrum of J0611. Top: TODCOR profile and contour lines. The maximum, marked with a cross in the center, shows the measured RVs. Middle and bottom: slices through the maximum cross correlation along either axis, showing the profiles for the RVs of each component. The vertical red lines show the measured RVs and the dashed lines show the $\pm 1\sigma$ uncertainties.}
    \label{fig:todcor_example}
\end{figure}

All of our targets have been observed by the Transiting Exoplanet Survey Satellite \citep[\TESS{},][]{Ricker15}. We use the \TESS{}-\Gaia{} Light Curve pipeline \citep[{\tt TGLC},][]{Han23} to extract aperture photometry light curves from the full-frame images. Since our targets have orbital periods that exceed the length of an individual \TESS{} sector, half of our targets do not have visible eclipses in \TESS{}, or have eclipses that are cut off by the orbit/sector gap. Four of our targets have \TESS{} observations where one or both eclipses are visible in a single sector (Table \ref{tab:summary_table}). Because the instrumental response and amount of blended light can vary between \TESS{} observations in different sectors, we only use one \TESS{} sector per target even if more than one sector is available. Figure \ref{fig:tglc_tess} shows the {\tt TGLC} light curves for these four targets.

\begin{figure*}
    \centering
    \includegraphics[width=\linewidth]{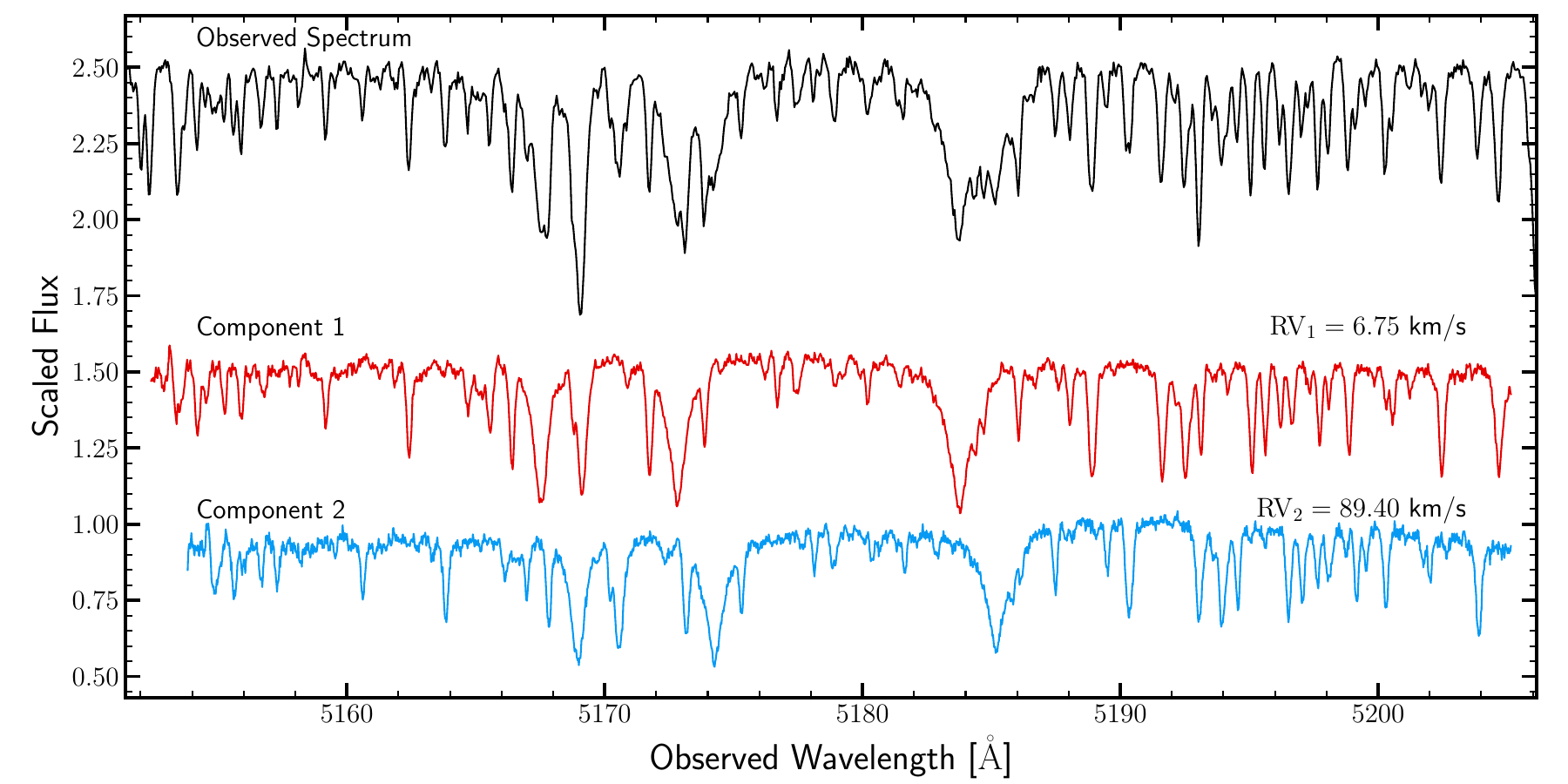}
    \caption{Example of spectral disentangling with {\tt FDBinary} for J0611. An observed PEPSI spectrum ($R\approx 43,000$) is shown in black on top. The disentangled components are shown in red and blue shifted to match the RVs determined from the observed spectrum.}
    \label{fig:spec_disentangling}
\end{figure*}

We also apply a detrending procedure to the {\tt TGLC} light curves. Since some of the trends observed in the {\tt TGLC} light curves are due to systematic instrumental effects, we use a smaller window of 5 days and the {\tt cosine} method in {\tt wotan}, which allows us to mask out the eclipses during detrending. We chose this window to be larger than the eclipse duration, and found that the choice of window did not significantly effect the detrended light curve. Figure \ref{fig:tglc_tess} shows the \TESS{} light curves and the trend models from {\tt wotan}. While this detrending process is fairly effective at flattening the light curve so it can be fit with \PHOEBE{} (Section \S\ref{sec:phoebe}), the large time windows masked during detrending, particularly for J1109 and J2236, limit our ability to remove systematic effects on timescales of $<5$~days. The detrending procedure could also remove astrophysical variability from low-amplitude elliposidal modulations, and we dicsuss this further in Section \S\ref{sec:phoebe}.

\section{Spectroscopic Observations and RV Measurements}\label{sec:observations}

To measure the radial velocities and other properties of our target stars, we collected 88 spectra of eight binaries using three different spectroscopic instruments. Here we briefly describe the three instruments. 

We obtained high-resolution ($R\approx 43,000$) spectra for four targets with the Potsdam Echelle Polarimetric and Spectroscopic Instrument \citep[PEPSI,][]{Strassmeier15} on the Large Binocular Telescope. The observations used the 300$\mu$m fiber and two cross-dispersers (CDs) covering 4758--5416~\AA{} (CD3) and 6244--7427~\AA{} (CD5). The CD5 data is not used for RV determination because of its significant overlap with telluric features. Exposure times ranged from 200 to 1000 seconds. The 2D echelle spectra are processed following the procedure described in \citet{Strassmeier18}. 

We obtained high-resolution ($R\approx 80,000$) spectra for four targets with the Automated Planet Finder (APF) Levy spectrograph on the Lick Observatory 2.4m \citep{Vogt14}. The observations used the $2\arcsec\times3\arcsec$ Decker-T slit. The APF spectra have a wavelength range of 3730--10206\AA{} and the raw 2D echelle spectra are reduced to 1D spectra through the California Planet Survey \citep[CPS,][]{Howard10} pipeline. The 1D spectra are then de-spiked to remove signals from cosmic rays and blaze corrected by fitting polynomials to the continuum in each order. The APF observations had a typical exposure time of 1700 seconds. We selected 33 orders spanning 4600--7813\AA{} for RV analysis, excluding orders affected by telluric lines. 

We obtained high-resolution spectra ($R\approx 28,000$) for four targets using CHIRON on the SMARTS 1.5m telescope \citep{Tokovinin13, Schwab12}. Spectra are taken in the fiber mode using $4\times4$~pixel binning and a Th-Ar comparison lamp. As with the APF observations, the extracted 1D spectra are de-spiked to remove signals from cosmic rays and blaze corrected by fitting polynomials to the continuum in each order. The CHIRON observations used a typical exposure time of 500 seconds. We use 36 orders spanning 4700--7792\AA{} for the RV analysis, avoiding regions containing telluric lines. 

\begin{table*}
    \centering
    \caption{Spectroscopic parameters for the template spectra used to derive radial velocities with TODCOR. We use a preliminary orbit model and {\tt FDBinary} to disentangle the spectra and fit the components with iSpec.}
    \sisetup{table-auto-round,
     group-digits=false}
    \setlength{\tabcolsep}{10pt}
    \begin{center}
        \begin{tabular}{l S[table-format=1.2] c c S[table-format=2.1] S[table-format=1.2] c c S[table-format=2.1] S[table-format=1.2]}
\toprule
{Target Name} &  {$\alpha$} &  {$T_{\rm{eff},1}$} &  {$\log g_1$} &  {$v \sin i_1$} &  {[Fe/H]$_1$} &  {$T_{\rm{eff},2}$} &  {$\log g_2$} &  {$v \sin i_2$} &  {[Fe/H]$_2$} \\
{} & {} & {$(K)$} & {} & {$(\rm{km/s})$} & {(dex)} & {$(K)$} & {} & {$(\rm{km/s})$} & {(dex)}\\ 
\midrule
        J0611 &    0.891805 &                5450 &          3.38 &             9.1 &          0.02 &                5410 &          3.32 &             4.0 &         -0.00 \\
        J0656 &    0.995272 &                5650 &          3.25 &            17.6 &          0.08 &                5460 &          3.46 &            17.6 &         -0.01 \\
        J1108 &    0.649893 &                5050 &          3.38 &            16.1 &          0.09 &                5190 &          3.42 &            16.2 &          0.27 \\
        J1109 &    0.939119 &                5200 &          3.46 &            18.2 &          0.07 &                5150 &          3.15 &            18.4 &          0.08 \\
        J1329 &    0.555899 &                5450 &          3.85 &            16.0 &          0.64 &                5570 &          3.61 &            15.7 &          0.52 \\
        J1705 &    0.923556 &                5490 &          3.50 &            14.2 &          0.26 &                5500 &          3.30 &            14.2 &          0.24 \\
        J2107 &    0.864818 &                5670 &          3.39 &            16.7 &          0.47 &                5130 &          3.21 &            15.0 &          0.01 \\
        J2236 &    0.970422 &                5610 &          3.42 &            17.2 &          0.15 &                5420 &          3.38 &            16.7 &         -0.03 \\
\bottomrule
\end{tabular}

    \end{center}
    \label{tab:template_table}
\end{table*}

\begin{figure*}[ht]
    \centering
    \begin{tabular}{cc}
        \includegraphics[width=0.5\textwidth]{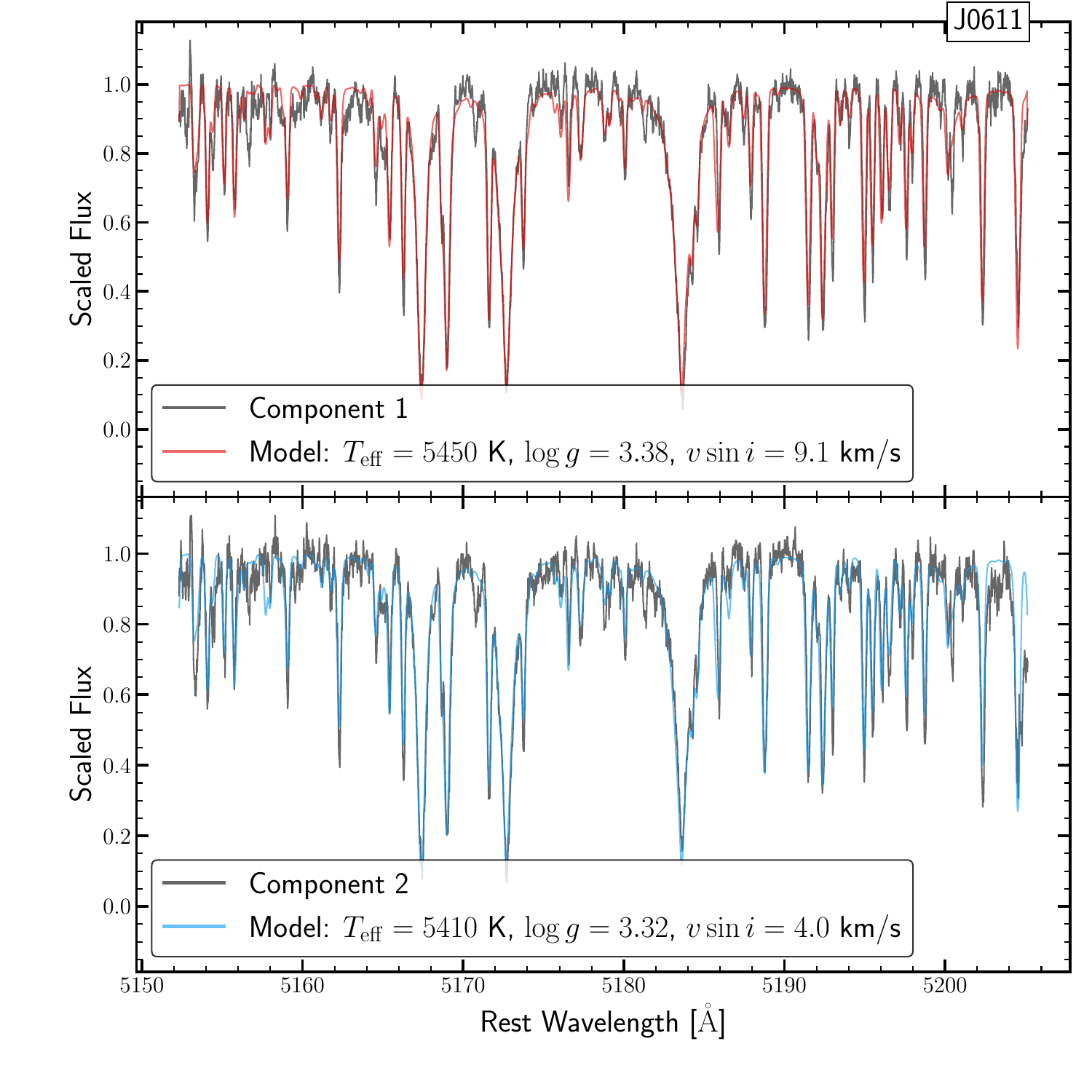} & \includegraphics[width=0.5\textwidth]{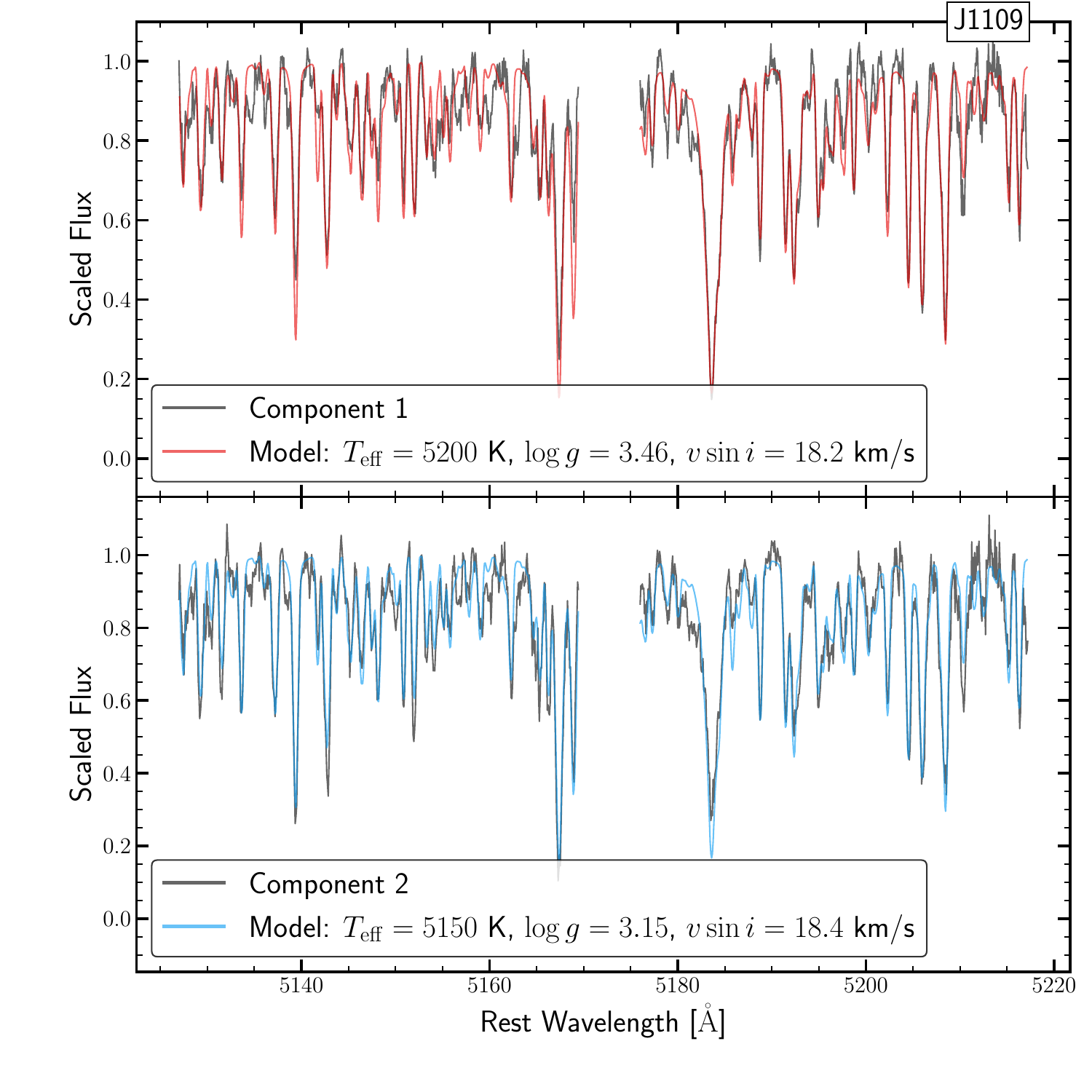} \\
    \end{tabular}
    \caption{Rest frame \iSpec{} model fits to the disentangled spectra to J0611 (left) and J1109 (right). The atmospheric parameters are reported in Table \ref{tab:template_table}, and we use these model fits as templates to derive the final RVs with TODCOR.}
    \label{fig:ispec_models}
\end{figure*}

\begin{table*}
    \centering
    \caption{Radial velocity measurements for J0611. The full table including all targets is available online in the ancillary material.}
    \sisetup{table-auto-round,
     group-digits=false}
    \setlength{\tabcolsep}{10pt}
    \renewcommand{\arraystretch}{1.2}
    \begin{center}
        \begin{tabular}{l S[table-format=3.3] S[table-format=2.2] S[table-format=1.2] S[table-format=2.2] S[table-format=1.2] c}
\toprule
{Target Name} &       {JD} &   {$RV_1$} &  {$\sigma_1$} &   {$RV_2$} &  {$\sigma_2$} & {Instrument} \\
{} & {$-2.46\times10^6$} & {$(\rm{km/s})$} & {$(\rm{km/s})$} & {$(\rm{km/s})$} & {$(\rm{km/s})$} & {}\\ 
\midrule
        J0611 & -16.274534 &   6.274531 &      0.234774 &  89.932425 &      0.225401 &          APF \\
        J0611 & -15.299910 &   6.753819 &      0.234279 &  89.398645 &      0.224909 &        PEPSI \\
        J0611 & 181.990074 &  16.411613 &      0.234623 &  80.341886 &      0.225204 &          APF \\
        J0611 & 182.992914 &  14.007200 &      0.235243 &  82.787445 &      0.225828 &          APF \\
        J0611 & 184.997137 &  10.295885 &      0.234606 &  86.541679 &      0.225156 &          APF \\
        J0611 & 193.959653 &   9.908363 &      0.234395 &  87.476276 &      0.224969 &          APF \\
        J0611 & 193.982533 &   9.901258 &      0.234412 &  87.484807 &      0.224996 &          APF \\
        J0611 & 198.965997 &  20.159348 &      0.234352 &  76.037435 &      0.224932 &          APF \\
        J0611 & 200.986810 &  26.590316 &      0.234324 &  70.043043 &      0.224904 &          APF \\
        J0611 & 204.026422 &  36.815554 &      0.234527 &  59.062772 &      0.225099 &          APF \\
        J0611 & 214.943924 &  75.290338 &      0.234418 &  19.826140 &      0.224996 &          APF \\
        J0611 & 220.894142 &  87.274648 &      0.234826 &   7.449408 &      0.225531 &          APF \\
        J0611 & 223.871828 &  88.744164 &      0.234863 &   5.074562 &      0.225478 &          APF \\
        J0611 & 226.879854 &  88.083815 &      0.234367 &   6.672516 &      0.225019 &          APF \\
        J0611 & 231.927324 &  79.801231 &      0.234358 &  15.778385 &      0.224982 &          APF \\
        J0611 & 250.884734 &  16.311573 &      0.235202 &  79.486536 &      0.225989 &        PEPSI \\
\bottomrule
\end{tabular}

    \end{center}
    \label{tab:rv_table}
\end{table*}

We measure radial velocities for both binary components using a two-dimensional cross-correlation function \citep[TODCOR,][]{Zucker94}. The TODCOR method generalizes the one-dimensional cross-correlation function (1D-CCF) by applying two template spectra to compute the correlation function over a two-dimensional grid of velocity shifts. With TODCOR, radial velocities for both components can be determined even with small radial velocity differences or when the flux ratio of the two components $F_2/F_1 \ll 1$. Many of our binaries have two well-separated velocity components where the 1D-CCF is effective at measuring accurate and precise RVs, but TODCOR is necessary for some systems and at some orbital phases. We therefore use TODCOR uniformly for all of our RV determinations. 

TODCOR requires two template spectra for the cross-correlation. For the best results, the template spectra should match the spectral types of the binary. We identified the best templates using a combination of spectral disentangling and empirical spectroscopic parameter estimation. First, we derive RVs using Solar-type templates (effective temperature $T_{\rm{eff}}=6000$, surface gravity $\log g=4.0$) using ATLAS model atmospheres \citep{Kurucz05} implemented in \texttt{iSpec} \citep{BlancoCuaresma14} for both stars. We calculate the two-dimensional cross-correlation function, $\mathcal{R}$, over a range of velocity shifts based on the 1D-CCF results. For the APF and CHIRON spectra, we apply TODCOR to each echelle order independently. We combine the TODCOR profiles from each order following the scheme described in \citet{Zucker03}. We determine the RVs by maximizing $\mathcal{R}$ with the Nelder-Mead algorithm as implemented in \citet{Gao12}. To determine RV uncertainties, we take slices through the maximum along each axis. We measure RV uncertainties as
\begin{equation} \label{eqn:rv_err}
    \sigma_{\rm{RV}}^2 = - \left( N \frac{C^{\prime\prime}(\hat{s})}{C(\hat{s})}\frac{C^2(\hat{s})}{1-C^2(\hat{s})}\right)^{-1},
\end{equation}
following \citet{Zucker03} where $C$ is the slice through $\mathcal{R}$, $\hat{s}$ is the value of velocities where $\mathcal{R}$ at maximum, and $N$ is the number of bins in the spectra. For the APF and CHIRON spectra where $\mathcal{R}$ is combined from all the echelle orders, Equation \ref{eqn:rv_err} is modified such that the factor $N$ is replaced by $NM$, where $M$ is the number of orders. Figure \ref{fig:todcor_example} shows an example of the TODCOR profile for one of the J0611 spectra.

After preliminary RVs have been derived using two Solar-templates, we fit a Keplerian orbit model of the form, 
\begin{equation} \label{eqn:orbit}
    \begin{split}
        \text{RV}_1(t) &= \gamma + K_1 \left[(\omega+f)+e\cos\omega\right]\\
        \text{RV}_2(t) &= \gamma - K_2 \left[(\omega+f)+e\cos\omega\right],
    \end{split}
\end{equation}
\noindent where $\gamma$ is the center-of-mass velocity, $K_1$ and $K_2$ are the radial velocity semiamplitudes, $f$ is the true anomaly, and $\omega$ is the argument of periastron. The true anomaly, $f$, is related to the eccentric anomaly, $E$, and the eccentricity, $e$ by
\begin{equation}
    \cos f = \frac{\cos E - e}{1-e \cos E},
\end{equation}
and the eccentric anomaly is
\begin{equation}
    E - e\sin E = \frac{2\pi(t-t_0)}{P}
\end{equation}
where $P$ is the period and $t_0$ is the time of periastron. We sample over the orbital parameters using Markov Chain Monte Carlo with {\tt emcee} \citep{ForemanMackey13}. The orbital period of the binaries is well-constrained from the ASAS-SN light curve, so we set a Gaussian prior on the orbital period with $\sigma=10^{-3} P$. We also include terms for the stellar jitter, $s$, of each component in the log-likelihood following {\tt TheJoker} \citep{PriceWhelan17}. This term is included to model the effects of intrinsic stellar variability and underestimated radial velocity uncertainties. We re-scale the RV errors based on the measured stellar-jitter from our RV orbit model to $\sigma^2 = \sigma_{\rm{RV}}^2 + s^2$, where $\sigma_{\rm{RV}}$ is the measured RV uncertainty from Equation \ref{eqn:rv_err}. 

After the preliminary RV orbit has been derived using Solar-type templates, we use {\tt FDBinary} \citep{Ilijic04} to disentangle the observed spectra into component spectra. {\tt FDBinary} can solve for both the RV orbit model and the spectra, but we fix the orbit at the solution from our MCMC model. The flux ratio, $\alpha$, is needed for the components to be disentangled, and we estimate the flux ratio using the TODCOR profile. As described in \citet{Zucker94}, the cross-correlation functions can be used to estimate the flux ratio, $\hat{\alpha}$, where the cross-correlation function is maximized. We take the median value of $\hat{\alpha}$ across all spectra, excluding those taken in eclipse, to be the flux ratio for disentangling. We then run {\tt FDBinary} separately for each echelle order.

Figure \ref{fig:spec_disentangling} shows an example of the disentangled spectral components of J0611 compared to one of the observed PEPSI spectra. The disentangled spectra have some sinusoidal variations in the continuum, which is a known artifact of the Fourier-based disentangling method \citep[e.g.,][]{Beck14}. We use \iSpec{} \citep{BlancoCuaresma14} to normalize the disentangled components with a second-degree spline and remove this signal. We then use \iSpec{} to estimate the effective temperature $T_{\rm{eff}}$, the surface gravity $\log g$, the metallicity [Fe/H], and the projected rotational velocity, $v \sin i$ of each disentangled component. The $\alpha$-element enhancement, microturblent velocity, and macroturbulent velocity are all set using default empirical relations within \iSpec{}. We fit the spectra in windows around $5150$--$5200$~\AA{} and $5125$--$5220$~\AA{} for APF/PEPSI and CHIRON targets, respectively. Table \ref{tab:template_table} reports the best-fit parameters that we adopt for our templates and Figure \ref{fig:ispec_models} shows an example of the disentangled spectra of J0611 and J1109.

\begin{figure}
    \centering
    \includegraphics[width=\linewidth]{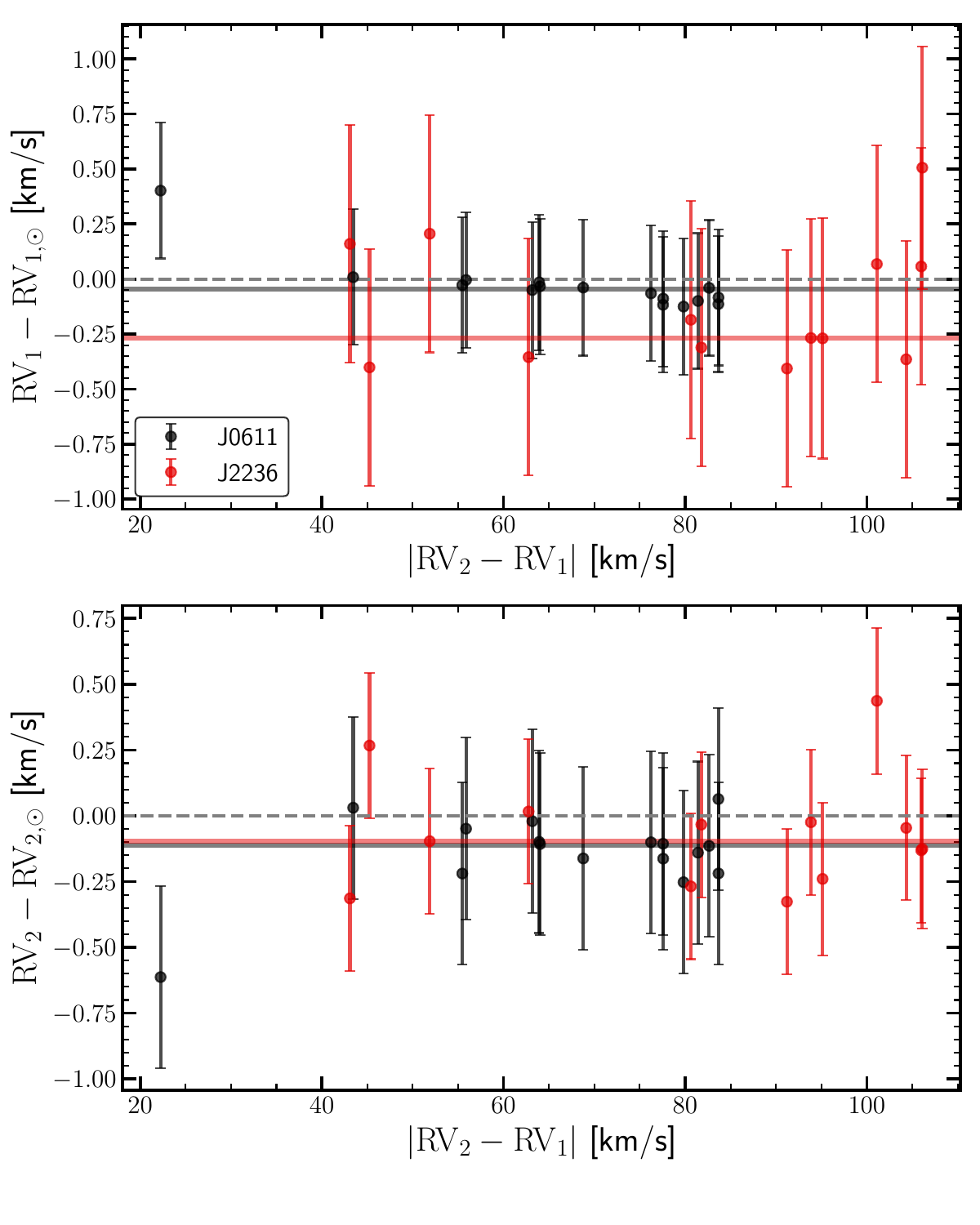}
    \caption{Comparison of the radial velocities derived using Solar templates ($\rm{RV}_{1,\odot}$ and $\rm{RV}_{2,\odot}$) to the velocities found using templates determined from spectral disentangling ($\rm{RV}_{1}$ and $\rm{RV}_{2}$) for the primary star (top) and secondary star (bottom). RV measurements for J0611 and J2236 are shown in black and red, respectively. The difference is shown as a function of the RV difference between the two components. For both targets we see that there is no strong trend with $\rm{RV}_{2}$-$\rm{RV}_{1}$, indicating that the choice of template is equally important near RV quadrature and superior/inferior conjunction. The solid horizontal lines show the median difference between the RV measurements using the different templates. Although the difference in the measured velocities is small compared to the uncertainties, we find that the choice of template is important at the $\sim 0.1$~km/s level. }
    \label{fig:template_comparison}
\end{figure}

We generate synthetic templates using the \iSpec{} model corresponding to the atmospheric parameters in Table \ref{tab:template_table} over the full wavelength range of the APF/PEPSI/CHIRON spectra. We then repeat the TODCOR RV determination process described above using these templates. Figure \ref{fig:template_comparison} compares the RVs derived with the best-fitting templates and the Solar type templates for two targets. While the difference in RVs is small, we find that the choice of template can introduce systematic effects on the final RV measurements at the $\sim 100$~m/s level. Table \ref{tab:rv_table} reports our RV measurements for all targets. 

\section{PHOEBE Models} \label{sec:phoebe}

\begin{table*}
    \centering
    \caption{MCMC posteriors for the \PHOEBE{} models of the eclipsing red giant binaries without \TESS{} data. The four systems with \TESS{} data are shown in Table \ref{tab:phoebe_table_tess}. Uncertainties are reported at the $1\sigma$ level. $T_0$ is the time of superior conjunction. Figure \ref{fig:corner_example} shows an example of the corner plot for J0656, and similar figures for the remaining targets are available online in the ancillary material. Figure \ref{fig:phoebe_models} shows the light curve and RV models for each target. We include models with and without third light ($l_3$), but the results are largely the same because the systems are viewed at near edge-on inclinations.}
    \sisetup{table-auto-round,
     group-digits=false}
    \setlength{\tabcolsep}{10pt}
    \renewcommand{\arraystretch}{2}
    \begin{center}
        \begin{tabular}{lcccc}
\toprule
\multicolumn{1}{c}{}& \multicolumn{2}{c}{J0656} & \multicolumn{2}{c}{J1329}  \\
\toprule
{} &                  without $l_3$ &                     with $l_3$ &                  without $l_3$ &                     with $l_3$ \\
\midrule
Period $(\rm{d})$               &  $41.4450^{+0.0003}_{-0.0003}$ &  $41.4453^{+0.0003}_{-0.0003}$ &  $37.3338^{+0.0002}_{-0.0002}$ &  $37.3338^{+0.0002}_{-0.0001}$ \\
$T_0-2.45\times10^6\ (\rm{d})$  &    $8221.252^{+0.01}_{-0.009}$ &      $8221.24^{+0.01}_{-0.01}$ &   $8094.747^{+0.007}_{-0.009}$ &   $8094.747^{+0.007}_{-0.007}$ \\
Ecc                             &   $0.0009^{+0.0009}_{-0.0004}$ &   $0.0013^{+0.0007}_{-0.0004}$ &      $0.142^{+0.003}_{-0.003}$ &      $0.142^{+0.003}_{-0.003}$ \\
$\omega\ (^{\circ})$            &               $50^{+30}_{-30}$ &               $40^{+20}_{-30}$ &          $253.9^{+0.3}_{-0.4}$ &          $253.9^{+0.4}_{-0.4}$ \\
Incl $(^{\circ})$               &        $85.43^{+0.07}_{-0.06}$ &           $88.6^{+0.5}_{-0.4}$ &           $90.1^{+0.7}_{-0.7}$ &           $90.1^{+0.7}_{-0.9}$ \\
$T_{\rm{eff,2}}/T_{\rm{eff,1}}$ &      $0.985^{+0.001}_{-0.001}$ &      $0.985^{+0.003}_{-0.002}$ &         $1.25^{+0.03}_{-0.05}$ &         $1.24^{+0.03}_{-0.04}$ \\
$l_3$                           &                              - &         $0.25^{+0.02}_{-0.02}$ &                              - &         $0.09^{+0.09}_{-0.06}$ \\
$\gamma\ (\rm{km/s})$           &           $48.2^{+0.1}_{-0.1}$ &           $48.1^{+0.1}_{-0.1}$ &       $-10.64^{+0.09}_{-0.09}$ &       $-10.63^{+0.09}_{-0.09}$ \\
RV Offset $(\rm{km/s})$         &           $-0.3^{+0.2}_{-0.2}$ &           $-0.3^{+0.2}_{-0.2}$ &                              - &                              - \\
$M_1\ (M_\odot)$                &         $2.71^{+0.02}_{-0.03}$ &         $2.68^{+0.02}_{-0.03}$ &         $1.12^{+0.02}_{-0.02}$ &         $1.12^{+0.02}_{-0.02}$ \\
$M_2\ (M_\odot)$                &         $2.69^{+0.03}_{-0.03}$ &         $2.67^{+0.03}_{-0.03}$ &         $1.01^{+0.01}_{-0.01}$ &         $1.01^{+0.01}_{-0.01}$ \\
$R_1\ (R_\odot)$                &           $12.1^{+0.4}_{-0.4}$ &           $12.3^{+0.2}_{-0.2}$ &            $3.1^{+0.1}_{-0.1}$ &            $3.1^{+0.2}_{-0.1}$ \\
$R_2\ (R_\odot)$                &           $12.1^{+0.4}_{-0.4}$ &           $12.1^{+0.2}_{-0.2}$ &         $0.97^{+0.05}_{-0.03}$ &         $1.05^{+0.08}_{-0.07}$ \\
\end{tabular}

        \begin{tabular}{lcccc}
\toprule
\multicolumn{1}{c}{}& \multicolumn{2}{c}{J1705} & \multicolumn{2}{c}{J2107}  \\
\toprule
{} &                  without $l_3$ &                     with $l_3$ &                  without $l_3$ &                     with $l_3$ \\
\midrule
Period $(\rm{d})$               &  $52.6154^{+0.0002}_{-0.0001}$ &  $52.6155^{+0.0006}_{-0.0002}$ &  $68.1391^{+0.0006}_{-0.0005}$ &  $68.1395^{+0.0006}_{-0.0006}$ \\
$T_0-2.45\times10^6\ (\rm{d})$  &   $8192.593^{+0.002}_{-0.001}$ &    $8192.592^{+0.002}_{-0.02}$ &      $8256.23^{+0.01}_{-0.01}$ &      $8256.22^{+0.01}_{-0.01}$ \\
Ecc                             &   $0.0005^{+0.0010}_{-0.0002}$ &    $0.0012^{+0.001}_{-0.0009}$ &   $0.0007^{+0.0005}_{-0.0003}$ &   $0.0010^{+0.0005}_{-0.0004}$ \\
$\omega\ (^{\circ})$            &              $340^{+60}_{-60}$ &               $280^{+14}_{-7}$ &               $10^{+40}_{-50}$ &              $-10^{+60}_{-40}$ \\
Incl $(^{\circ})$               &           $87.9^{+0.1}_{-0.2}$ &           $88.0^{+0.2}_{-0.1}$ &        $83.33^{+0.05}_{-0.06}$ &           $92.9^{+0.5}_{-0.4}$ \\
$T_{\rm{eff,2}}/T_{\rm{eff,1}}$ &      $1.004^{+0.001}_{-0.001}$ &      $1.004^{+0.001}_{-0.001}$ &      $0.971^{+0.001}_{-0.001}$ &      $0.973^{+0.002}_{-0.002}$ \\
$l_3$                           &                              - &         $0.03^{+0.02}_{-0.01}$ &                              - &         $0.29^{+0.02}_{-0.03}$ \\
$\gamma\ (\rm{km/s})$           &       $-51.80^{+0.06}_{-0.08}$ &       $-51.85^{+0.06}_{-0.08}$ &       $-18.69^{+0.07}_{-0.07}$ &       $-18.69^{+0.07}_{-0.06}$ \\
RV Offset $(\rm{km/s})$         &                              - &                              - &        $-0.31^{+0.06}_{-0.06}$ &        $-0.31^{+0.06}_{-0.06}$ \\
$M_1\ (M_\odot)$                &         $2.21^{+0.01}_{-0.02}$ &         $2.20^{+0.01}_{-0.01}$ &         $3.54^{+0.02}_{-0.01}$ &         $3.49^{+0.02}_{-0.02}$ \\
$M_2\ (M_\odot)$                &         $2.27^{+0.02}_{-0.02}$ &         $2.26^{+0.02}_{-0.02}$ &         $3.38^{+0.01}_{-0.01}$ &         $3.33^{+0.01}_{-0.01}$ \\
$R_1\ (R_\odot)$                &            $8.4^{+0.3}_{-0.2}$ &            $8.5^{+0.2}_{-0.2}$ &           $20.5^{+0.7}_{-0.3}$ &           $21.2^{+0.5}_{-0.5}$ \\
$R_2\ (R_\odot)$                &            $9.8^{+0.1}_{-0.2}$ &            $9.7^{+0.1}_{-0.2}$ &           $22.1^{+0.4}_{-0.6}$ &           $21.4^{+0.4}_{-0.5}$ \\
\bottomrule
\end{tabular}

    \end{center}
    \label{tab:phoebe_table_asassn}
\end{table*}

\begin{table*}
    \centering
    \caption{Same as Table \ref{tab:phoebe_table_asassn}, but for systems with ASAS-SN and  \TESS{} data.}
    \sisetup{table-auto-round,
     group-digits=false}
    \setlength{\tabcolsep}{8pt}
    \renewcommand{\arraystretch}{2}
    \begin{center}
        \begin{tabular}{lcccccc}
\toprule
\multicolumn{1}{c}{}& \multicolumn{3}{c}{J0611} & \multicolumn{3}{c}{J1108}  \\
\toprule
{} &              ASAS-SN no $l_3$ &            ASAS-SN with $l_3$ &   {ASAS-SN $+$ \textit{TESS}} &                  ASAS-SN no $l_3$ &                ASAS-SN with $l_3$ &       {ASAS-SN $+$ \textit{TESS}} \\
\midrule
Period $(\rm{d})$               &    $69.002^{+0.002}_{-0.002}$ &    $69.003^{+0.001}_{-0.002}$ &    $69.004^{+0.001}_{-0.001}$ &  $32.39913^{+0.00007}_{-0.00007}$ &  $32.39909^{+0.00008}_{-0.00009}$ &  $32.39872^{+0.00006}_{-0.00007}$ \\
$T_0-2.45\times10^6\ (\rm{d})$  &     $8240.35^{+0.03}_{-0.03}$ &     $8240.34^{+0.04}_{-0.03}$ &     $8240.34^{+0.02}_{-0.02}$ &      $8095.416^{+0.004}_{-0.004}$ &      $8095.420^{+0.004}_{-0.004}$ &      $8095.434^{+0.004}_{-0.004}$ \\
Ecc                             &  $0.0027^{+0.0008}_{-0.0007}$ &  $0.0027^{+0.0008}_{-0.0008}$ &  $0.0014^{+0.0007}_{-0.0006}$ &      $0.2622^{+0.0004}_{-0.0004}$ &      $0.2623^{+0.0005}_{-0.0004}$ &      $0.2633^{+0.0001}_{-0.0001}$ \\
$\omega\ (^{\circ})$            &              $10^{+30}_{-30}$ &              $20^{+30}_{-30}$ &             $40^{+290}_{-60}$ &             $191.0^{+0.4}_{-0.5}$ &             $190.9^{+0.5}_{-0.5}$ &             $191.7^{+0.1}_{-0.1}$ \\
Incl $(^{\circ})$               &          $87.4^{+0.9}_{-0.5}$ &            $87.7^{+2}_{-0.8}$ &          $86.3^{+0.2}_{-0.2}$ &              $89.0^{+0.5}_{-0.3}$ &                $91.0^{+0.4}_{-1}$ &            $90.38^{+0.07}_{-0.1}$ \\
$T_{\rm{eff,2}}/T_{\rm{eff,1}}$ &        $1.04^{+0.01}_{-0.01}$ &        $1.04^{+0.02}_{-0.02}$ &      $1.019^{+0.01}_{-0.008}$ &         $1.085^{+0.002}_{-0.002}$ &         $1.087^{+0.003}_{-0.003}$ &       $1.0996^{+0.002}_{-0.0004}$ \\
$l_3$                           &                             - &         $0.16^{+0.1}_{-0.10}$ &        $0.26^{+0.05}_{-0.03}$ &                                 - &            $0.04^{+0.04}_{-0.03}$ &         $0.162^{+0.005}_{-0.005}$ \\
$\gamma\ (\rm{km/s})$           &       $47.90^{+0.04}_{-0.05}$ &       $47.89^{+0.05}_{-0.05}$ &       $47.89^{+0.05}_{-0.05}$ &          $-12.10^{+0.08}_{-0.08}$ &          $-12.07^{+0.07}_{-0.09}$ &          $-12.10^{+0.09}_{-0.09}$ \\
RV Offset $(\rm{km/s})$         &       $-0.20^{+0.06}_{-0.07}$ &       $-0.21^{+0.07}_{-0.07}$ &          $-0.4^{+0.1}_{-0.1}$ &                                 - &                                 - &                                 - \\
$M_1\ (M_\odot)$                &     $2.143^{+0.009}_{-0.009}$ &     $2.142^{+0.009}_{-0.009}$ &     $2.150^{+0.008}_{-0.008}$ &         $1.055^{+0.008}_{-0.007}$ &         $1.056^{+0.006}_{-0.007}$ &         $1.053^{+0.007}_{-0.008}$ \\
$M_2\ (M_\odot)$                &     $2.084^{+0.008}_{-0.009}$ &     $2.082^{+0.008}_{-0.009}$ &     $2.090^{+0.008}_{-0.009}$ &         $1.049^{+0.008}_{-0.007}$ &         $1.050^{+0.007}_{-0.008}$ &         $1.047^{+0.007}_{-0.007}$ \\
$R_1\ (R_\odot)$                &           $9.8^{+0.6}_{-0.7}$ &           $9.5^{+0.8}_{-0.9}$ &          $10.3^{+0.1}_{-0.2}$ &            $4.55^{+0.08}_{-0.07}$ &            $4.52^{+0.09}_{-0.08}$ &            $4.38^{+0.01}_{-0.01}$ \\
$R_2\ (R_\odot)$                &           $2.9^{+0.2}_{-0.3}$ &           $3.1^{+0.4}_{-0.4}$ &           $4.5^{+0.2}_{-0.2}$ &            $2.04^{+0.06}_{-0.05}$ &            $2.12^{+0.06}_{-0.06}$ &         $2.219^{+0.006}_{-0.006}$ \\
\end{tabular}

        \begin{tabular}{lcccccc}
\toprule
\multicolumn{1}{c}{}& \multicolumn{3}{c}{J1109} & \multicolumn{3}{c}{J2236}  \\
\toprule
{} &               ASAS-SN no $l_3$ &             ASAS-SN with $l_3$ &       {ASAS-SN $+$ \textit{TESS}} &               ASAS-SN no $l_3$ &             ASAS-SN with $l_3$ &       {ASAS-SN $+$ \textit{TESS}} \\
\midrule
Period $(\rm{d})$               &  $31.7548^{+0.0002}_{-0.0002}$ &  $31.7547^{+0.0001}_{-0.0001}$ &  $31.75487^{+0.00010}_{-0.00009}$ &  $36.8363^{+0.0001}_{-0.0001}$ &  $36.8365^{+0.0001}_{-0.0001}$ &  $36.83631^{+0.00007}_{-0.00006}$ \\
$T_0-2.45\times10^6\ (\rm{d})$  &    $8303.065^{+0.005}_{-0.01}$ &   $8303.067^{+0.005}_{-0.003}$ &      $8303.001^{+0.005}_{-0.005}$ &   $8216.332^{+0.006}_{-0.005}$ &   $8216.334^{+0.005}_{-0.005}$ &      $8216.325^{+0.004}_{-0.004}$ \\
Ecc                             &    $0.0046^{+0.003}_{-0.0005}$ &   $0.0044^{+0.0004}_{-0.0002}$ &         $0.014^{+0.001}_{-0.002}$ &   $0.0004^{+0.0005}_{-0.0003}$ &   $0.0007^{+0.0003}_{-0.0002}$ &      $0.0022^{+0.0006}_{-0.0009}$ \\
$\omega\ (^{\circ})$            &              $150^{+30}_{-30}$ &              $180^{+20}_{-20}$ &             $269.9^{+0.3}_{-0.4}$ &              $140^{+90}_{-50}$ &              $180^{+40}_{-40}$ &                   $281^{+7}_{-3}$ \\
Incl $(^{\circ})$               &           $83.4^{+0.1}_{-0.1}$ &           $86.1^{+0.4}_{-0.3}$ &              $88.6^{+0.2}_{-0.1}$ &        $87.20^{+0.06}_{-0.05}$ &           $88.4^{+0.2}_{-0.2}$ &           $90.00^{+0.02}_{-0.02}$ \\
$T_{\rm{eff,2}}/T_{\rm{eff,1}}$ &         $1.07^{+0.01}_{-0.01}$ &      $1.061^{+0.004}_{-0.006}$ &         $1.080^{+0.003}_{-0.004}$ &    $1.0015^{+0.0010}_{-0.001}$ &      $1.001^{+0.001}_{-0.001}$ &      $0.9990^{+0.0003}_{-0.0003}$ \\
$l_3$                           &                              - &         $0.21^{+0.02}_{-0.02}$ &         $0.278^{+0.002}_{-0.002}$ &                              - &         $0.09^{+0.01}_{-0.01}$ &         $0.170^{+0.002}_{-0.002}$ \\
$\gamma\ (\rm{km/s})$           &            $3.8^{+0.2}_{-0.2}$ &            $3.7^{+0.1}_{-0.1}$ &               $3.7^{+0.1}_{-0.1}$ &       $-32.32^{+0.03}_{-0.04}$ &       $-32.34^{+0.03}_{-0.03}$ &          $-32.33^{+0.03}_{-0.03}$ \\
RV Offset $(\rm{km/s})$         &                              - &                              - &                                 - &        $-0.33^{+0.04}_{-0.04}$ &        $-0.33^{+0.04}_{-0.04}$ &           $-0.55^{+0.09}_{-0.10}$ \\
$M_1\ (M_\odot)$                &         $1.43^{+0.02}_{-0.02}$ &         $1.41^{+0.02}_{-0.02}$ &            $1.37^{+0.02}_{-0.02}$ &      $2.326^{+0.007}_{-0.006}$ &      $2.321^{+0.006}_{-0.006}$ &         $2.316^{+0.006}_{-0.005}$ \\
$M_2\ (M_\odot)$                &         $1.42^{+0.02}_{-0.02}$ &         $1.40^{+0.02}_{-0.02}$ &            $1.37^{+0.02}_{-0.02}$ &       $2.322^{+0.01}_{-0.010}$ &       $2.318^{+0.009}_{-0.01}$ &         $2.312^{+0.009}_{-0.009}$ \\
$R_1\ (R_\odot)$                &           $11.5^{+0.2}_{-0.2}$ &           $11.7^{+0.1}_{-0.1}$ &           $11.69^{+0.05}_{-0.05}$ &           $10.8^{+0.1}_{-0.2}$ &           $10.9^{+0.1}_{-0.1}$ &           $11.24^{+0.02}_{-0.02}$ \\
$R_2\ (R_\odot)$                &            $9.8^{+0.3}_{-0.2}$ &            $9.7^{+0.2}_{-0.2}$ &            $9.61^{+0.05}_{-0.05}$ &           $10.3^{+0.2}_{-0.2}$ &           $10.5^{+0.1}_{-0.1}$ &           $11.22^{+0.02}_{-0.02}$ \\
\bottomrule
\end{tabular}

    \end{center}
    \label{tab:phoebe_table_tess}
\end{table*}

\begin{figure*}[p]
    \centering
    \begin{tabular}{cc}
        \includegraphics[width=\linewidth]{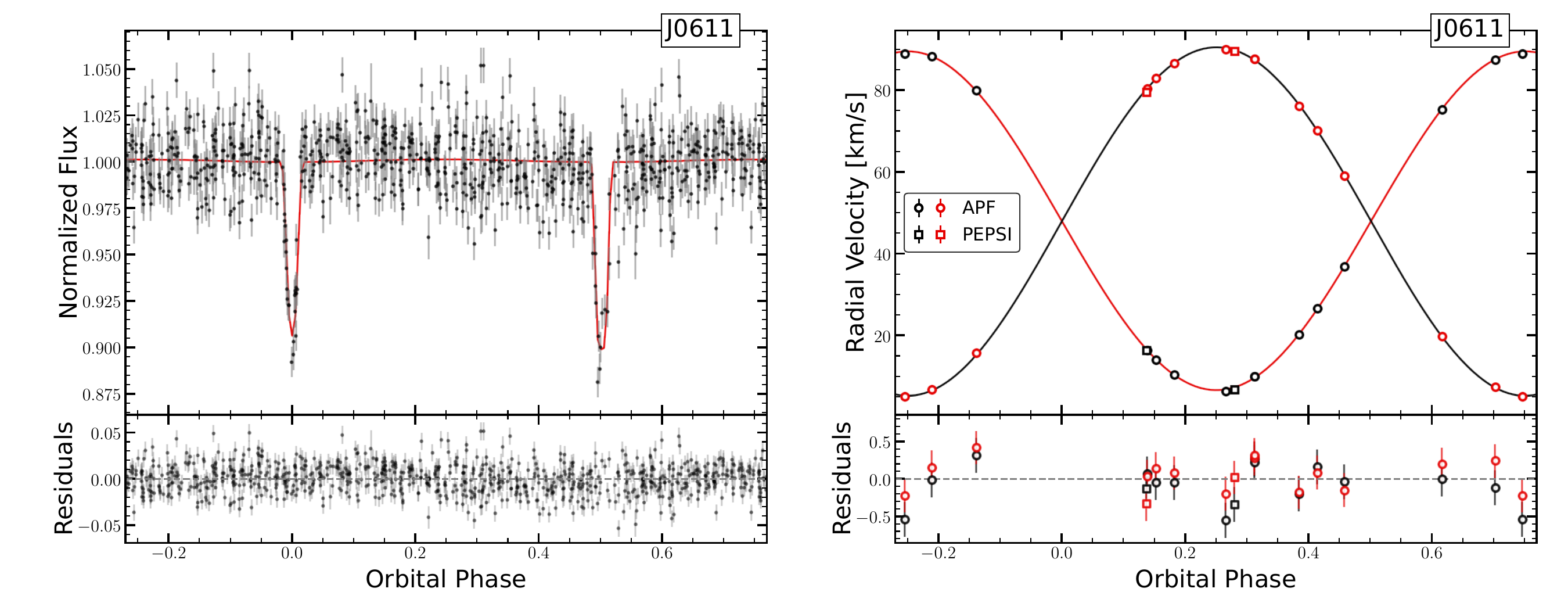} \\
        \includegraphics[width=\linewidth]{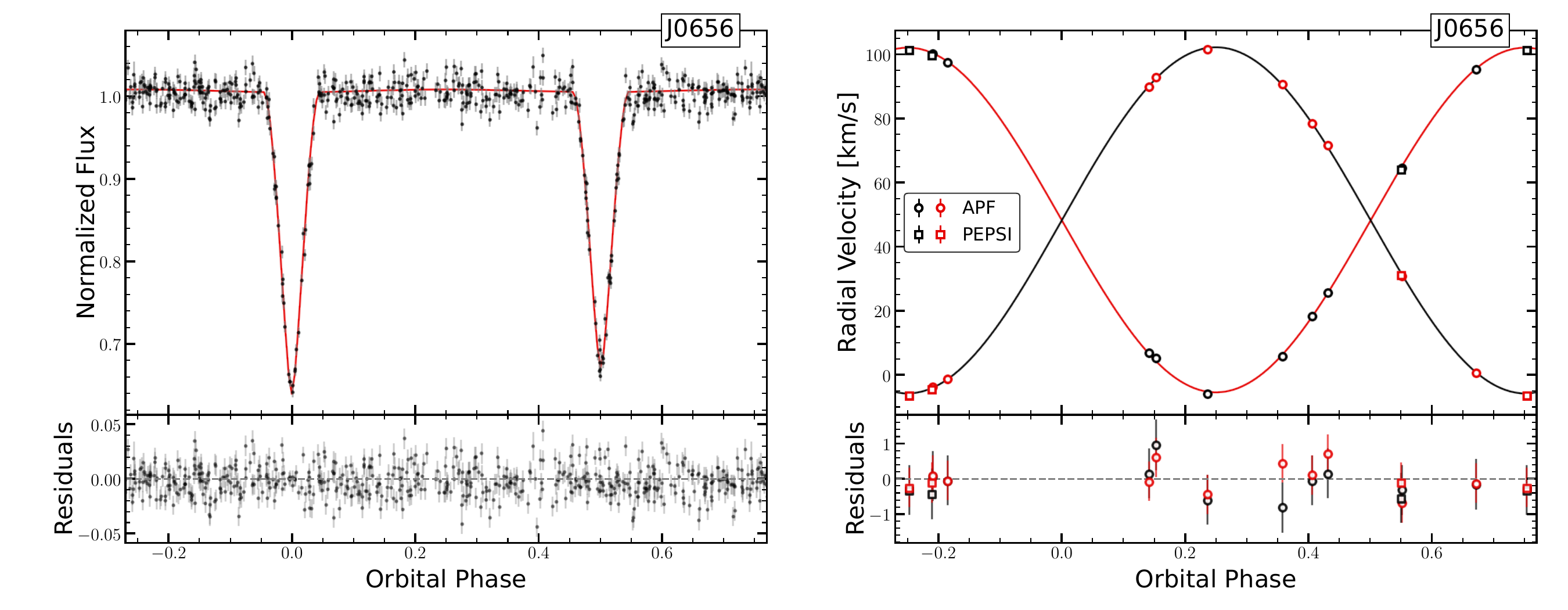} \\
        \includegraphics[width=\linewidth]{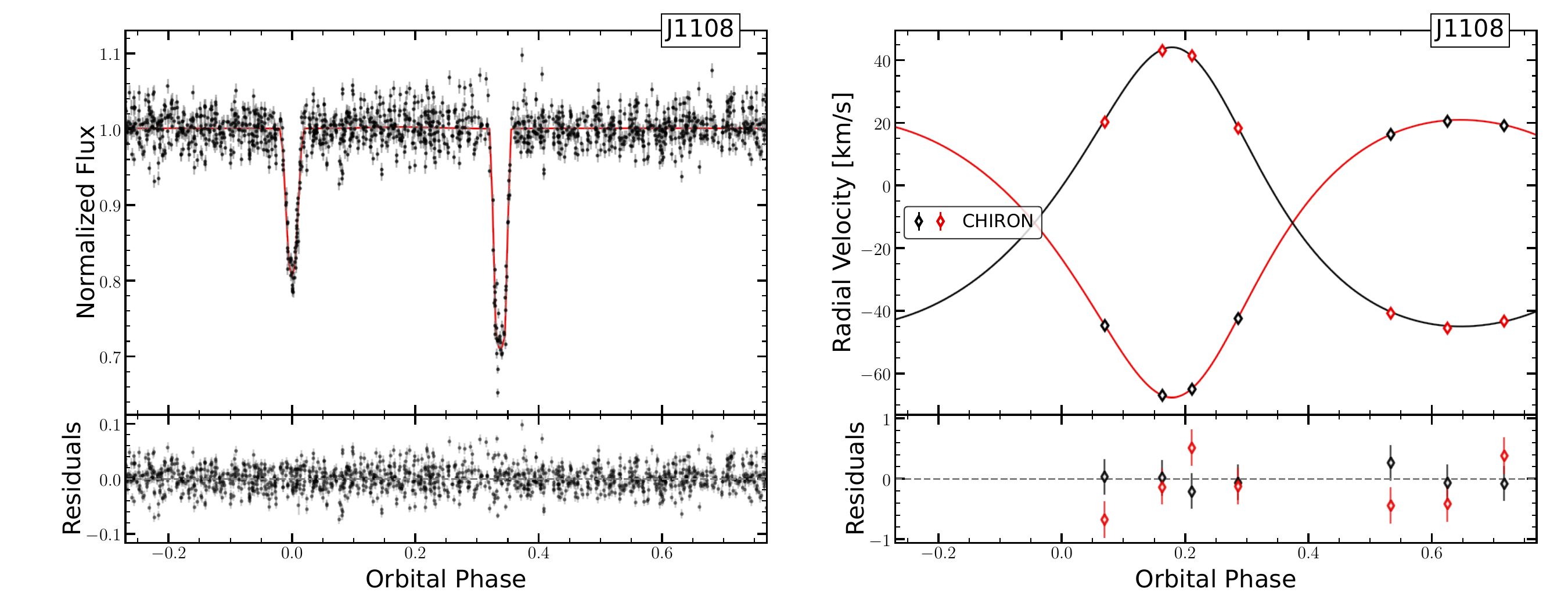} \\
    \end{tabular}
    \caption{\PHOEBE{} light curve and radial velocity fits. The RVs of the primary and secondary are the black and red points, respectively. In each panel, the \PHOEBE{} model is shown as twenty random samples from the MCMC posteriors. The smaller panels show the light curve and radial velocity residuals.}
    \label{fig:phoebe_models}
\end{figure*}

We use the Physics Of Eclipsing BinariES \citep[\PHOEBE{},][]{Prsa05, Conroy20} modeling tool to simultaneously fit the ASAS-SN light curves and the RV observations to measure masses and radii. \PHOEBE{} has been used extensively for eclipsing binaries of various morphologies, including contact binary systems \citep[e.g.,][]{Li21}, semi-detached binaries \citep[e.g.,][]{Xiong24}, detached binaries \citep[e.g.,][]{Kovalev23}, and ellipsoidal variables \citep[e.g.,][]{Rowan24}. 

We start by using the light curve and radial velocity geometry estimators within \PHOEBE{} to get an initial guess for the orbital and stellar parameters. Then, we use Nelder-Mead optimization method within \PHOEBE{} to optimize the solution and determine the starting point for our MCMC walkers. 

We sample over orbital parameters ($P$, $t_0$, $e$, $i$, $\omega$, $\gamma$) and stellar parameters ($M_1$, $q$, $R_1$, $R_2$, $T_{\rm{eff,2}}/T_{\rm{eff},1}$). We also sample over the $T_{\rm{eff},1}$ to allow the models to account for temperature effects in limb-darkening, but do not report the posteriors on $T_{\rm{eff},1}$ since the individual temperatures are only well constrained when fitting light curves in multiple filters that cover different wavelength ranges. Similarly, we sample over the \PHOEBE{} passband luminosity, which controls the scaling of the absolute fluxes computed by \PHOEBE{} to the normalized fluxes \citep{Conroy20}. For targets with both APF and PEPSI observations, we include an RV offset parameter to account for any differences in the RV zeropoint of the two spectrographs. Finally, if there is additional light in the photometry from a physical tertiary companion, blended light from nearby stars, or a poorly estimated sky background, the light curve can be ``diluted'' and the inclination underestimated. To test for the possible impact of this ``third light'' in the system, we fit two sets of models with and without a fractional third parameter, $l_3$. We discuss possible sources of third light in more detail for each target below. 

We sample over these $n=13$--$15$ parameters, depending on if $l_3$ and an RV offset are included, and use $2n$ walkers. We start by running MCMC for 20,000 iterations with the {\tt ellc} backend \citep{Maxted16}. We visually inspect the walker probabilities to select an appropriate burn-in period for each target. Burn-in periods are chosen to be between three and five times the maximum autocorrelation time, which corresponds to $\sim$5000 iterations. For three targets, we manually remove one or two walkers that have low probability and have not converged. We resample from this MCMC run and run another 5,000 iterations using the {\tt PHOEBE} backend, which is more accurate but more computationally expensive. Table \ref{tab:phoebe_table_asassn} reports the median \PHOEBE{} posterior values and $1\sigma$ uncertainties for the four binaries without \TESS{} data. The light curve fits are shown in Figure \ref{fig:phoebe_models} and we show an example of the MCMC posteriors in Figure \ref{fig:corner_example}.

For the four systems with eclipses in the \TESS{} light curves, we run \PHOEBE{} models that also include the \TESS{} data. We initialize the MCMC walkers continuing from the posteriors obtained with just the ASAS-SN and RV data (Table \ref{tab:phoebe_table_tess}). We then add in the detrended \TESS{} light curve, binning the out-of-eclipse observations to reduce the computational cost. Since we are using the detrended \TESS{} light curves, long-term variability signals like ellipsoidal modulations are removed. However, the the \PHOEBE{} model does not assume that the stars are spherical, and we do not see evidence for elliposidal modulations in the ASAS-SN light curves, so any effects from detrending the \TESS{} light curves are small. We add two additional parameters for the passband luminosity and fractional third-light in the \TESS{} $T$-band. Here we use the {\tt ellc} backend in \PHOEBE{}, rather than the default {\tt PHOEBE} backend, since the later is computationally expensive with the high-cadence \TESS{} light curves. We run the MCMC for 10,000 iterations, but find that the walkers converge quickly since they are starting from the ASAS-SN MCMC solution. 

Table \ref{tab:phoebe_table_tess} reports the MCMC posteriors for the ASAS-SN and ASAS-SN~$+$~\TESS{} fits to four EBs and Figure \ref{fig:phoebe_models_tess} shows the \TESS{} light curves fits. There are some statistically significant differences between the two models, which we discuss for each individual system below. Unsurprisingly, adding in the \TESS{} light curves generally decreases the uncertainty on the stellar radii, since the eclipse ingress and egress times are very well-constrained with the high-cadence \TESS{} light curves. Figure \ref{fig:tess_compare_corner_example} shows an example of a corner plot comparing the MCMC posteriors using just the ASAS-SN light curve and after adding in the \TESS{} light curve for J0611. 

\begin{figure*}[p]
    \centering
    \begin{tabular}{cc}
        \includegraphics[width=\linewidth]{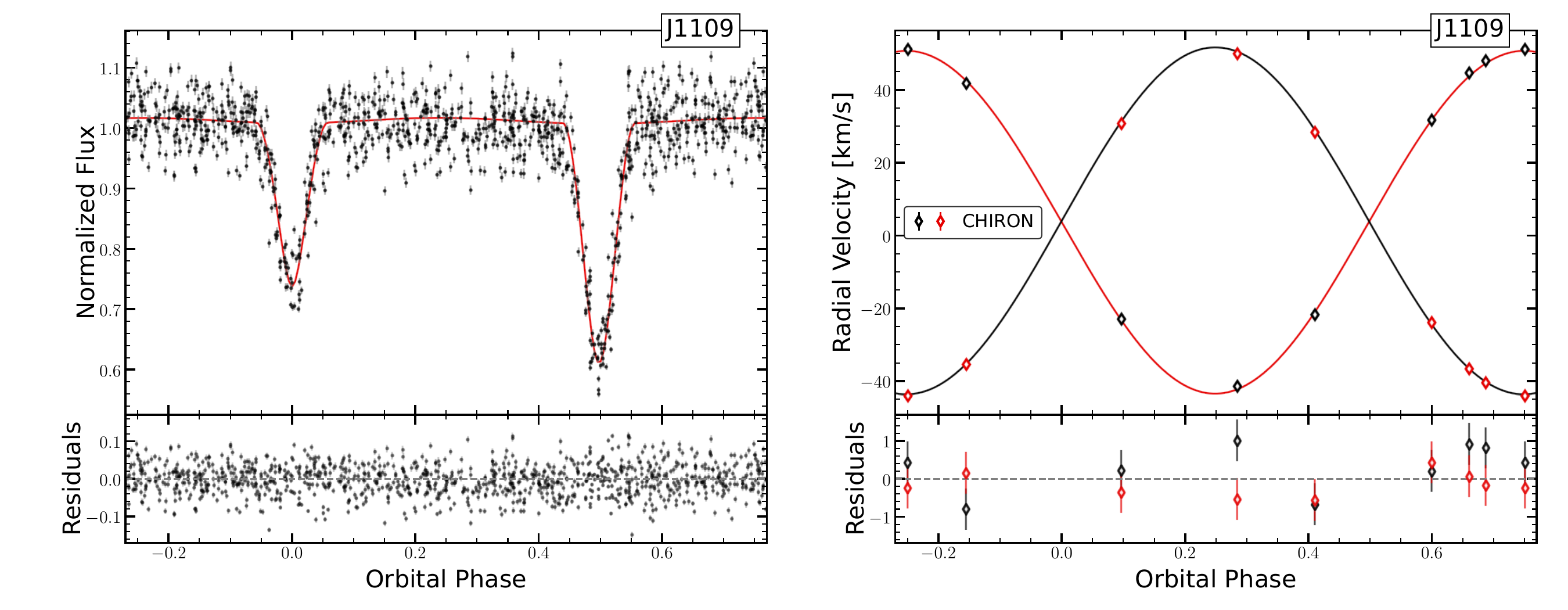} \\
        \includegraphics[width=\linewidth]{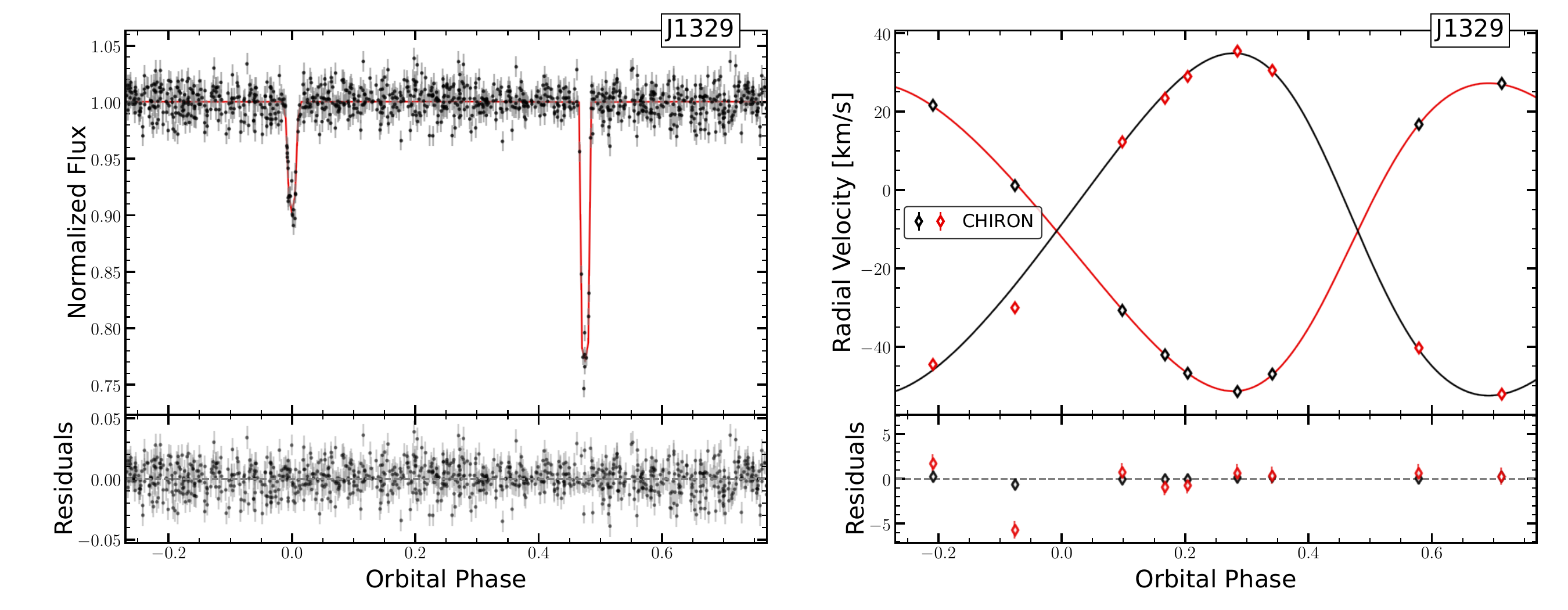} \\
        \includegraphics[width=\linewidth]{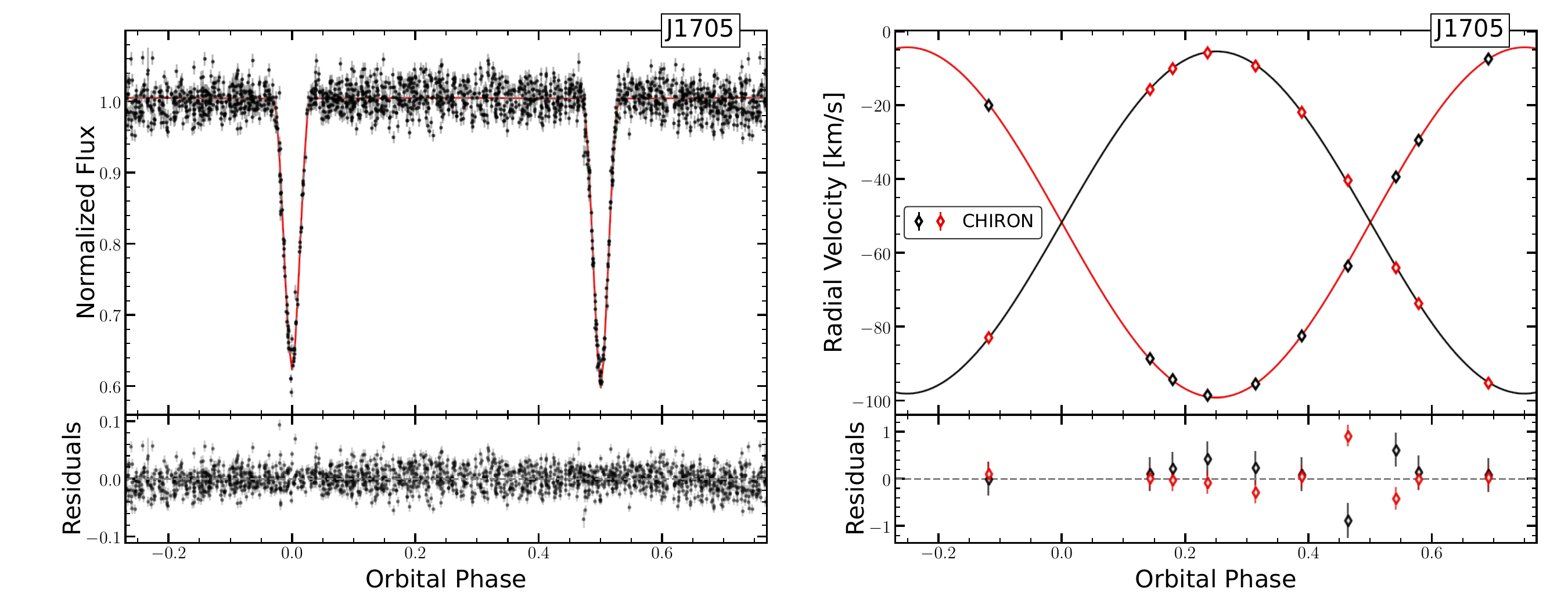} \\
    \end{tabular}
    \caption{Same as Figure \ref{fig:phoebe_models} for J1109, J1329, and J1705.}
    \label{fig:phoebe_models_2}
\end{figure*}

\begin{figure*}
    \centering
    \begin{tabular}{cc}
        \includegraphics[width=\linewidth]{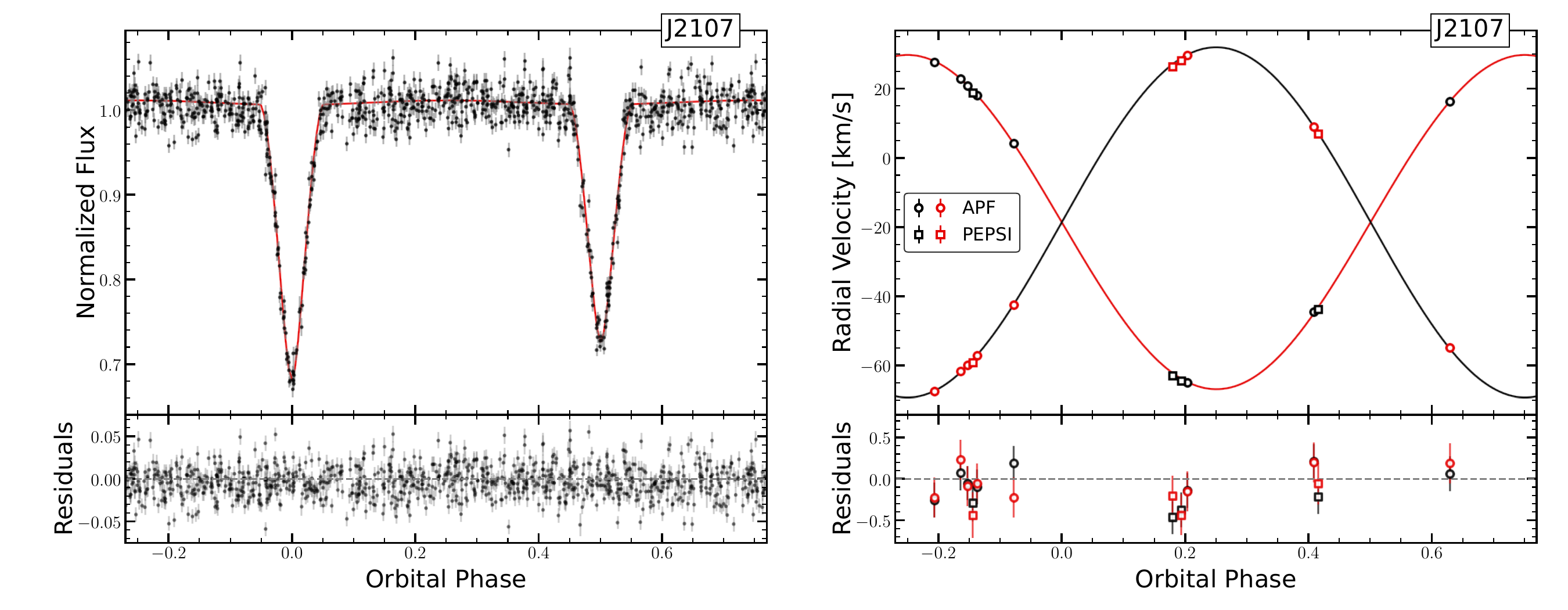} \\
        \includegraphics[width=\linewidth]{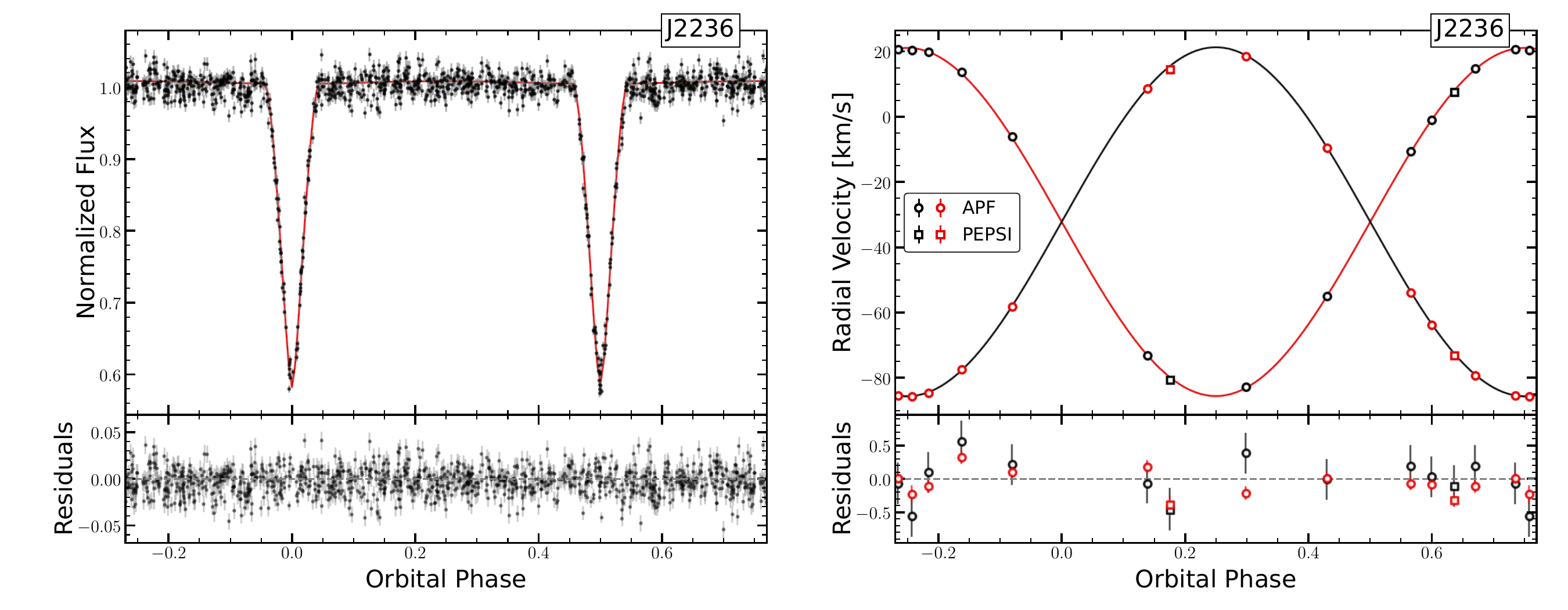} \\
    \end{tabular}
    \caption{Same as Figure \ref{fig:phoebe_models} for J2107 and J2236.}
    \label{fig:phoebe_models_3}
\end{figure*}

\begin{figure*}[p]
    \centering
    \includegraphics[width=\linewidth]{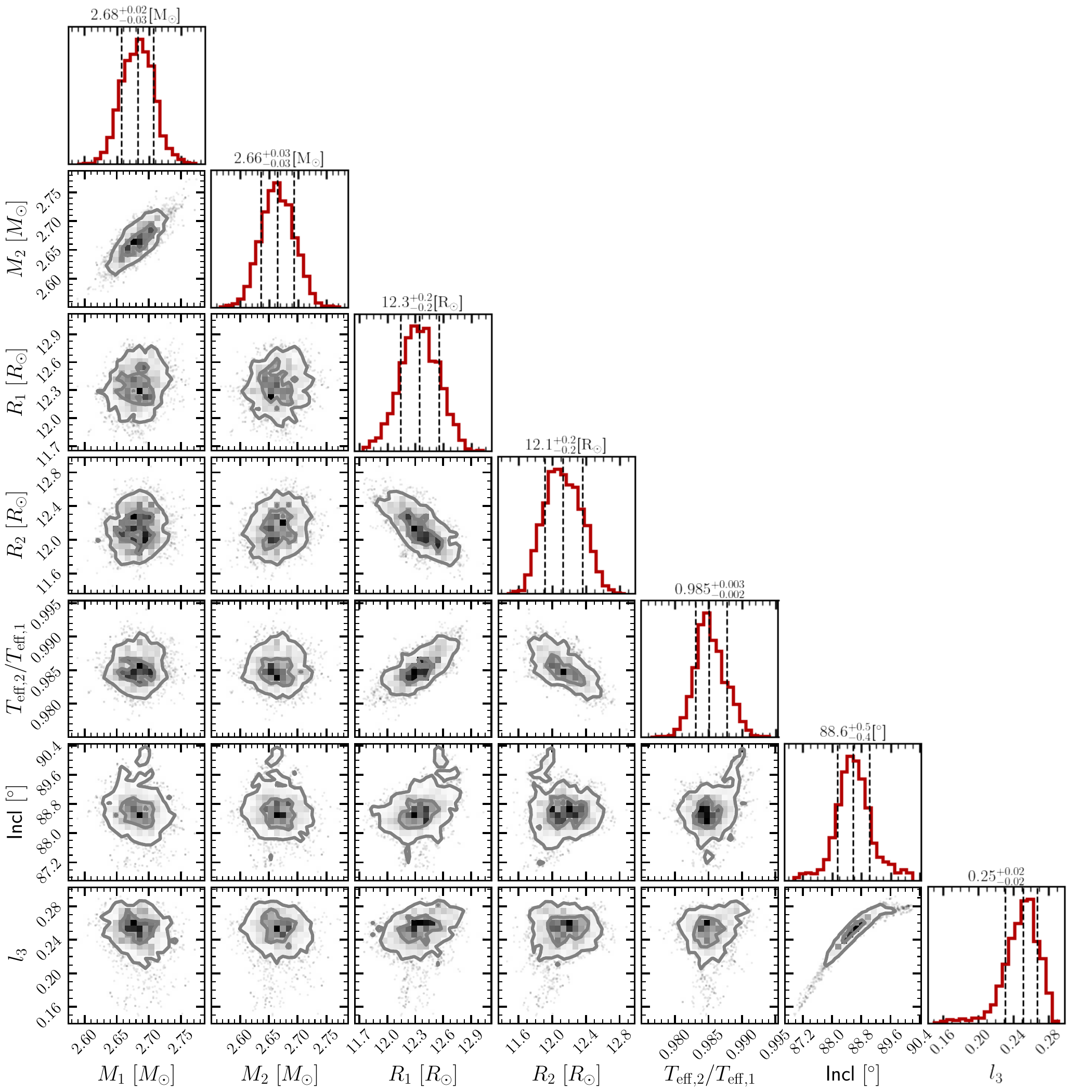}
    \caption{Example corner plot showing the MCMC posteriors for the ASAS-SN$+$RV fit to \Gaia{} J0656. Similar plots for the other seven binaries are available in the ancillary material. Table \ref{tab:phoebe_table_asassn} and \ref{tab:phoebe_table_tess} report the \PHOEBE{} MCMC posteriors and uncertainties for all targets.}
    \label{fig:corner_example}
\end{figure*}

\begin{figure*}[ht]
    \centering
    \begin{tabular}{cc}
        \includegraphics[width=0.5\textwidth]{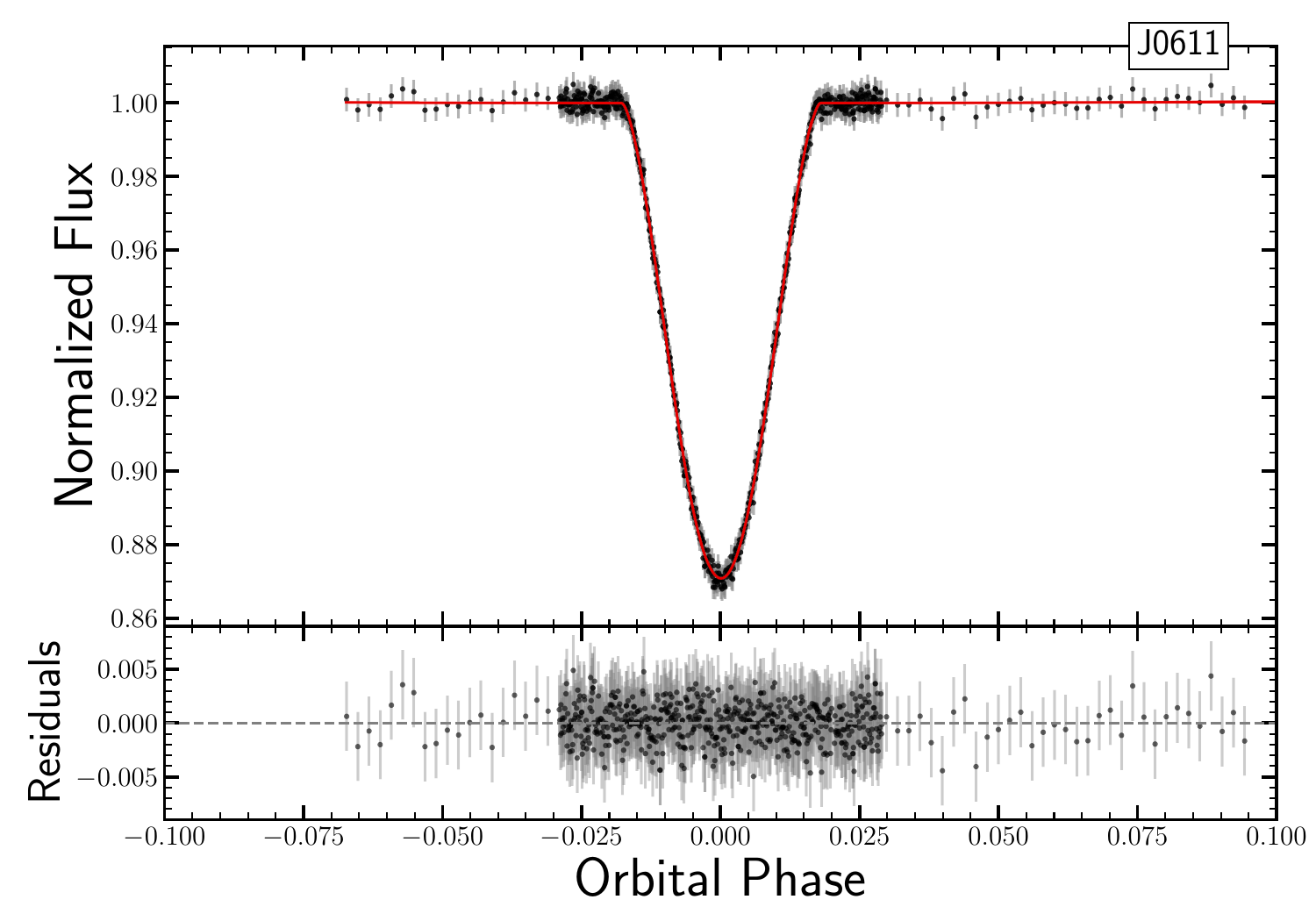} & \includegraphics[width=0.5\textwidth]{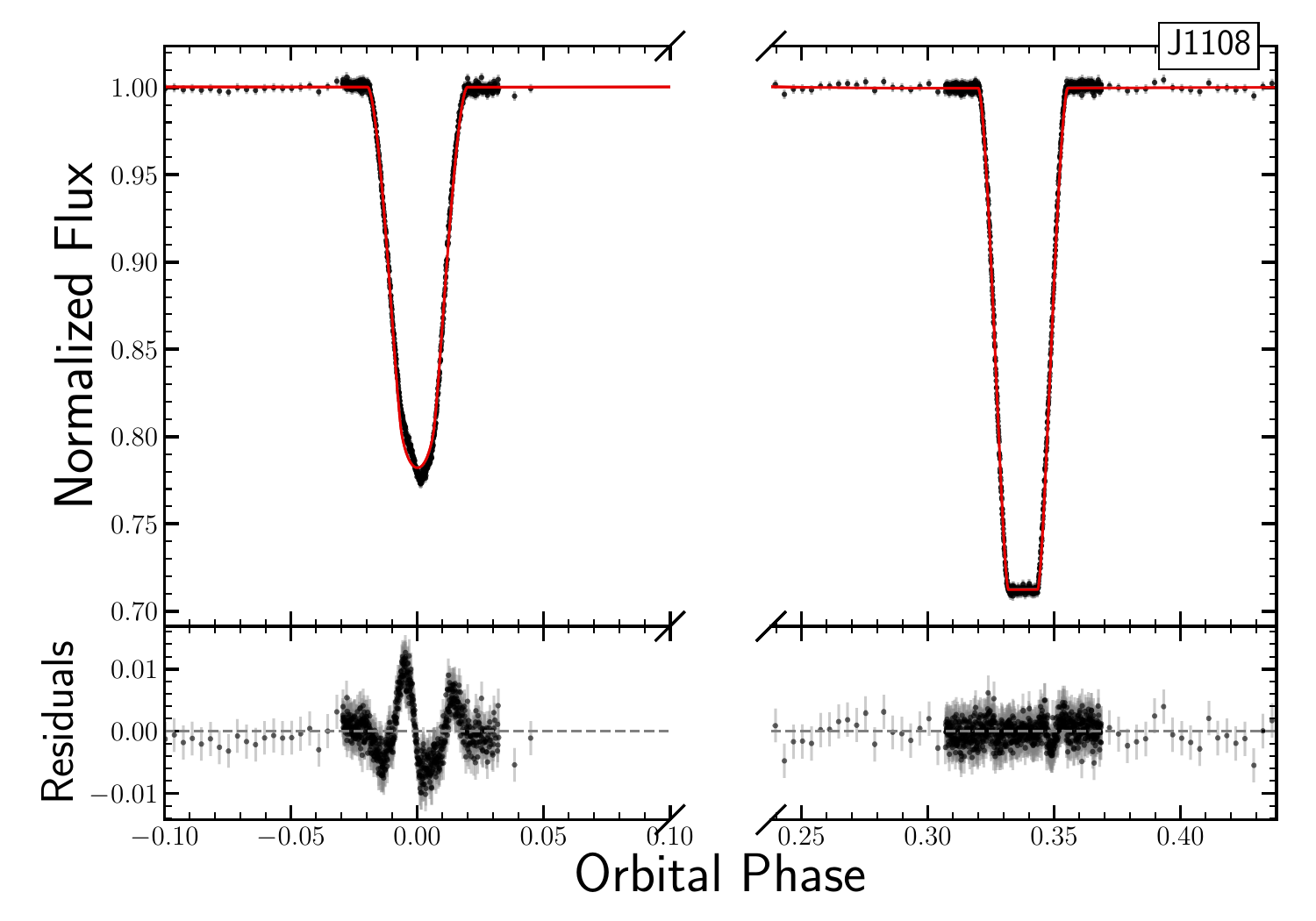} \\
        \includegraphics[width=0.5\textwidth]{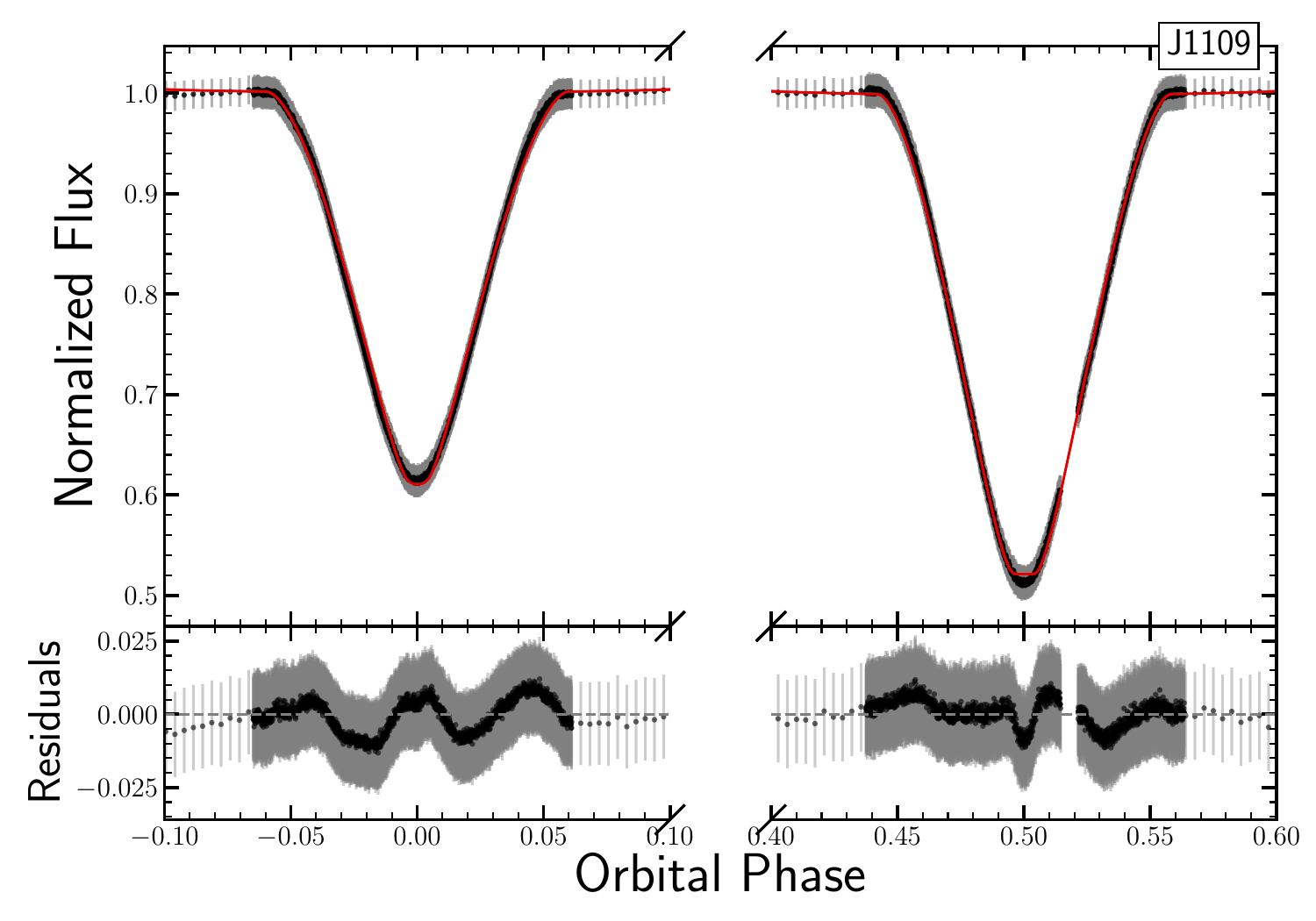} & \includegraphics[width=0.5\textwidth]{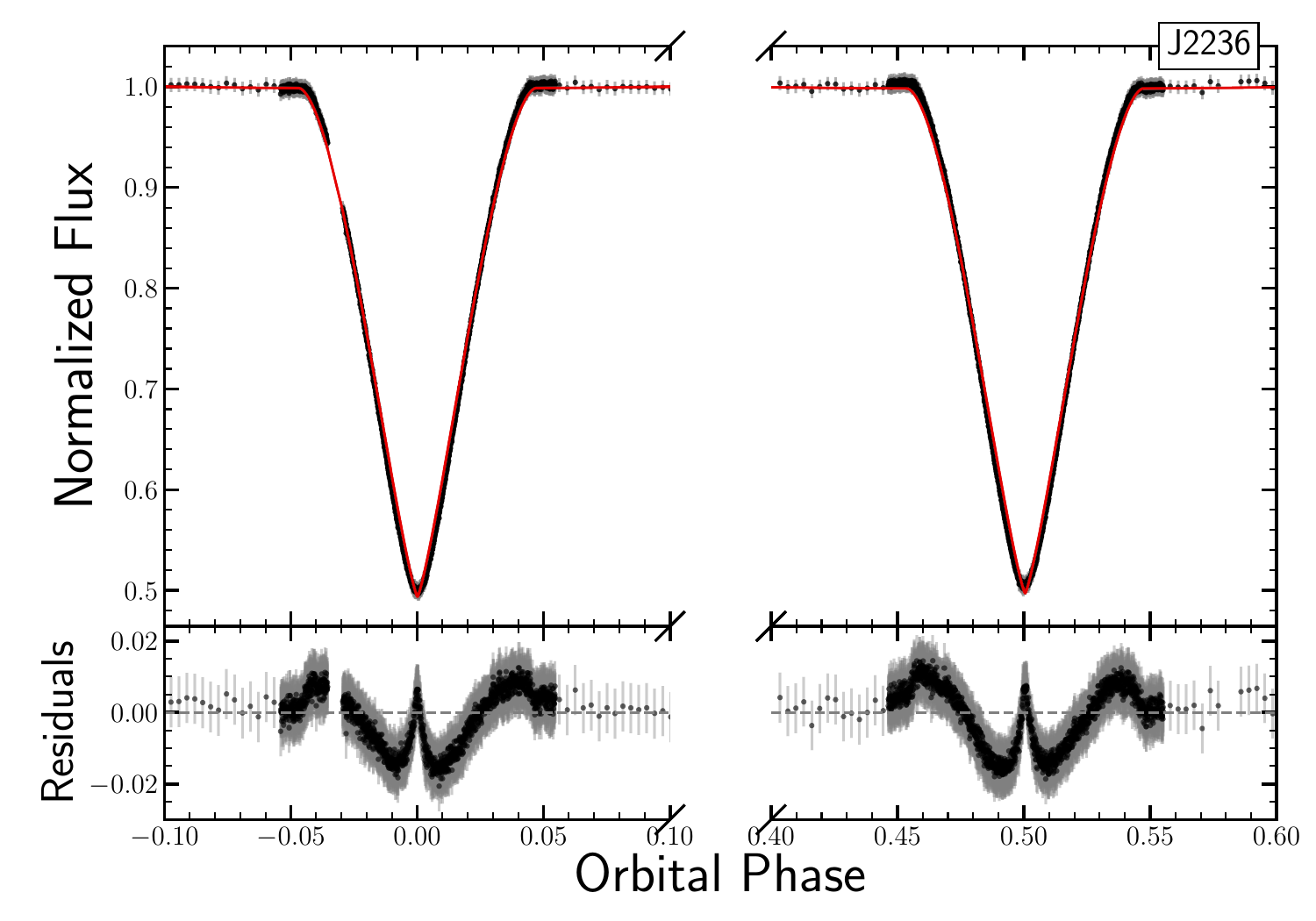} \\
    \end{tabular}
    \caption{\PHOEBE{} light curve fits to the \TESS{} data. In each panel the \PHOEBE{} model corresponding to the median of the MCMC posteriors is shown in red and black. The smaller panels show the light curve and radial velocity residuals. Three systems show low-amplitude correlated residuals in the \TESS{} light curves. For J1108 and J1109, this could be due to spots. For J1109 and J2236, $\gtrsim 20\%$ of the \TESS{} light curve is covered by eclipses, which could affect the detrending process (Section \S\ref{sec:target_selection}).}
    \label{fig:phoebe_models_tess}
\end{figure*}

\section{Comparisons to Evolutionary Tracks} \label{sec:tracks}

Unless a binary has formed though interactions in dense stellar environments, we expect the two stellar components to have the same age and metallicity. Here, we verify that this is the case by comparing our measured masses and radii to theoretical evolutionary tracks from MIST \citep{Dotter16, Choi16}. We download MIST evolutionary tracks\footnote{\url{https://waps.cfa.harvard.edu/MIST/interp_tracks.html}} for stars corresponding to our measured masses. 

For each binary component, we draw 1000 mass and radius samples from our \PHOEBE{} posteriors (Tables \ref{tab:phoebe_table_asassn} and \ref{tab:phoebe_table_tess}). Since not all targets have \TESS{} data, and spots and detrending complicate the analysis of the \TESS{} light curves, we adopt the \PHOEBE{} posteriors corresponding to the ASAS-SN$+$RV model with third-light as our final mass and radius measurements. For each sample, we determine the age when a star of mass $M$ will have radius $R$. If there are multiple ages for a given $M$ and $R$ sample, we randomly select one and construct age posteriors. We fit a one or two component Gaussian model to the age distribution to estimate the age and uncertainties. We then compare the age posteriors of the two binary components. We do this both for Solar metallicity and the [Fe/H] estimated for the RV templates (Table \ref{tab:template_table}). We assume negligible mass loss since the expected mass loss is smaller than our mass uncertainties. 

Figures \ref{fig:evolutionary_tracks} and \ref{fig:evolutionary_tracks2} show the evolutionary tracks for each target. The smaller panels show the age posteriors for each component. For stars where the measured mass and radius could be consistent with either the first ascent up the red giant branch (RGB) or the core He-burning sequence, there are two possible stellar ages. Although the relative amplitudes of the corresponding peaks in the age posteriors differ, we do not know \textit{a priori} whether a given star is a first ascent RGB star or a core He-burning star, and both have equal probability. For some systems, we can determine a more specific evolutionary state by combining information on the ages from both components. We discuss each individual system below. 

For systems that may be in the core He-burning stage, it is important to consider if they could have interacted in the past, even if the systems are currently observed as detached binaries. Some of our stars are also $>1.8\ M_\odot$, where He burning ignites under non-degenerate conditions at smaller stellar radii. In these systems, He-core burning can begin in smaller orbits without a common-envelope phase. We use the MIST evolutionary tracks and Eggleton approximation \citep{Eggleton83} to estimate the Roche-lobe radius, 
\begin{equation} \label{eqn:eggleton}
    \frac{R_{\rm{Roche}}}{a} \approx \frac{0.49 q^{-2/3}}{0.6 q^{-2/3} + \log(1+q^{-1/3})}
\end{equation}
where $q$ is the mass ratio and $a$ is the binary semimajor axis. The filling factor $f$ is then $f=R/R_{\rm{Roche}}$, where $f>1$ indicates that the star has overflowed its Roche lobe. For the systems that could have core He-burning stars, we compute the maximum $f$ the stars reach when ascending the giant branch in the MIST models and estimate how long a star has $f>1.0$. 

\section{Discussion of Individual Targets} \label{sec:results}

\begin{table*}[]
    \centering
    \caption{Summary of key measured properties for the eight binary systems.}
    \setlength{\tabcolsep}{6pt}
    \renewcommand{\arraystretch}{2}
    \begin{center}
        \begin{tabular}{c c c c c c c c c c}
        \toprule
        Source & Name & Period & Ecc & $M_1$ & $M_2$ & $R_1$ & $R_2$ \\
        & & [d] & & [$M_\odot$] & [$M_\odot$] & [$R_\odot$] & [$R_\odot$] \\
        \midrule
        3328584192518301184 & J0611 & $69.003^{+0.001}_{-0.002}$ & $0.0027^{+0.0008}_{-0.0008}$ & $2.142^{+0.009}_{-0.009}$ & $2.082^{+0.008}_{-0.009}$ & $9.5^{+0.8}_{-0.9}$ & $3.1^{+0.4}_{-0.4}$ \\ 
        3157581134781556480 & J0656 & $41.4453^{+0.0003}_{-0.0003}$ & $0.0013^{+0.0007}_{-0.0004}$ & $2.68^{+0.02}_{-0.03}$ & $2.67^{+0.03}_{-0.03}$ & $12.3^{+0.2}_{-0.2}$ & $12.1^{+0.2}_{-0.2}$ \\ 
        5388654952421552768 & J1108 & $32.39909^{+0.00008}_{-0.00009}$ & $0.2623^{+0.0005}_{-0.0004}$ & $1.056^{+0.006}_{-0.007}$ & $1.050^{+0.007}_{-0.008}$ & $4.52^{+0.09}_{-0.08}$ & $2.12^{+0.06}_{-0.06}$ \\ 
        5347923063144824448 & J1109 & $31.7547^{+0.0001}_{-0.0001}$ & $0.0044^{+0.0004}_{-0.0002}$ & $1.41^{+0.02}_{-0.02}$ & $1.40^{+0.02}_{-0.02}$ & $11.7^{+0.1}_{-0.1}$ & $9.7^{+0.2}_{-0.2}$ \\ 
        6188279177469245952 & J1329 & $37.3338^{+0.0002}_{-0.0001}$ & $0.142^{+0.003}_{-0.003}$ & $1.12^{+0.02}_{-0.02}$ & $1.01^{+0.01}_{-0.01}$ & $3.1^{+0.2}_{-0.1}$ & $1.05^{+0.08}_{-0.07}$ \\ 
        5966976692576953216 & J1705 & $52.6155^{+0.0006}_{-0.0002}$ & $0.0012^{+0.001}_{-0.0009}$ & $2.20^{+0.01}_{-0.01}$ & $2.26^{+0.02}_{-0.02}$ & $8.5^{+0.2}_{-0.2}$ & $9.7^{+0.1}_{-0.2}$ \\ 
        1969468871480562560 & J2107 & $68.1395^{+0.0006}_{-0.0006}$ & $0.0010^{+0.0005}_{-0.0004}$ & $3.49^{+0.02}_{-0.02}$ & $3.33^{+0.01}_{-0.01}$ & $21.2^{+0.5}_{-0.5}$ & $21.4^{+0.4}_{-0.5}$ \\ 
        2002164086682203904 & J2236 & $36.8365^{+0.0001}_{-0.0001}$ & $0.0007^{+0.0003}_{-0.0002}$ & $2.321^{+0.006}_{-0.006}$ & $2.318^{+0.009}_{-0.01}$ & $10.9^{+0.1}_{-0.1}$ & $10.5^{+0.1}_{-0.1}$ \\ 
        \bottomrule
        \end{tabular}
    \end{center}
    \label{tab:phoebe_summary}
\end{table*}

Figure \ref{fig:mass_radius} shows the mass and radius measurements for the 16 stars compared to the \citet{Torres10} catalog and a sample of oscillating eclipsing red giants from \citet{Gaulme16}, \citet{Themebl18}, and \citet{Brogaard22}. Table \ref{tab:phoebe_summary} summarizes our measurements and the key properties of each target. Seven of the binaries have two evolved stars, and one system has a secondary that is still on the main sequence. In all binaries, the stars are nearly the same mass, and the maximum mass difference is $\Delta M = 0.11\ M_\odot$ for J1329. As compared to previous samples of red giant EBs from OGLE \citep[e.g.,][]{Graczyk20} or \textit{Kepler} \citep[e.g.,][]{Themebl18}, our systems are generally on smaller, more circular orbits. 

\subsection{J0611: GDR3 3328584192518301184} \label{sec:3328584192518301184}
\newcommand{\source}{3328584192518301184}
\newcommand{\shortname}{J0611}

\shortname{} (ASASSN-V J061119.27$+$082957.4) has components \mbox{\primarymass{\source},} \primaryradius{\source} and \mbox{\secondarymass{\source},} \mbox{\secondaryradius{\source}.} Despite the small difference in mass between the two components ($\sim 0.06\ M_\odot$), the radius difference is large $\sim 6.4\ R_\odot$. However, as compared to the other targets, \shortname{} has shallower eclipses, leading to larger uncertainties on the radii. This target is also the faintest in the sample (Table \ref{tab:summary_table}), but the relatively poor light curve fit is due to the lower inclination of $\sim 87^{\circ}$ rather than photometric uncertainties. 

The \PHOEBE{} model that includes third light prefers a solution with $l_3=0.16\pm0.10$, but both models predict masses and radii that are consistent within their uncertainties. We use the ATLAS All-Sky Reference Catalog \citep[ATLAS-REFCAT2,][]{Tonry18refcat} to search for nearby stars that could contribute to ASAS-SN flux and dilute the observed light curve. There is a nearby star (GDR3 3328584196815079168), $4\farcs5$ from \shortname{}, that has $g=16.8$~mag, and there is another $g=16.5$~mag star (GDR3 3328584231177668736) that is separated by $11\farcs0$. These nearby stars, which are $\Delta m \approx 3.8$~mag fainter than \shortname{}, are too faint to explain the estimated fractional third light contribution of $l_3 = 0.16$ in the $g$-band, but we note that the uncertainties on $l_3$ are large and consistent with zero at $\sim 1.5\sigma$.

\shortname{} also has a TESS light curve (TIC 166929994). The Sector 33 light curve shown in Figure \ref{fig:tglc_tess} contains only one eclipse. In general, both primary and secondary eclipses are needed to measure both masses and radii in an EB, but here we simultaneously fit the \TESS{} data with the ASAS-SN light curve, which covers both eclipses. The high-cadence \TESS{} light curve gives more precise eclipse times and improves the fit to the shape of the eclipse. This is especially important for this target, which has the shallowest eclipses in the sample. Figure \ref{fig:tess_compare_corner_example} shows a comparison of the MCMC posteriors between the models with and without the \TESS{} light curve. The addition of the \TESS{} light curve improves the orbital inclination determination, which improves the mass and radius posteriors. Figure \ref{fig:phoebe_models_tess} shows the \TESS{} light curve fit, which has residuals consistent with noise.  

Figure \ref{fig:evolutionary_tracks}a shows the evolutionary tracks corresponding to the \PHOEBE{} models of J0611. The evolutionary models are consistent with a system where the primary is on the first ascent of the RGB and the secondary has just evolved off the main~sequence. 

\subsection{J0656: GDR3 3157581134781556480} \label{sec:3157581134781556480}
\renewcommand{\source}{3157581134781556480}
\renewcommand{\shortname}{J0656}

The two components of \shortname{} (ASASSN-V J065618.52$+$092626.8) have very similar masses, \primarymass{\source} and \mbox{\secondarymass{\source}.} The radii (\primaryradius{\source} and \secondaryradius{\source}) differ by $\sim 0.2\ R_\odot$, but agree within the uncertainties. The ASAS-SN eclipses are deep ($>30\%$) and have roughly equal depth, so the effective temperature ratio is close to one. 

This target has a large fractional third light that is significantly greater than zero, $l_3 = 0.25\pm0.02$. There is a nearby star, GDR3 3157581139074431488, that is separated by $4\farcs15$ with $g=15.8$~mag. Even if a star of that magnitude was entirely under the ASAS-SN PSF, it would only contribute $\sim 6\%$ to the total flux. The next nearest star is $8\farcs8$ away and is $g=19.6$~mag. There is also no evidence for a wide, bound companion in \Gaia{} DR3. The \Gaia{} renormalized unit weight error is $\texttt{RUWE}=1.2$, and an additional resolved companion was not identified in any observations ($\texttt{ipd\_frac\_multi\_peak}=0$). There is no published \Gaia{} astrometric orbit solution. Additional RV observations could be used to search for evidence of a tertiary companion in the RV orbit model residuals. The \PHOEBE{} model run without a third light component predicts a smaller inclination ($88.6^{\circ}$ versus $85.4^{\circ}$, which increases the masses of both components, but the two sets of mass measurements are consistent within $1\sigma$. 

Figure \ref{fig:evolutionary_tracks}b shows the evolutionary tracks of J0656. This is a twin system, so the evolutionary tracks overlap for the entire evolution. Based on the measured stellar radii, the system could be either on the first ascent of the RGB or a core He-burning star. If the system is a core He-burning star, the radii were previously much larger and mass transfer could have occurred. We use the MIST evolutionary tracks to compute the Roche-lobe filling factor $f$ (Equation \ref{eqn:eggleton}). We find that both components would have filled their Roche-lobes immediately before He burning began with maximum $f=1.05$. However, this period of mass transfer would be brief, lasting $\lesssim 0.3$~Myr. More detailed binary evolution models would be necessary to determine how mass transfer could alter the evolution of this system.

\subsection{J1108: GDR3 5388654952421552768} \label{sec:5388654952421552768}
\renewcommand{\source}{5388654952421552768}
\renewcommand{\shortname}{J1108}

\begin{figure*}
    \centering
    \includegraphics[width=0.8\linewidth]{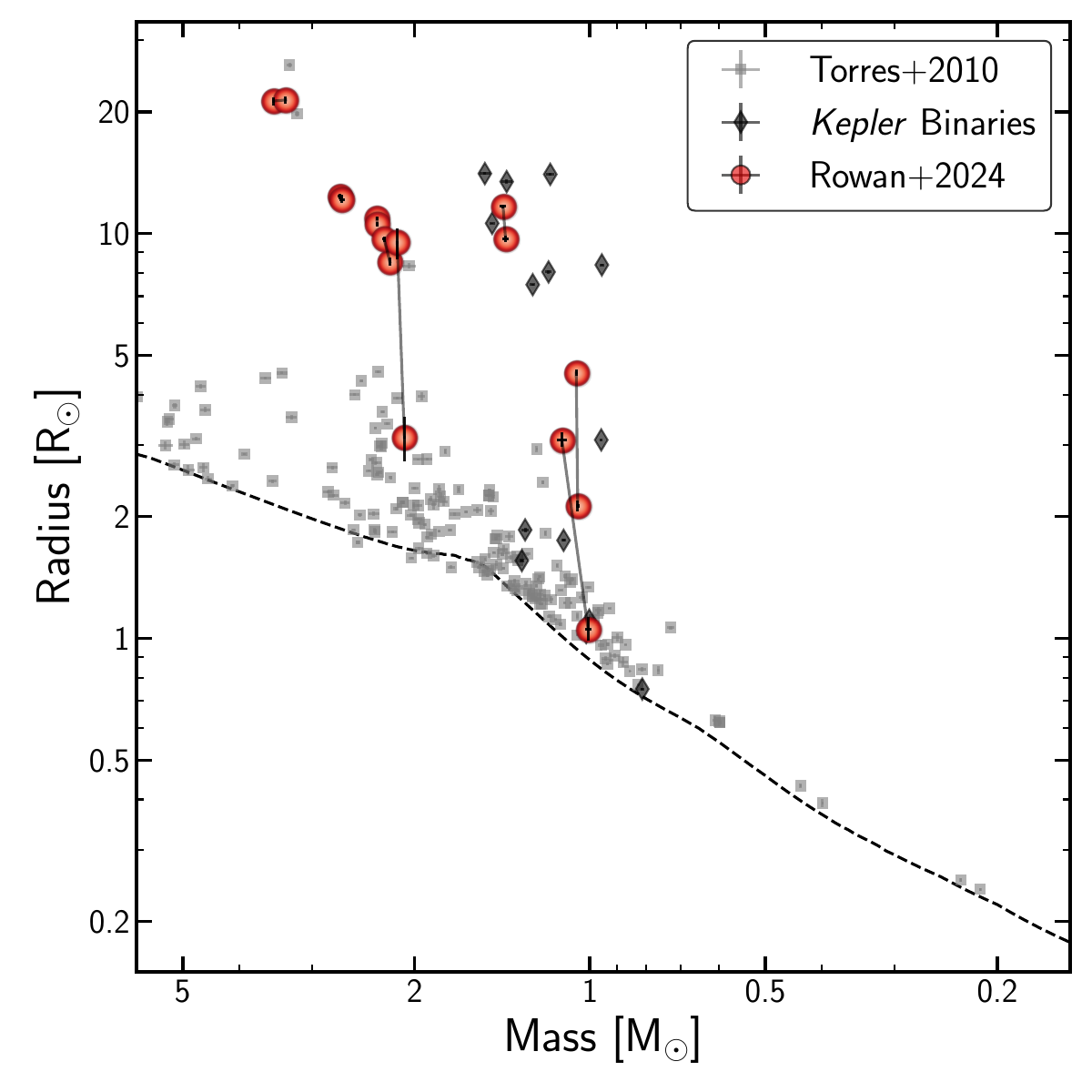}
    \caption{Mass and radius results for the eight binary systems (red). The solid lines connect binary companions. The \citet{Torres10} catalog is shown in gray, and the oscillating eclipsing red giants from \textit{Kepler} are shown in black \citep{Gaulme16, Themebl18, Brogaard22}. The dashed black line shows the zero-age main sequence at Solar metallicity from MIST evolutionary tracks \citep{Choi16, Dotter16}.}
    \label{fig:mass_radius}
\end{figure*}

\begin{figure*}
    \centering
    \includegraphics[width=\linewidth]{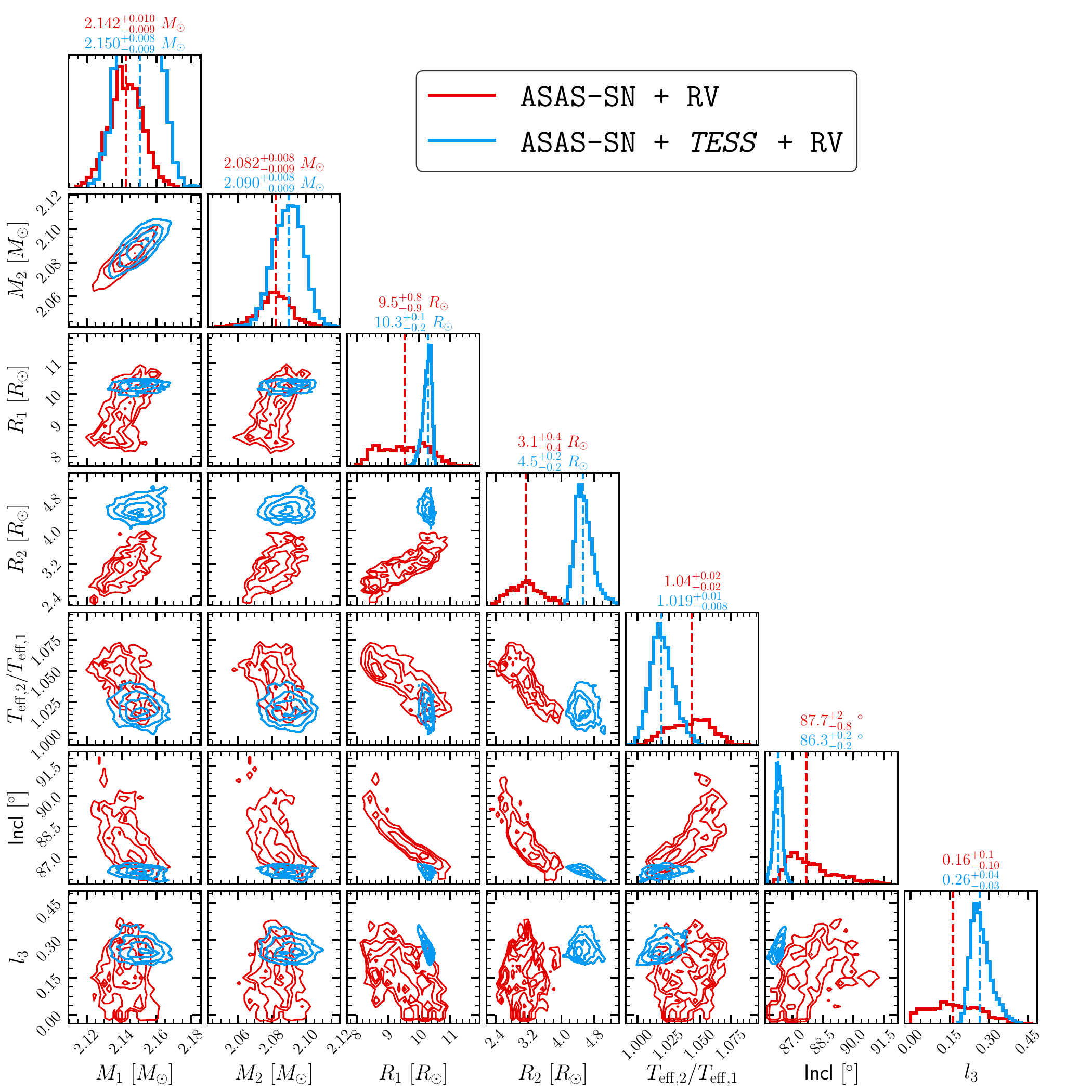}
    \caption{Comparison of the \PHOEBE{} MCMC posteriors for J0611 using just the ASAS-SN light curve and the RVs (red) and adding the \TESS{} light curve (blue).}
    \label{fig:tess_compare_corner_example}
\end{figure*}

\shortname{} (ASASSN-V J110800.86$-$440658.9) is one of two eccentric binaries in the sample ($e=0.26$). The primary and secondary masses are consistent within $1\sigma$, with \primarymass{\source} and \mbox{\secondarymass{\source}}. However, the radii differ by more than a factor of two with \primaryradius{\source} and \mbox{\secondaryradius{\source}}. The temperatures of the stars are also different by $\sim 9\%$, with the secondary being the hotter of the pair. Based on the location on the \Gaia{} CMD (Figure \ref{fig:cmd}) and the mass-radius figure (Figure \ref{fig:mass_radius}), the binary has just evolved off of the main sequence, so it is not surprising that a small difference between the masses of the two components results in a large difference in radius.

\shortname{} is also one of the three targets with additional rotational variability in the light curve (Figure \ref{fig:unfolded_lightcurves}). This target is detected as an X-ray source in \textit{eROSITA} \citep{Merloni24} with a separation of $4\farcs4$ and is classified as a coronal emitting source \citep{Freund24}. \shortname{} was observed by the RAdial Velocity Experiment \citep[RAVE,][]{Steinmetz06} and found to be chromospherically active based on the CaII triplet \citep{Zerjal17}.

We use the \PHOEBE{} geometry estimator to mask out the eclipses and search for periodicity in the non-detrended ASAS-SN light curve. There is periodic variability at $P=65.46$~d, which is $\sim 2$ times the orbital period. This suggests that the stars are tidally synchronized even if the orbit is not tidally circularized. For a $q=1$ binary at $P=32.4$~days, the difference between the tidal circularization and tidal synchronization timescales is only $\approx 10^4$~years \citep{Zahn77}. As the giants continue to evolve and $R$ increases, the circularization timescale will decrease further. 

\shortname{} (TIC 71877648) was observed by \TESS{} in four sectors. Figure \ref{fig:tglc_tess} shows the Sector 63 \TESS{} light curve, which includes both eclipses. One of the eclipses occurs close to the end of the Sector, which could introduce systematic effects on the eclipse shape. The \TESS{} light curve (Figure \ref{fig:phoebe_models_tess}) shows that the \PHOEBE{} model fits the deeper eclipse well. The flat-bottomed eclipse, indicating a total eclipse, is much more apparent in the \TESS{} light curve than in the ASAS-SN data (Figure \ref{fig:phoebe_models}). As with J0611, the \TESS{} light curve improves the inclination constraint, leading to better-constrained masses and radii (Table \ref{tab:phoebe_table_tess}).

The shallower eclipse shows a clear asymmetry in the \TESS{} light curve, which introduces correlated residuals on either side of the minimum. This is likely due to star spots, which create asymmetric eclipse profiles as the star transits the non-uniform stellar disk. This is consistent with the evidence for chromospheric activity from \textit{eROSITA} and RAVE. We investigated the other three \TESS{} sectors and find that the eclipse changes shape, which is expected since the spots evolve over time, although it is difficult to disentangle this astrophysical asymmetry origin from systematic effects in the \TESS{} detrending procedure. While it may be possible to simultaneously fit for the spot parameters and improve the \PHOEBE{} fit to this eclipsing feature, we do not do so here because of the degeneracies in modeling the light curves of spotted stars \citep[e.g.,][]{Luger21}.

Figure \ref{fig:evolutionary_tracks}c shows the evolutionary tracks of \shortname{}. The evolutionary tracks indicate that both stars have only recently evolved off of the main sequence, which is consistent with their CMD position. At $\rm{[Fe/H]}=0.18$, the MIST tracks predict an age $\approx 10.5$~Gyr. At Solar metallicity, the binary age is predicted to be slightly younger, $\approx 8.9$~Gyr. This is one of the two low mass ($\sim 1\ M_\odot$) binaries in our sample, and it is unsurprising that it is older than the higher mass systems despite being relatively less evolved. Based on the dynamical mass and radius, $\log g=3.15$, so the primary has likely just completed first dredge-up \citep[see Fig. 1,][]{Roberts24}.

\subsection{J1109: GDR3 5347923063144824448} \label{sec:5347923063144824448}
\renewcommand{\source}{5347923063144824448}
\renewcommand{\shortname}{J1109}

\shortname{} (ASASSN-V J110949.25$-$521017.0) has two stars of similar mass, \primarymass{\source} and \mbox{\secondarymass{\source}} in a circular orbit. The radii of the two stars are \primaryradius{\source} and \mbox{\secondaryradius{\source}}. The phase folded light curves show significant scatter even after removing long-term variations in the ASAS-SN light curve (Figure \ref{fig:unfolded_lightcurves}). Unlike J1108 and J1705, where the {\tt wotan} trend shows periodic variability on the timescale of the orbital period, the photometry of \shortname{} is dominated by a long-term trend and a sudden jump in brightness around $\rm{JD}=2459650$. Since this target is $g\sim11.4$~mag, this trend is likely due to systematic effects for stars approaching the ASAS-SN saturation limit ($g<11.9$~mag, see Figure 3 of \citetalias{Rowan22}). However, like J1108, this source is listed as a chromospherically active star in RAVE \citep{Zerjal17} and detected as an X-ray coronal source in \textit{eROSITA} \citep{Freund24} with a separation of $0\farcs4$. Whether the scatter is from systematic effects in ASAS-SN or chromospheric activity, the orbital and stellar parameters are well-constrained with fractional uncertainties of $\lesssim 1\%$. 

\shortname{} is the brightest target in our sample and is also the only one to be included in the \Gaia{} DR3 catalog of SB2 orbit models \citep{GaiaCollaboration2022, Arenou23}. The \Gaia{} SB2 RV orbit model fits 13 epochs and finds $K_1=52.4\pm1.1$~km/s and $K_2=44.8\pm0.8$~km/s, so $q=0.85$. This is significantly less than the $q=1$ we measure from our RV observations and \PHOEBE{} model. \citet{Rowan23} used the \Gaia{} RV orbit model with the ASAS-SN light curve to measure masses and radii of 61 binaries, including \shortname{}, and consequently reported mass and radius measurements that disagree with those we report here. Unfortunately, only the RV orbit model is included in \Gaia{} DR3, and the epoch RV measurements from \Gaia{} are unavailable, limiting our ability to make a more direct comparison between the two sets of measurements. However, \citet{Rowan23} also found that $\sim 50\%$ of the \Gaia{} RV orbit models for ASAS-SN eclipsing binaries had incorrect periods or eccentricities, so it may not be surprising that some RV orbit models have inaccurate velocity semi-amplitudes as well. 

The MCMC posteriors suggest a large fractional third light, $l_3 = 0.21$. There is a nearby star, GDR3 5347923097493539072 that is separated by $3\farcs7$. This nearby star has $g=13.1$~mag, which cannot contribute $21\%$ extra flux in the ASAS-SN photometry. There is no evidence for a tertiary companion to \shortname{} in the \Gaia{} RUWE or {\tt ipd\_frac\_multi\_peak} statistics. The \PHOEBE{} model that does not include third light predicts a lower inclination ($83.4^{\circ}$ versus $86.1^{\circ}$), but the masses are only increased slightly and are consistent within their uncertainties.

\shortname{} (TIC 81462274) has been observed in five \TESS{} sectors. Figure \ref{fig:tglc_tess} shows the Sector 63 light curve where both eclipses were observed. As compared to the \PHOEBE{} fits that only used the ASAS-SN light curve, the combined ASAS-SN and \TESS{} \PHOEBE{} model prefers a higher inclination ($88.6^{\circ}$ versus $86.1^{\circ}$). As a result, both component masses decrease by $\sim 0.03\ M_\odot$. As compared to J0611 and J1108, the eclipses of this target are wider, which could introduce systematic effects in the detrending process since the eclipses are masked. Figure \ref{fig:phoebe_models_tess} shows the \TESS{} light curve and \PHOEBE{} model. We find there are some correlated residual features, but the scale of these residuals is small ($<2.5\%$) relative to the photometric errors, so it is reasonable to conclude that MCMC walkers have converged on this solution. 

Figure \ref{fig:evolutionary_tracks}d shows the evolutionary tracks for \shortname{}. The components have masses consistent with each other within their uncertainties and the evolutionary tracks overlap for the full age range. The evolutionary tracks show that a star of mass $\sim 1.4\ M_\odot$ does not shrink back to $\sim 11\ R_\odot$ following the He-flash, so the components of J1109 are probably first ascent RGB stars. 

\Needspace{10\baselineskip}
\subsection{J1329: GDR3 6188279177469245952} \label{sec:6188279177469245952}
\renewcommand{\source}{6188279177469245952}
\renewcommand{\shortname}{J1329}

\shortname{} (ASASSN-V J132912.67$-$283324.7) is the only one of our systems where one component is still on the main sequence. This is also the only other target in the sample besides J1108 to have an eccentric orbit with $e=0.142$. The photometric primary has \primarymass{\source} and \mbox{\primaryradius{\source}.} The secondary is a Sun-like star with \secondarymass{\source} and \mbox{\secondaryradius{\source}.} It follows that this binary has the faintest absolute magnitude out of the sample, $M_G=2.6$~mag (Table \ref{tab:summary_table}) and is near the boundary between subgiant and giant binaries based on the criteria used in \citetalias{Rowan22} (Figure \ref{fig:cmd}). 

This target was included in the \Gaia{} DR3 catalog of single-lined spectroscopic binaries \citep{Arenou23}. The \Gaia{} RV orbit solution is $P=37.3\pm0.03$~km/s, $e=0.16\pm0.02$, and $K_1=38.4\pm0.4$~km/s. The \Gaia{} orbital period and eccentricity are consistent with our solution to within $1\sigma$, and the \Gaia{} velocity semi-amplitude is consistent to within~$2\sigma$.

J1329 (Figure \ref{fig:evolutionary_tracks2}a) has the largest difference between the age estimates of the two components, though they do agree within $1\sigma$. The primary of this system has just evolved off the main sequence and the companion is towards the end of its main sequence lifetime. As a result, the age posterior on the secondary is fairly broad. Figure \ref{fig:evolutionary_tracks2} shows the evolutionary tracks at super-Solar metallicity $\rm{[Fe/H]}=+0.5$, as measured from the {\tt iSpec} disentangled spectra (Table \ref{tab:template_table}). If we instead use evolutionary tracks at Solar metallicity, the age posteriors do not agree within $1\sigma$. Based on the surface gravity of the photometric primary, $\log g\approx 3.5$ from the dynamical mass and radius, the star has likely only just begun first-dredge up \citep{Roberts24}, so it is unlikely that additional constraints on its age could be determined from surface abundances. As with J1108, this system must be old ($>8$~Gyr) in order to observe the low mass binary near the start of its first ascent up the RGB. We would naively expect older stars to be metal poor, but RAVE reports a metallicity $\rm{[Fe/H]}=0.24$ \citep{Kunder17}, and \Gaia{} uses a super-Solar metallicity template ($\rm{[Fe/H]}=0.25$) for their RV measurements, supporting the higher metallicity we measure for the RV templates.

\subsection{J1705: GDR3 5966976692576953216} \label{sec:5966976692576953216}
\renewcommand{\source}{5966976692576953216}
\renewcommand{\shortname}{J1705}

\shortname{} (V603 Sco) is a twin system with masses \primarymass{\source} and \secondarymass{\source} and radii \primaryradius{\source} and \mbox{\secondaryradius{\source}}. This is also the third and final target to show additional variability in the light curve, though at much lower amplitude than the other two targets (Figure \ref{fig:unfolded_lightcurves}). We mask out the eclipses in the non-detrended light curve and find a periodic signal at $P=51.42$~days. This is only slightly less than the orbital period of the binary $P=52.62$~days, suggesting the binary is nearly tidally synchronized. This target was detected by \textit{eROSITA} in the $0.2$--$2.3$~keV band with a separation of $2\farcs6$ \citep{Merloni24}, but it is not included in the \textit{eROSITA} coronal source catalog.

The ASAS-SN binary stars catalog \citetalias{Rowan22} incorrectly reports the orbital period for this system to be $\sim 26.3$~days, which is roughly half of the orbital period we report here. The estimates of the other orbital and stellar parameters reported in \citetalias{Rowan22} are also likely unreliable for this target. 

The evolutionary tracks of J1705 (Figure \ref{fig:evolutionary_tracks2}b) show that our measurements are inconsistent with both stars being first ascent RGB stars. Instead, both components are more likely core He-burning stars. The MIST evolutionary tracks show that if both stars are core He-burning, the maximum Roche-lobe filling factors of the primary and secondary on the first ascent of the RGB would be $f\approx0.83$ and $f\approx0.77$ for the primary and secondary star, respectively, so mass transfer is unlikely to have occurred. 

\subsection{J2107: GDR3 1969468871480562560} \label{sec:1969468871480562560}
\renewcommand{\source}{1969468871480562560}
\renewcommand{\shortname}{J2107}

\shortname{} (ASASSN-V J210726.63$+$421401.3) is the highest mass binary in our sample, with \primarymass{\source} and \secondarymass{\source} and it has the brightest absolute $G$-band magnitude (Figure \ref{fig:cmd}). Despite the mass difference of $\sim 0.17\ M_\odot$, the \PHOEBE{} model prefers two stars with similar radii, \primaryradius{\source} and \mbox{\secondaryradius{\source}}.

The \PHOEBE{} model also prefers a large fractional third light in the $g$-band, $l_3=0.29^{+0.02}_{-0.03}$. The nearest star is separated by $6\farcs6$ with $g\sim18.8$~mag, which is $\Delta g = 5.4$~mag fainter than the target. While the RV model has larger residuals for this target compared to other binaries, the ASAS-SN light curve shows deep, sharp eclipses and the \PHOEBE{} model light curve residuals are consistent with noise. If the third light contribution is from a bound tertiary companion, we might expect to see long-term trends in the RV residuals. The PEPSI observations, which were all taken between 120 and 300 days after the APF observations, do all have negative residuals, but additional observations would be needed to model the dynamical effects from a potential third body.

The \PHOEBE{} model run without including third light finds a lower inclination and masses that are larger by $>2\sigma$. In this model, the radii are no longer equal, with the less massive secondary having a larger radius. We investigate evolutionary tracks for both sets of mass and radius measurements. Figure \ref{fig:evolutionary_tracks2}c shows the evolutionary tracks for the models including $l_3$ at $\rm{[Fe/H]}=+0.5$. This shows our mass and radius measurements are consistent with a system where the primary star has started core-He burning and the secondary is on the first ascent of the RGB. This is also true when we use the mass and radius measurements from the model that does not include $l_3$, but the age estimates of the two stars are only consistent at the $\sim 1.5\sigma$ level. In both cases, the MIST evolutionary tracks suggest that the primary likely filled its Roche-lobe when it was first expanding on the RGB. The filling factor of the primary reached a maximum $f\approx1.1$ and was $f>1.0$ for $\sim 0.75$~Myr.

Finally, \shortname{} was detected by \textit{Chandra} with an X-ray-optical source separation of $0\farcs3$ \citep{Evans10} and it is classified as a high-mass star with classification probability $P_{\rm{class}}=0.5$ in \citet{Yang22} based on a combination of X-ray features and optical/near-IR photometry. Unlike the other X-ray sources detected by \textit{eROSITA}, \shortname{} does not show any large-amplitude, long-term variability in the full ASAS-SN light curve (Figure \ref{fig:unfolded_lightcurves}). 

\subsection{J2236: GDR3 2002164086682203904} \label{sec:2002164086682203904}
\renewcommand{\source}{2002164086682203904}
\renewcommand{\shortname}{J2236}

\setlength{\fboxsep}{1.5pt} 
\setlength{\fboxrule}{1pt} 
\newcommand{\xpos}{7.6}
\newcommand{\ypos}{96}
\begin{figure*}[ht]
    \centering
    \begin{tabular}{cc}
        \begin{overpic}[width=0.5\textwidth]{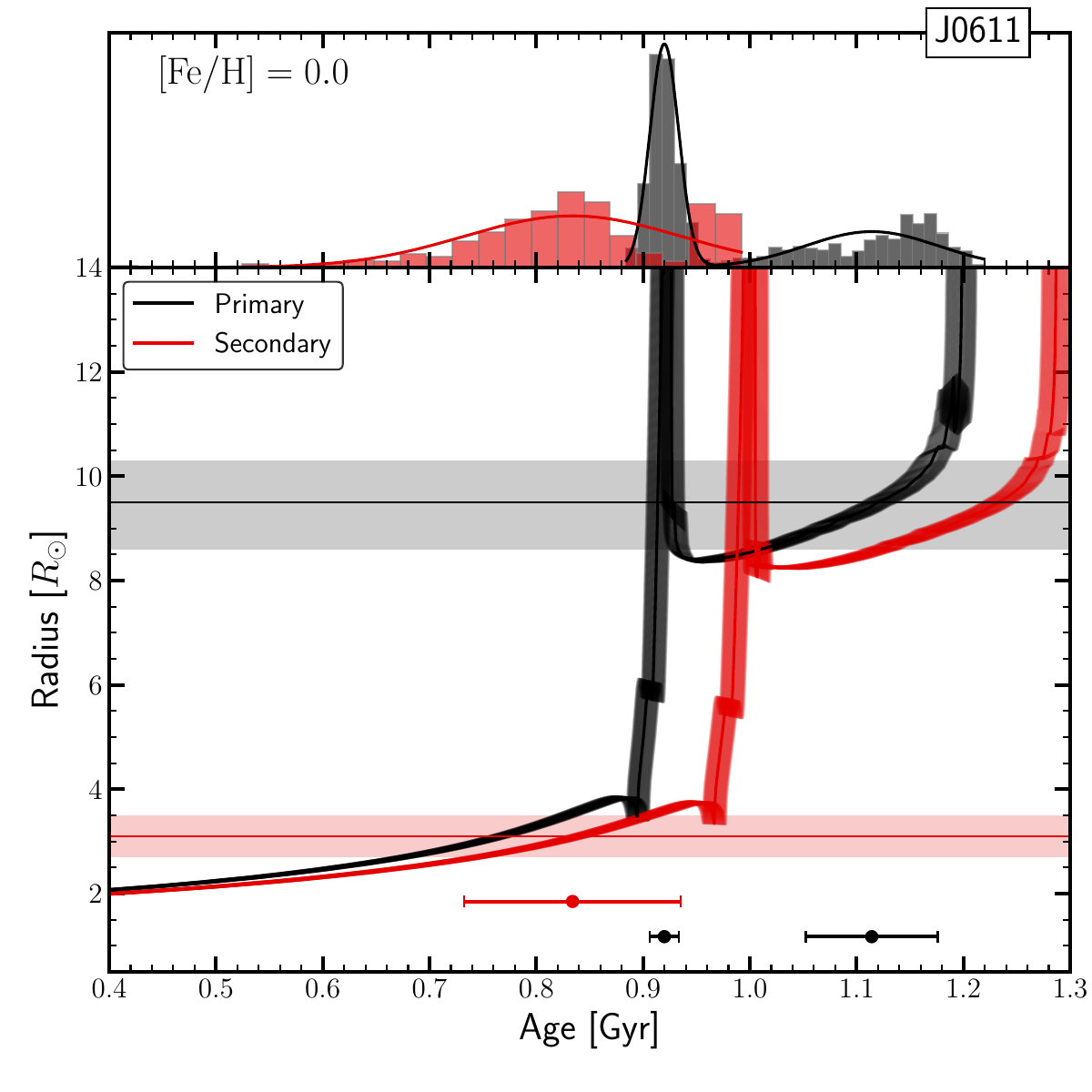} 
            \put(\xpos, \ypos){\Large\fcolorbox{black}{white}{\textbf{a}}}
        \end{overpic} & 
        \begin{overpic}[width=0.5\textwidth]{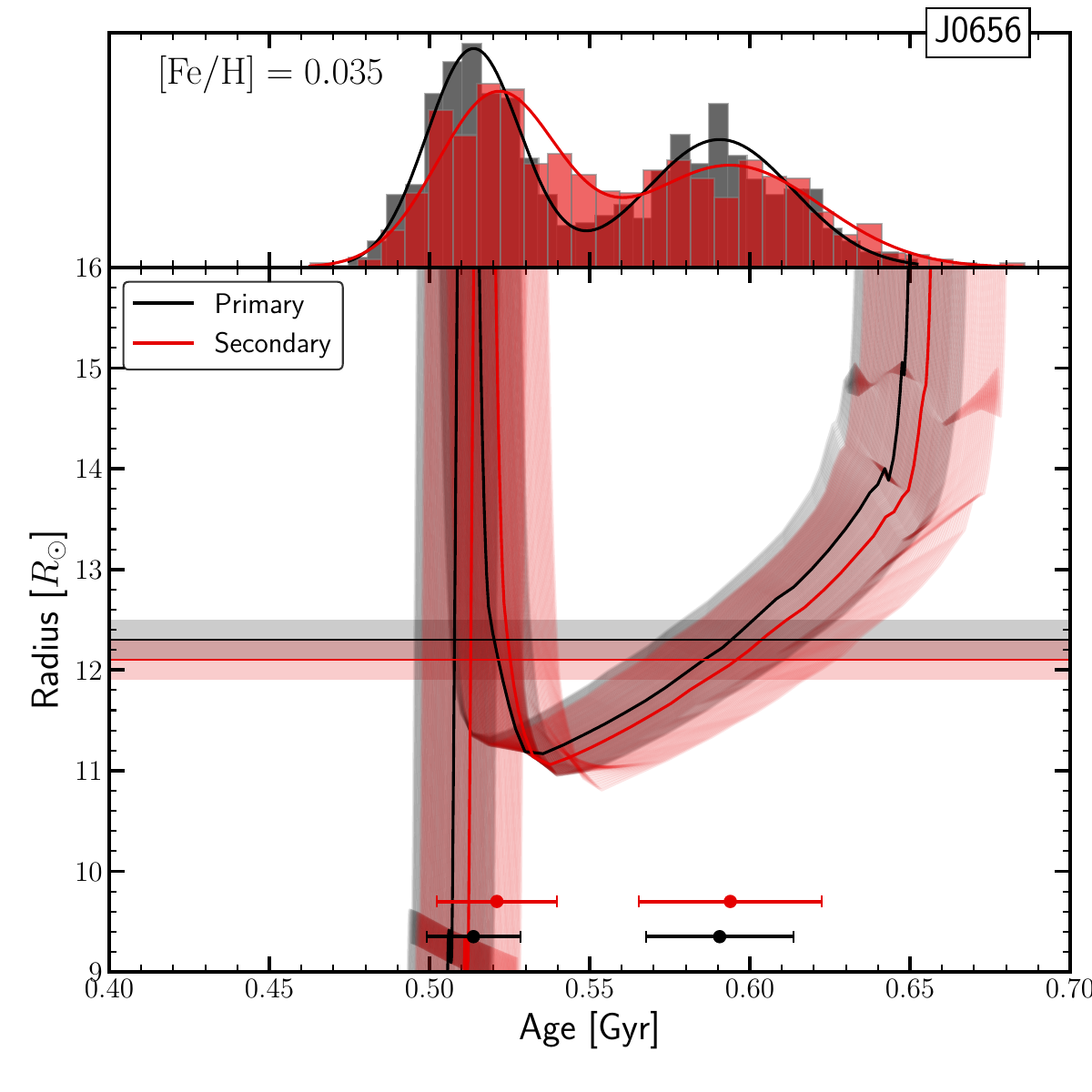}
            \put(\xpos, \ypos){\Large\fcolorbox{black}{white}{\textbf{b}}}
        \end{overpic} \\
        \begin{overpic}[width=0.5\textwidth]{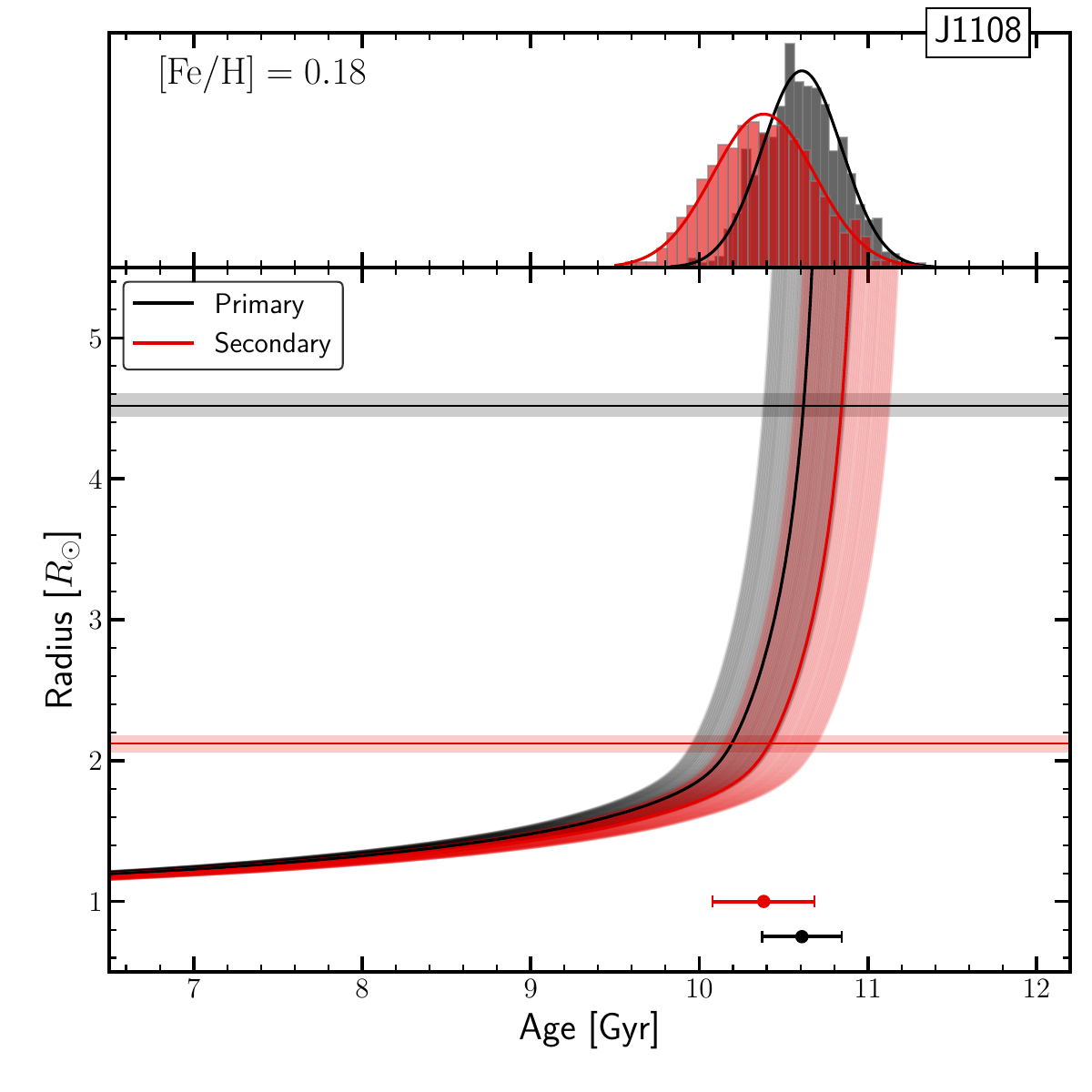}
            \put(\xpos, \ypos){\Large\fcolorbox{black}{white}{\textbf{c}}}
        \end{overpic} &
        \begin{overpic}[width=0.5\textwidth]{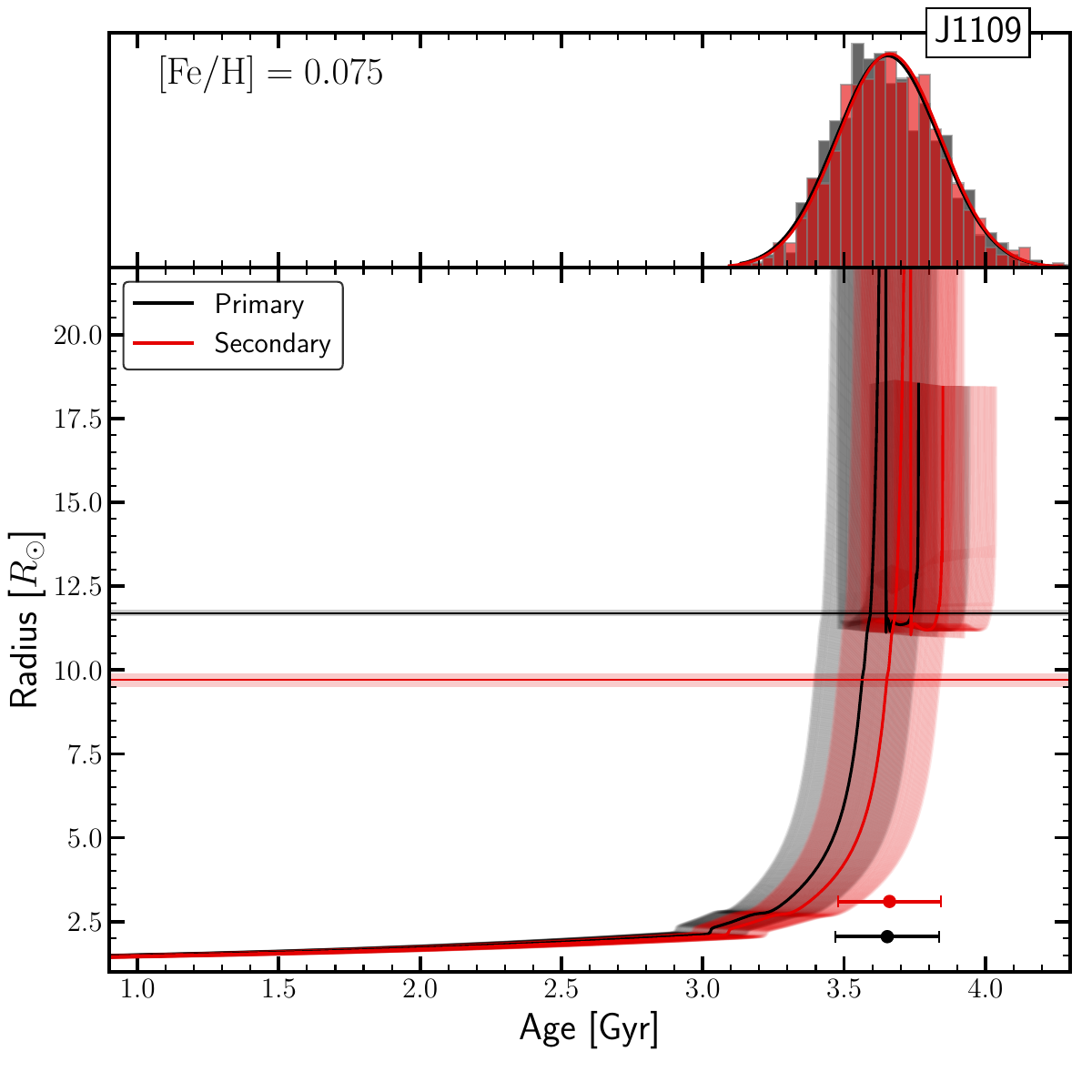}
            \put(\xpos, \ypos){\Large\fcolorbox{black}{white}{\textbf{d}}}
        \end{overpic} \\
    \end{tabular}
    \caption{MIST evolutionary tracks for the first four binary systems. In each panel we show the evolutionary tracks for the primary (black) and secondary (red). The horizontal bands show the radius measurements from the \PHOEBE{} models. The smaller upper panels show the age posteriors determined from sampling over our mass and radius posteriors for each component. We fit a one or two component Gaussian model to estimate the age of each component given our mass and radius measurements. For systems with two components in the age posteriors, the first corresponds to the first ascent of the RGB and the second corresponds to the core He-burning stage. Even though the amplitude of the latter is smaller, the similar area under each curve indicates that the two evolutionary states are equally likely. The mean and standard deviation of these Gaussians are shown in the larger panels. Ages where the red and black age posteriors overlap indicate allowed ages for the binary system. For all systems we find the ages agree within $1\sigma$, and in some cases (J0656 and J2236) there are two possible ages corresponding to the first ascent of the RGB and the core He-burning stage. }
    \label{fig:evolutionary_tracks}
\end{figure*}

\begin{figure*}[ht]
    \centering
    \begin{tabular}{cc}
        \begin{overpic}[width=0.5\textwidth]{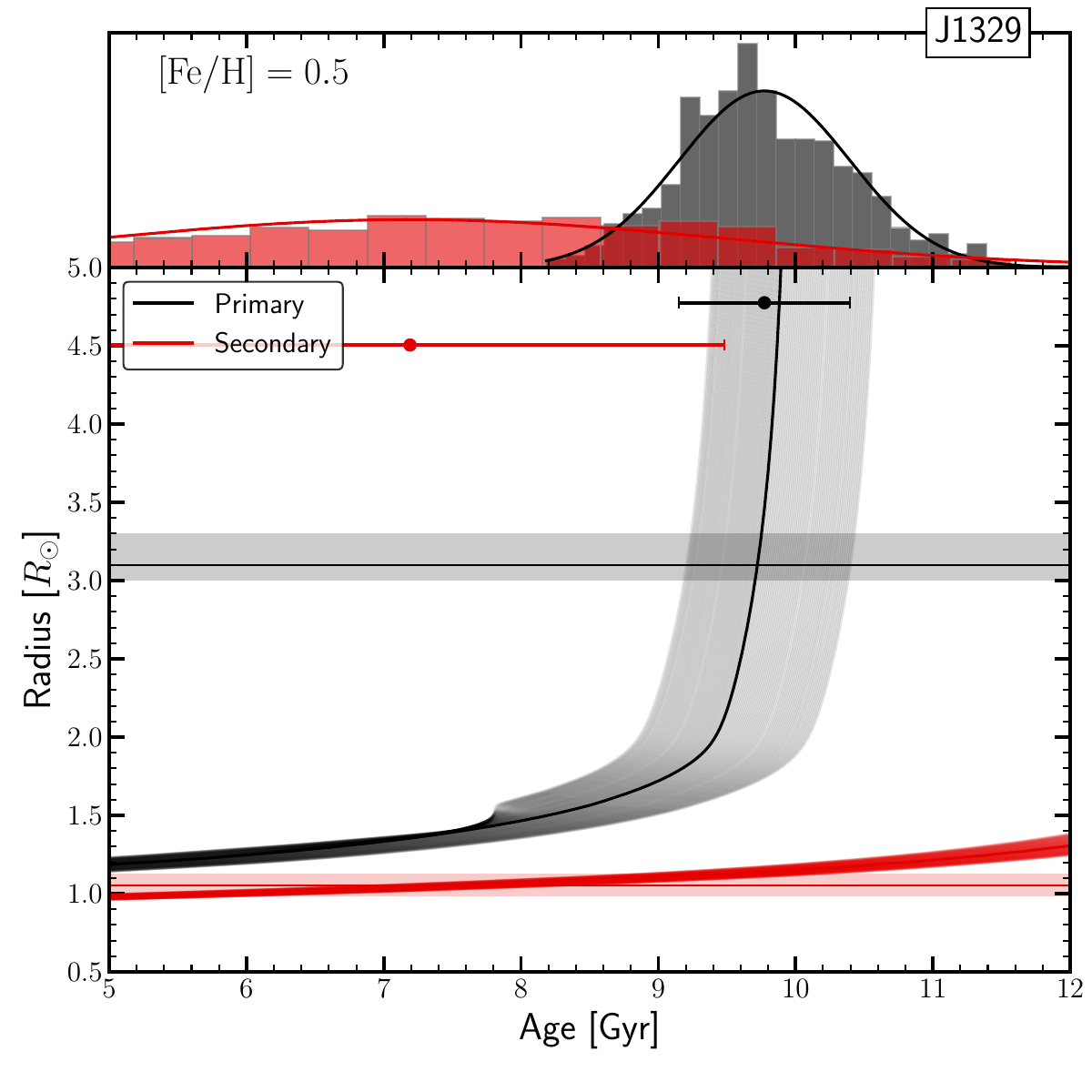} 
            \put(\xpos, \ypos){\Large\fcolorbox{black}{white}{\textbf{a}}}
        \end{overpic} & 
        \begin{overpic}[width=0.5\textwidth]{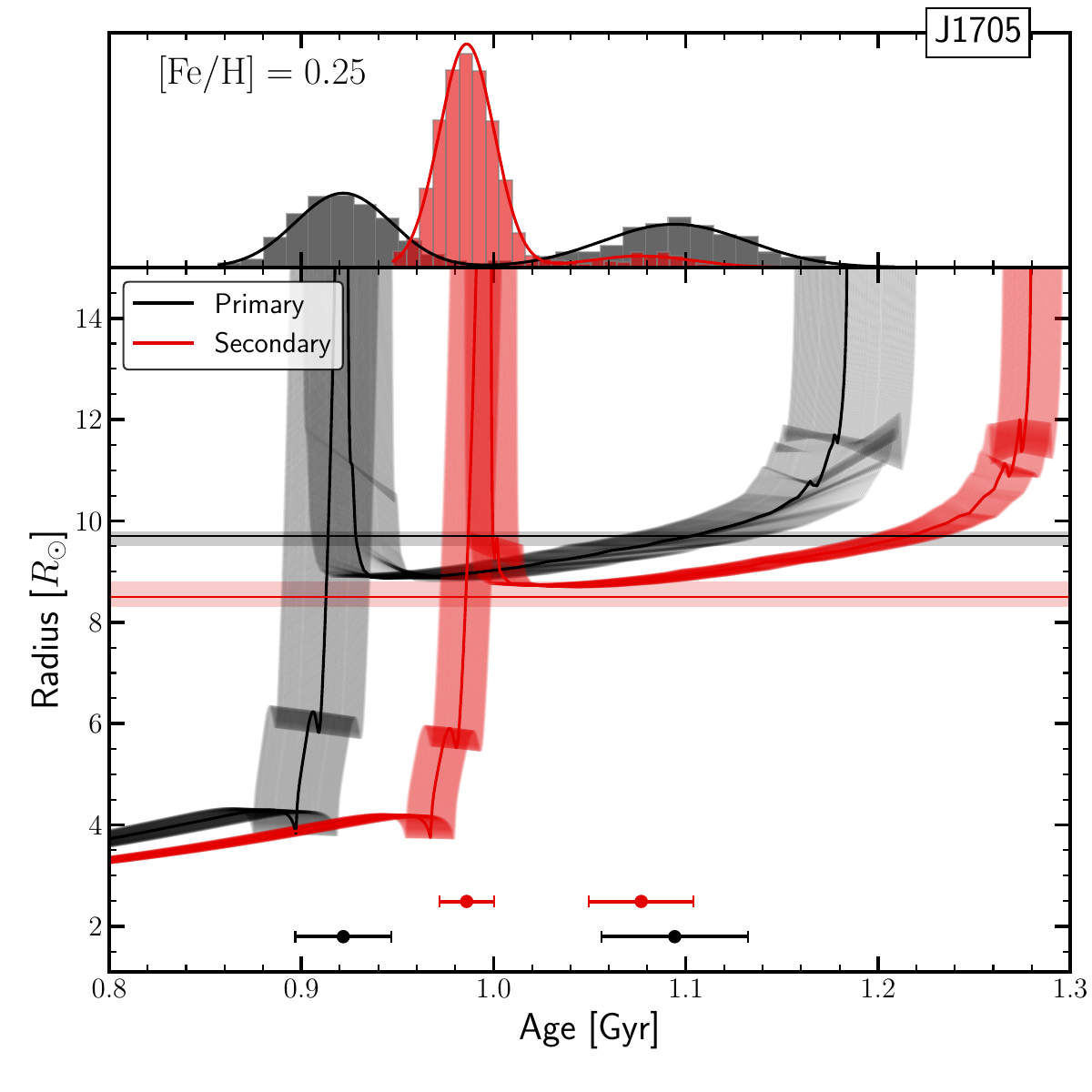}
            \put(\xpos, \ypos){\Large\fcolorbox{black}{white}{\textbf{b}}}
        \end{overpic} \\
        \begin{overpic}[width=0.5\textwidth]{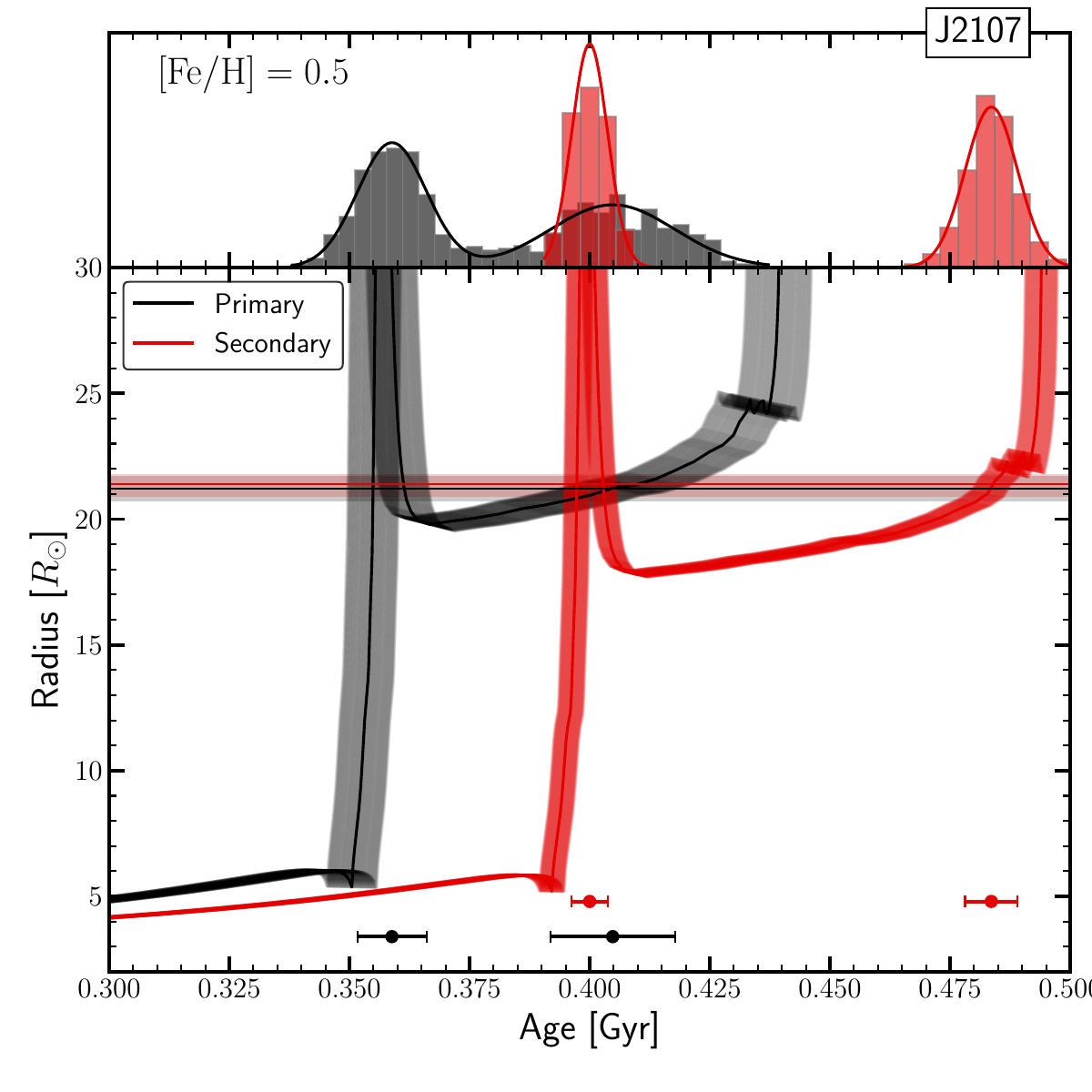}
            \put(\xpos, \ypos){\Large\fcolorbox{black}{white}{\textbf{c}}}
        \end{overpic} &
        \begin{overpic}[width=0.5\textwidth]{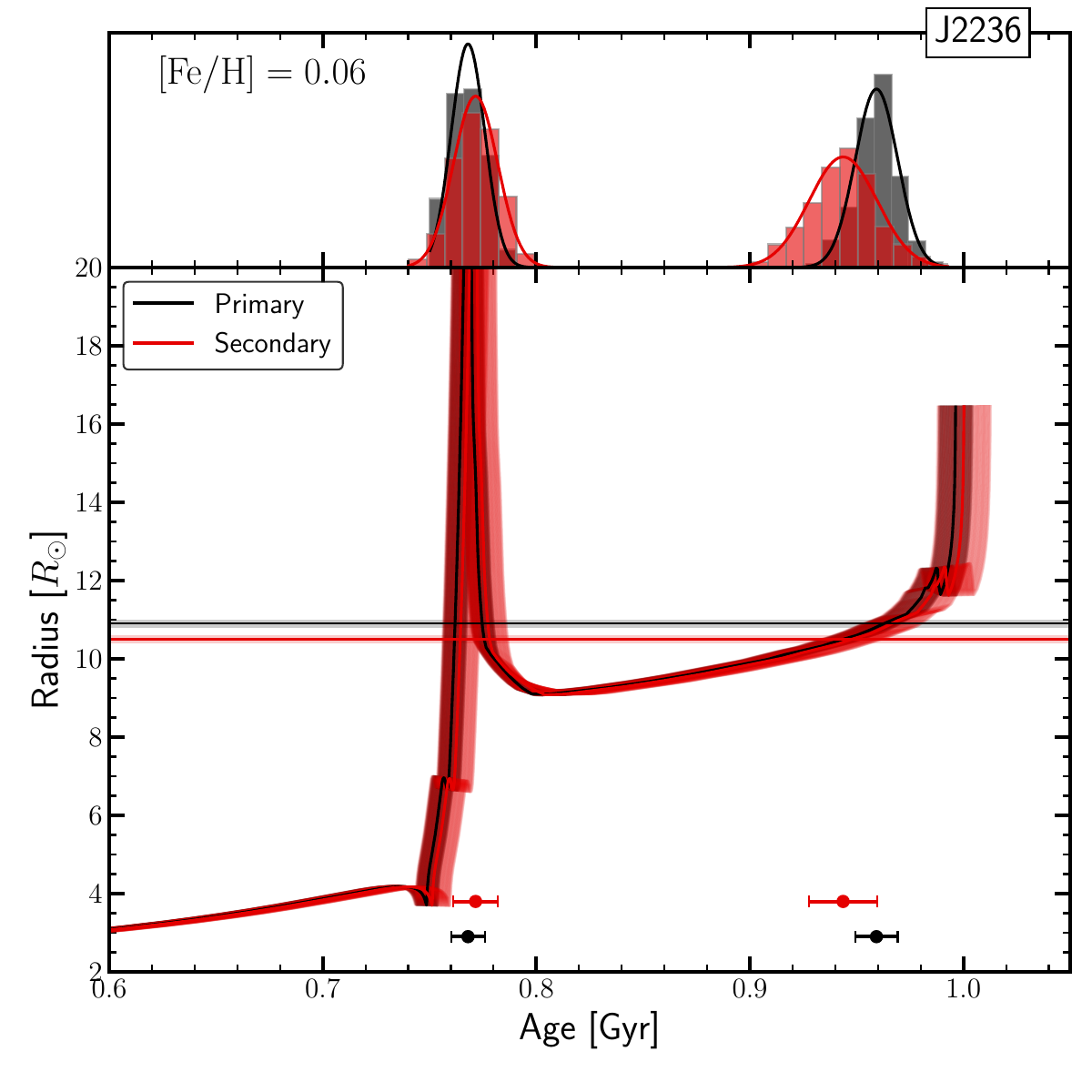}
            \put(\xpos, \ypos){\Large\fcolorbox{black}{white}{\textbf{d}}}
        \end{overpic} \\
    \end{tabular}
    \caption{Same as Figure \ref{fig:evolutionary_tracks} but for J1329, J1705, J2107, and J2236.}
    \label{fig:evolutionary_tracks2}
\end{figure*}

Finally, \shortname{} (AK Lac) is a twin system with nearly equal mass components of \primarymass{\source} and \mbox{\secondarymass{\source}}. The radii differ by $\sim 0.4\ R_\odot$, and the temperature ratio is consistent with unity. This is also the most circular orbit, with $e=0.0007^{+0.0003}_{-0.0002}$. 

The fractional third light is $l_3=0.09\pm0.01$. There are three nearby stars separated by $<10\farcs0$, but all have $g>18.8$~mag and are unlikely to contaminate the ASAS-SN flux, so the third light is likely from the sky background.

\shortname{} (TIC 428064231) was observed in six \TESS{} sectors and Figure \ref{fig:tglc_tess} shows the {\tt TGLC} light curve for Sector 76, which includes both eclipses. The detrended \TESS{} light curve (Figure \ref{fig:phoebe_models_tess}) shows eclipses with a fractional depth of $0.5$ in flux, compared to $0.42$ in ASAS-SN (Figure \ref{fig:phoebe_models}). As a result, the \PHOEBE{} model that includes the \TESS{} light curve prefers a completely edge-on inclination, $i=90.00^{\circ}\pm0.02^{\circ}$. The slightly higher inclination pushes the masses to decrease and both radii to increase. This total eclipse indicated by the \TESS{} light curve also requires a much higher third light in the $g$-band to effectively dilute the flux and produce the observed eclipse. 

There are large, correlated residuals in the \TESS{} light curve model of \shortname{}, but we are unable to find a better solution with \PHOEBE{}. It is possible that systematic effects in the light curve pipeline or detrending procedure have altered the shape of the eclipse, producing nearly symmetric residuals on each side of the minima.

Figure \ref{fig:evolutionary_tracks2}d shows the evolutionary tracks of J2236. The measured radii are consistent with these stars either being first ascent RGB stars or core He-burning stars. If the stars are core He-burning, they likely filled their Roche-lobes when they expanded up the RGB. The MIST evolutionary tracks indicate that they reached maximum Roche-lobe filling factors of $f\approx 1.09$ and had $f>1$ for $\sim 0.77$~Myr. Binary stellar evolution models could be used to determine how much mass transfer could have occurred in this system and how the subsequent evolution differs from standard single-star evolution.

\section{Discussion and Conclusions} \label{sec:discussion}

Here we have characterized eight eclipsing binary systems on the giant branch selected from the ASAS-SN eclipsing binaries catalog (Figure \ref{fig:cmd}; \citetalias{Rowan22}). We use PEPSI, APF, and CHIRON to obtain multi-epoch spectra of these targets and then use a combination of spectral disentangling (Figure \ref{fig:spec_disentangling}) and two-dimensional cross-correlations (Figure \ref{fig:todcor_example}) to measure the velocities of both stellar components. 

We then use \PHOEBE{} to simultaneously fit the ASAS-SN light curves and the radial velocities to measure both masses and radii with fractional uncertainties of $\lesssim 3\%$. For four systems, we also model the \TESS{} light curves (Figures \ref{fig:tglc_tess} and \ref{fig:phoebe_models_tess}), which improves the constraints on the stellar radii by more than a factor of two (Table \ref{tab:phoebe_table_tess}). The \TESS{} light curves do introduce additional challenges. For J1108, J1109, and J2236, we find correlated residuals from the \PHOEBE{} light curve model. For the J1108, this could be due to spots which produce an asymmetric eclipse shape. For J1109 and J2236, the eclipses last for $\gtrsim 20\%$ of the \TESS{} Sector, which could introduce systematic errors from the detrending process. We report our final mass and radius measurements for all systems from the ASAS-SN$+$RV model that includes the fractional third light parameter. 

Out of our eight systems, six are on circular orbits. Both of the eccentric systems are lower mass (Figure \ref{fig:mass_radius}), which could reflect the difference in tidal circularization timescales between stars with convective and radiative envelopes. Three of our systems (Sections \S\ref{sec:5388654952421552768}, \S\ref{sec:5347923063144824448}, and \S\ref{sec:5966976692576953216}) also show evidence for chromospheric activity in the ASAS-SN light curves (Figure \ref{fig:unfolded_lightcurves}). All three are detected \textit{eROSITA} X-ray observations, and two have chromospheric emission line features in RAVE. For J1109, we also see evidence for asymmetry during the eclipse in the \TESS{} light curve, which we attribute to star spots. We report the projected rotational velocity measured from the disentangled spectra in Table \ref{tab:template_table}. For the six binaries on circular orbits, we find that the projected rotational velocity of our fit RV templates is consistent with $2\pi R/P$ within $\sim 5$~km/s, so the binaries are likely tidally locked. For the two eccentric systems, J1108 and J1329, the measured $v\sin i$ is $\sim 10$--15~km/s larger than $2\pi R/P$, and these systems are therefore neither synchronized nor circularized. 

\begin{figure}
    \centering
    \includegraphics[width=\linewidth]{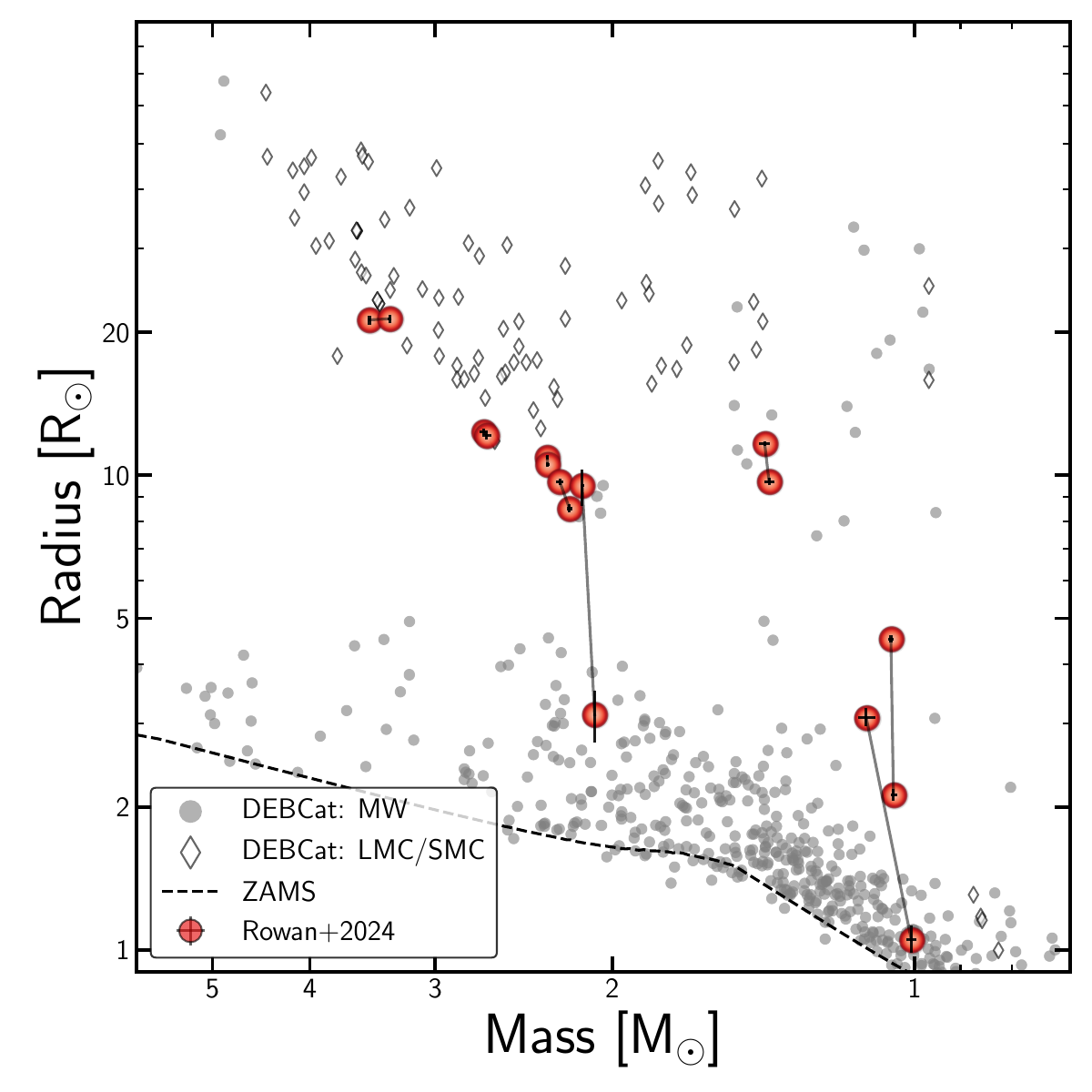}
    \caption{Same as Figure \ref{fig:mass_radius}, but showing the DEBCat sample of EBs \citep{Southworth15} for comparison. Targets in the Magellanic Clouds are marked as diamonds, and Milky Way targets are circles. }
    \label{fig:mass_radius_debcat}
\end{figure}

Figure \ref{fig:mass_radius} shows our mass and radius measurements compared to the \citet{Torres10} catalog. The majority of our systems have evolved substantially off of the main sequence, and only one binary (J1329, Section \S\ref{sec:6188279177469245952}) has a component still firmly on the main sequence. Since the release of the catalog in \citet{Torres10}, the majority of dynamical measurements of evolved EBs have come from the Magellanic Clouds. Figure \ref{fig:mass_radius_debcat} shows our systems compared to the updated sample from \citet{Southworth15}. The vast majority of the evolved stars are in the LMC/SMC \citep{Pietrzynski11, Graczyk12, Pietrzynski13, Pilecki13, Graczyk14, Gieren15, Graczyk18, Suchomska19, Suchomska22}. These binaries were primarily targeted to determine precise distances to the LMC/SMC, but are also probes of stellar evolution at lower metallicities. However, measurements of stellar parameters at a range of metallicities are needed to make complete comparisons to stellar theory. Figure \ref{fig:mass_radius_debcat} shows how our measurements probe a new part of the parameter space, not only for evolved stars at near-Solar metallicities, but also at smaller stellar radii. The \citet{Southworth15} catalog also includes Galactic eclipsing red giants identified in \textit{Kepler} \citep{Helminiak15, Gaulme16, Brogaard18, Themebl18}, some of which have been asteroseismically characterized as well. The \textit{Kepler} eclipsing red giants are mostly less massive ($\sim 1$--$1.5\ M_\odot$) than our sample (Figure \ref{fig:mass_radius_debcat}). 

We compare our mass and radius measurements to MIST evolutionary tracks in Figures \ref{fig:evolutionary_tracks} and \ref{fig:evolutionary_tracks2}. For all systems we find that the ages predicted from evolutionary tracks given our mass and radius measurements agree for both components. In some cases, we can distinguish between systems that are core-helium burning based on the measured radii. Since the exact ages depend strongly on the metallicity of the targets, which we only roughly estimate here for the purpose of RV template determination, we do not report our age posteriors and leave more detailed comparisons to theoretical models to future study. Two of our systems (J1108 and J1329) must be at least $\gtrsim 8$~Gyr old, given that they are low mass ($\sim 1\ M_\odot$) and have evolved off the main sequence. We use {\tt Banyan$\Sigma$} \citep{Gagne18} and the criteria for thin disk, thick disk, and halo membership from \citet{Ramirez07} to compute membership probabilities for J1108 and J1329. Both systems are consistent with being part of the Galactic thin disk with probabilities $>99\%$.

The evolutionary tracks of four of our systems (J0656, J1705, J2107, and J2236) show that one or both binary components could be core He-burning stars or first ascent RGB stars. If these stars are core He-burning, their radii were larger earlier in their evolution. Therefore, even if the systems are all observed as detached, non-interacting binaries now, they could have undergone mass transfer in the past. We use the MIST evolutionary tracks and the Eggleton approximation for Roche-lobe radii to determine if these stars could have filled their Roche-lobes and transferred mass. We find that three systems (J0656, J2107, and J2236) likely interacted in the past if they are currently core-He burning stars rather than first ascent RGB stars. We estimate the amount of time the stars had Roche-lobe filling factors $f>1$ and find it was likely brief ($\sim 0.3$--$0.8$~Myr). More detailed binary evolution models would be needed to see how possible mass transfer could have altered the evolutionary pathways of the stars in these systems. Additional age estimates from asteroseismology could also be useful to independently determine whether these stars have started He-burning. For example, red clump stars can be discriminated from RGB stars based on the period spacing and frequency spacing measured from asteroseismology \citep{Bedding11, Mosser14}.

We are continuing to target more ASAS-SN eclipsing binaries to expand the sample of evolved stars with mass and radius measurements. Large spectroscopic surveys can also be used to characterize eclipsing binaries and measure masses and radii on a large scale. Spectra from the Apache Point Observatory Galaxy Evolution Explorer \citep[APOGEE,][]{Majewski17} have been used to identify $>$7,000 SB2s \citep{Kounkel21}, and with future data releases from Milky Way Mapper \citep{Kollmeier17}, we can expect to have enough epochs ($>6$) for tens to hundreds of eclipsing binaries. \Gaia{} DR3 also includes $\sim 5000$ RV orbit models for SB2s, and the next data release is expected to include epoch RV measurements, allowing for simultaneous fitting of the \Gaia{} RVs and ASAS-SN and \TESS{} photometry. The number of detached EBs with precise mass and radius measurements is small relative to their importance as direct calibrators of stellar models. Large photometric and spectroscopic surveys are a promising path to expand this sample considerably, especially for parts of the parameter space where few previous measurements exist.

\section*{Acknowledgements}

We thank the anonymous reviewer for their helpful comments which improved the quality of this paper. We thank Jack Roberts, Casey Lam, and Marc Pinsonneault for helpful discussions. DMR is supported by the OSU Presidential Fellowship. CSK and KZS are supported by NSF grants AST-1907570, 2307385, and 2407206. 

We thank Las Cumbres Observatory and its staff for their continued support of ASAS-SN. ASAS-SN is funded by Gordon and Betty Moore Foundation grants GBMF5490 and GBMF10501 and the Alfred P. Sloan Foundation grant G-2021-14192. 

This paper includes data collected by the TESS mission. Funding for the TESS mission is provided by the NASA's Science Mission Directorate. 

This work presents results from the European Space Agency space mission Gaia. Gaia data are being processed by the Gaia Data Processing and Analysis Consortium (DPAC). Funding for the DPAC is provided by national institutions, in particular the institutions participating in the Gaia MultiLateral Agreement.

The LBT is an international collaboration among institutions in the United States, Italy, and Germany. LBT Corporation partners are: The University of Arizona on behalf of the Arizona Board of Regents; Istituto Nazionale di Astrofisica, Italy; LBT Beteiligungsgesellschaft, Germany, representing the Max-Planck Society, The Leibniz Institute for Astrophysics Potsdam, and Heidelberg University; The Ohio State University, representing OSU, University of Notre Dame, University of Minnesota, and University of Virginia. PEPSI was made possible by funding through the State of Brandenburg (MWFK) and the German Federal Ministry of Education and Research (BMBF) through their Verbundforschung grants 05AL2BA1/3 and 05A08BAC. 

\bibliographystyle{mnras}
\bibliography{rgebs} 

\appendix

\section{Appendix: Single-Lined Spectroscopic Binaries} \label{sec:sb1s}

\begin{figure}
    \centering
    \includegraphics[width=\linewidth]{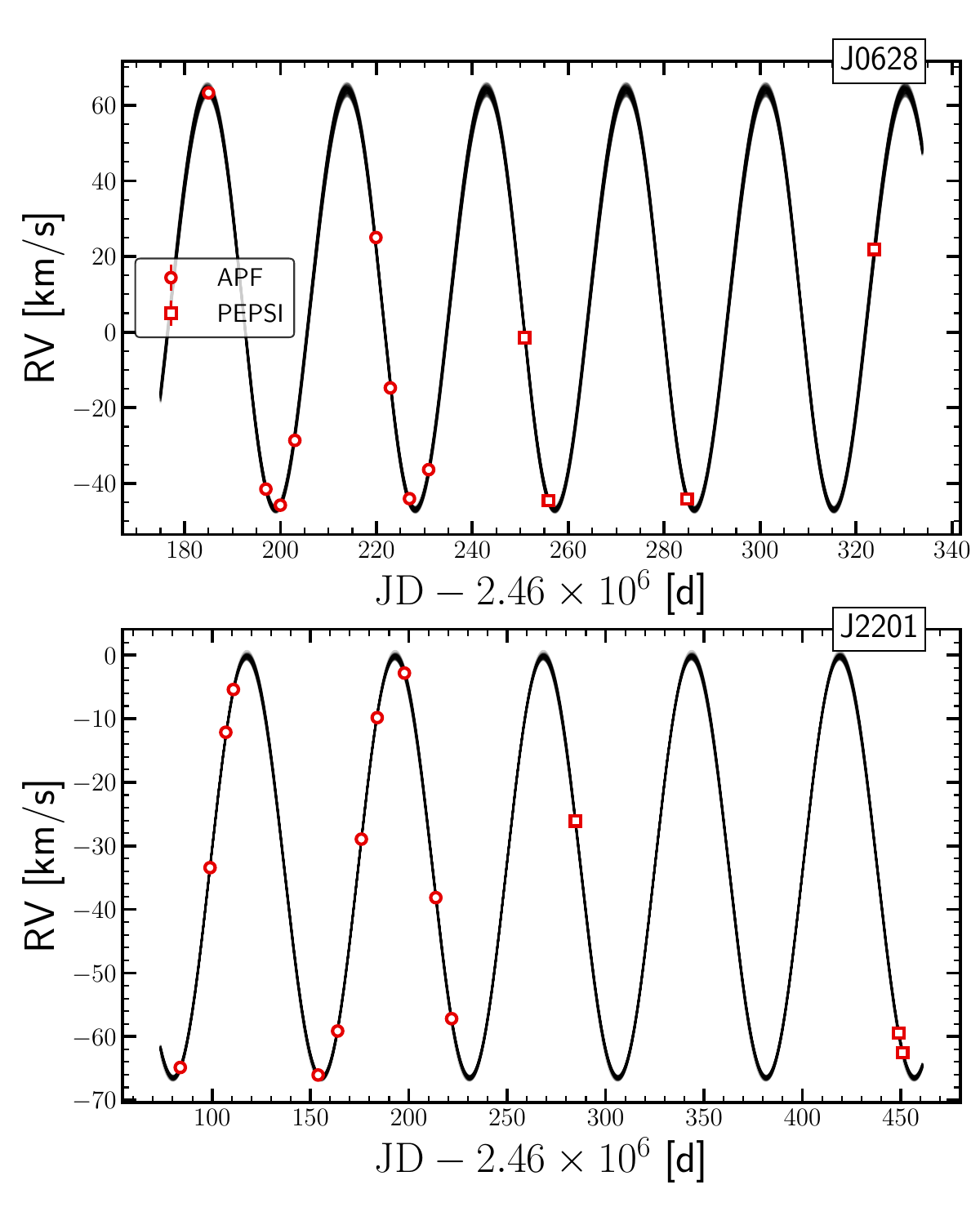}
    \caption{RV orbits for the two EBs with identified as single-lined spectroscopic binaries.}
    \label{fig:sb1_orbits}
\end{figure}

In addition to the eight systems identified as SB2s, we observed two systems that were SB1s. In order to measure masses and radii, the velocities of both components must be measured so the mass ratio can be determined directly. Table \ref{tab:sb1s} reports the parameters of these systems and Figure \ref{fig:sb1_orbits} shows their RV orbits. We also report the mass functions
\begin{equation} \label{eqn:massfunction}
    f(M) = \frac{P K^3}{2\pi G}\left(1-e^2\right)^{3/2} = \frac{M_2^3 \sin^3(i)}{(M_1+M_2)^2},
\end{equation}
\noindent where $P$ is the orbital period, $K$ is the velocity semiamplitude, and $e$ is the orbital eccentricity. The mass function represents the minimum mass of the unseen secondary star. If we estimate that both systems have photometric primaries with masses $\sim 1.5\ M_\odot$ and edge-on orbital inclinations, the companions would have to be $\sim 1.77$ and $\sim 1.3\ M_\odot$ for J0628 and J2201, respectively. Both of these systems are slightly bluer than the SB2s on the color-magnitude diagram (Figure \ref{fig:cmd}). The ratio of effective temperatures are \teffratio{}$=0.42$ and \teffratio{}$=0.68$ for J0628 and J2201, respectively \citepalias{Rowan22}, suggesting that the companions are cooler main sequence stars and the flux ratio $F_2/F_1 \ll 1$. While it may be possible to identify the spectral signatures of the companion with more careful disentangling or more spectra, we focus our analysis here on the most clear SB2 systems. 

\newpage
\begin{table}
    \centering
    \caption{Same as Table \ref{tab:summary_table}, but for the two single-lined spectroscopic binaries in our sample.}
    \sisetup{table-auto-round,
     group-digits=false}
    \setlength{\tabcolsep}{10pt}
    \renewcommand{\arraystretch}{1.2}
    \begin{center}
        \begin{tabular}{lrr}
\toprule
{} & 3324223082726657920 & 1976106210861223424 \\
\midrule
Short Name         &               J0628 &               J2201 \\
RA (deg)           &             97.0590 &            330.4318 \\
DEC (deg)          &              6.1262 &             47.6866 \\
$G$ (mag)          &                12.0 &                11.0 \\
Period (d)         &            29.08895 &            75.33522 \\
$f(M)\ [M_\odot]$  &       $0.52\pm0.01$ &     $0.284\pm0.003$ \\
$N_{\rm{RV}}$      &                12.0 &                14.0 \\
Distance (pc)      &              1637.1 &              1537.0 \\
$M_G$ (mag)        &                -0.5 &                -0.4 \\
\textit{TESS}      &              \xmark &              \xmark \\
\bottomrule
\end{tabular}

    \end{center}
    \label{tab:sb1s}
\end{table}

\label{lastpage}

\end{document}